\documentclass[reprint,superscriptaddress,onecolumn,aps,pre]{revtex4-2}
\usepackage{float}
\usepackage{amsmath}
\usepackage{amssymb}
\usepackage{subcaption}
\usepackage{siunitx}
\usepackage{xcolor}
\usepackage{hyperref}
\usepackage{graphicx}
\DeclareMathOperator{\erf}{erf}
\DeclareMathOperator{\erfc}{erfc}

\begin{document}

\title{Lagrangian flow statistics in experimental homogeneous isotropic turbulence}

\author{Cheng Wang}
\affiliation{Ens de Lyon, CNRS, Laboratoire de physique, UMR 5672, F-69342 Lyon, France}

\author{Sander G. Huisman}
\affiliation{Physics of Fluids Department, J.M. Burgers Center for Fluid Dynamics, University of Twente, P.O. Box 217, 7500AE Enschede, The Netherlands}

\author{Thomas Basset}
\affiliation{Ens de Lyon, CNRS, Laboratoire de physique, UMR 5672, F-69342 Lyon, France}

\author{Romain Volk}
\author{Micka\"el Bourgoin}
\email{mickael.bourgoin@ens-lyon.fr}
\affiliation{Ens de Lyon, CNRS, Laboratoire de physique, UMR 5672, F-69342 Lyon, France}

\begin{abstract}
We report on Lagrangian flow statistics from experimental measurements of homogeneous isotropic turbulence. The investigated flow is driven by 12 impellers inside an icosahedral volume. Seven impeller rotation rates are considered resulting in seven Reynolds numbers with $205 \leq \text{Re}_\lambda \leq 602$. We perform high-speed imaging using 3 cameras to record a total of $8.2\times10^6$ frames, and using high resolution three dimensional (3D) particle tracking velocimetry position, velocity, and acceleration of particle tracks are obtained in the vicinity of the center of the device. From these tracks, we obtain the Eulerian and Lagrangian flow statistics, mainly based on second-order structure functions and autocorrelation functions. The universal constants $C_0^*$ (for the Lagrangian second order structure function), $C_\varepsilon$ (for the energy injection rate) and $a_0$ (for the acceleration fluctuations) are determined, as well as all the relevant Lagrangian and Eulerian flow time scales. Analytical relations between these constants and time scales are experimentally verified. 

\end{abstract}

\maketitle

\section{Introduction}
 The characterization of statistical properties of turbulence is a long-term undertaking of fluid mechanicists. Despite the existence of the Navier--Stokes equations for two centuries, the results directly extracted from these equations are few. Thus, alternative frameworks emerged to describe turbulence as a random process with specific statistics through space and time, e.g.~with the seminal works of Richardson \cite{richardson1922weather} and Kolmogorov \cite{kolmogorov1941local}. The concept of a turbulent cascade was introduced to describe how, in three dimensional (3D) turbulence, energy is injected at large scale through large eddies which break up into smaller and smaller eddies to reach the dissipative scales. Mainly based on dimensional analysis with the introduction of constants to determine, this cascade phenomenology was verified with experiments (see for example \cite{bourgoin2018investigation}) and simulations \cite{she1991structure}, with a large success for second- and third- order statistics in homogeneous isotropic turbulence (HIT).

Nevertheless, this approach was historically developed for a Eulerian description of turbulence. Experimentally, the development for one century of well-resolved anemometers such as Pitot tubes, hot wires \cite{comtebellot1976hotwire}, or more recently of optical velocimetry like particle image velocimetry (PIV) \cite{raffel2018piv,adrian2011particle}, has made it possible to accurately study turbulence from an Eulerian point of view. At the same time, numerous direct numerical simulations (DNS) of the Navier--Stokes equations have provided a characterization of fine structure of turbulence \cite{ishihara2007small,bib:khurshid2023_PRE}.

Pioneering work by Taylor \cite{taylor1922diffusion} and Richardson \cite{richardson1926diffusion} revealed the importance of a Lagrangian description of turbulence, in particular when it comes to addressing turbulent dispersion and transport properties. However, the Lagrangian characterization of turbulence has long remained more challenging than the Eulerian description. Experimentally, the optical tracking of particles appeared in the late eighties \cite{sato1987lagrangian, maas1993particle, malik1993particle, virant1997ptv}, and has known since then a constant growth and improvement, accelerated in the past decade by the rapid development of high speed / high resolution digital imaging, to the point that modern methods nowadays are able to track tens of thousand particles simultaneously \cite{schroder2023lagrangian}. However, such experiments remain quite complex and the main Lagrangian results of homogeneous isotropic turbulence were offered by highly resolved DNS \cite{yeung2002lagrangian, buaria2016lagrangian, ishihara2007small}.

{Besides the Lagrangian metrological difficulties, a fundamental} experimental limitation for the study of turbulence is the ability to generate experimentally a homogeneous and isotropic turbulent flow. This condition of homogeneity and isotropy is particularly critical for a Lagrangian study, because a Eulerian inhomogeneity directly implies a non stationary dynamics in a Lagrangian description. {Most of the reference fundamental knowledge for Lagrangian turbulence gathered so far in experiments has been obtained in non-homogeneous flows, particularly in so called von K\'arm\'an flows}, i.e.~flows between two counter-rotating disks in a closed vessel~\cite{voth2002measurement,mordant2001measurement,ouellette2006small,toschi2009lagrangian}. {Such closed flows are particularly practical to deploy Lagrangian measurements (compared for instance to wind-tunnel experiments where the tracking cameras must move with the wind~\cite{ayyalasomayajula2006lagrangian})}. However, von K\'arm\'an flows are known to {exhibit quite specific properties, often far from ideal HIT, such as a persistent signature of the large scale anisotropy~\cite{ouellette2006small,huck2019lagrangian}, non-trivial inhomogeneity effects due to the presence of stagnation points~\cite{huck2017production} and helicity~\cite{angriman2021broken} and in many cases the existence of complex temporal instationarities~\cite{bib:deLaTorre2007_PRL,bib:machicoane2016_PRE}. These complexities make it quite difficult to unambiguously measure characteristic universal constants of the Lagrangian description of HIT and to explore their eventual trend with Reynolds number. For instance, Lagrangian statistics have been shown to exhibit a non-universal persistent signature across scales of the large scale anisotropy of the von K\'arm\'an flow~\cite{ouellette2006small}}

Recent alternative experimental set-ups have been developed to generate nearly HIT flows. A first family of set-ups is based on jet random forcing: the random-jet-stirred turbulence tank \cite{variano2008random, laplace2022phd} or random jet arrays \cite{carter2016generating}. They present large measurement volumes of tens of centimeters but residual mean flows. In the present study, we use a second family of set-ups based on a symmetrization of the von K\'arm\'an flow: the Lagrangian Exploration Module (LEM). A twenty-faced tank is stirred with twelve impellers which are positioned isotropically on the faces. This `generalized' von K\'arm\'an flow tends to an isotropic forcing and eliminates any preferential direction or region of the turbulent flow. Thus the flows exhibits a central region far from the impellers. The measurement volume is smaller, only a few centimeters, but no residual mean flow is observed, and turbulent fluctuations have been shown to reproduce HIT with high fidelity~\cite{zimmermann2010lagrangian}.

Based on Lagrangian experiments run in the LEM for seven different values of the Reynolds number, we propose an experimental Lagrangian characterization of HIT. Our goal is to produce a reference characterization for Lagrangian HIT for velocity and acceleration with a main focus on second-order statistics. Our data set is well resolved in time to extract precise statistics, and the Taylor-based Reynolds numbers ranging from 205 to 602 covers a wide enough range to address Reynolds number dependencies.

In section~\ref{sec:experimental_methods}, we describe in details the experimental set-up and the particle tracking. In section~\ref{sec:characterisation_flow}, we propose a classical Eulerian characterization to show that the generated turbulence is homogeneous and isotropic. We also provide classical Eulerian characterization of turbulence quantities such as the mean energy dissipation rate and the Eulerian integral scale. We are also taking particular care to account for any possible experimental and statistical bias. Finally, we propose a full Lagrangian characterization in section~\ref{sec:lem_lagrangian_stats} with one-point/two-time statistics for velocity and acceleration. Several relevant universal constants and time scales are presented and discussed for the seven Reynolds numbers considered. Main conclusions are summarized in section~\ref{sec:conclusion}.

\section{Experimental methods}\label{sec:experimental_methods}

\subsection{Experimental set-up}\label{subsec:setup}
\begin{figure}
\includegraphics[width=0.4\textwidth]{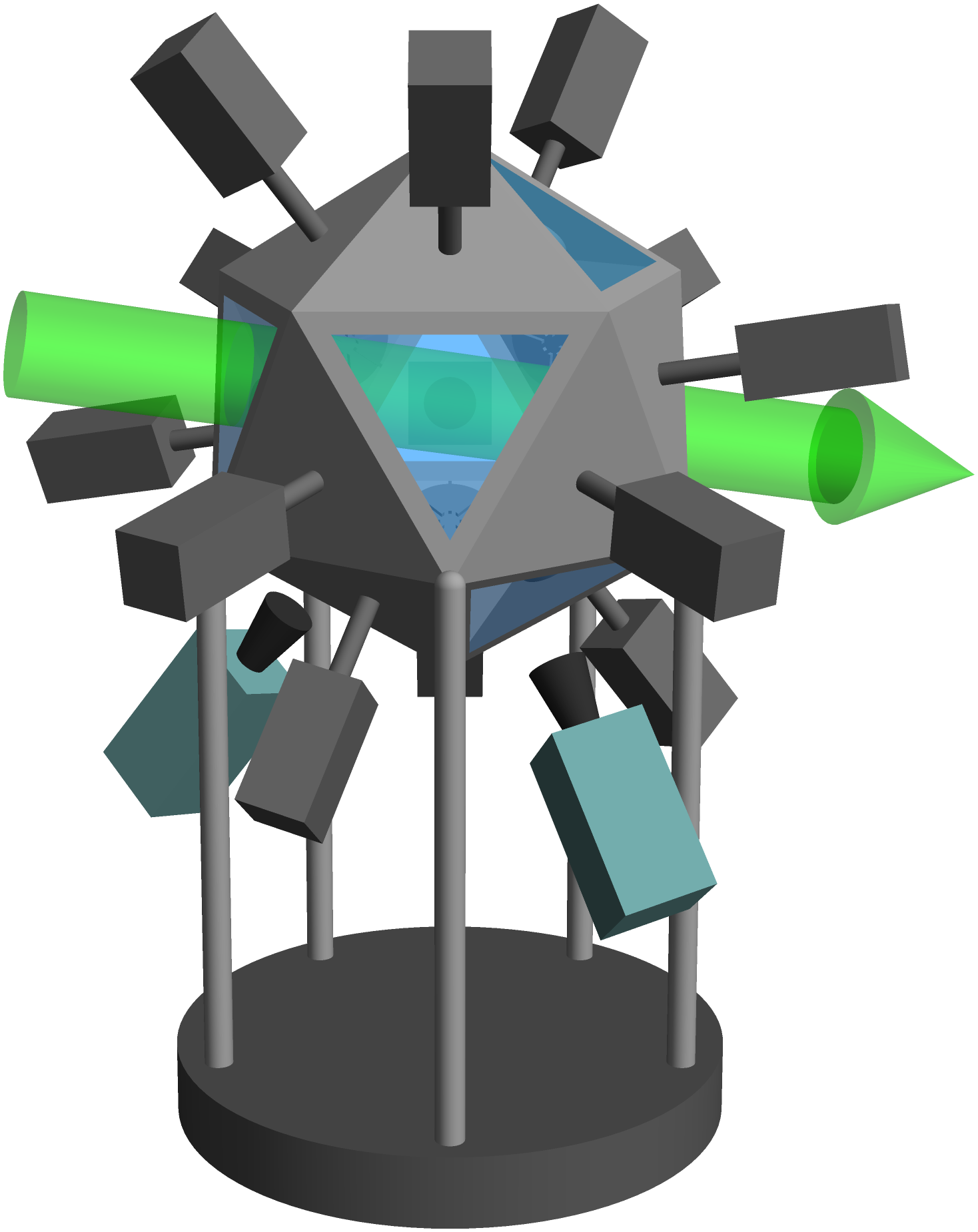}
\hspace{0.5cm}\raisebox{0.8\height}{\includegraphics[width=0.45\textwidth]{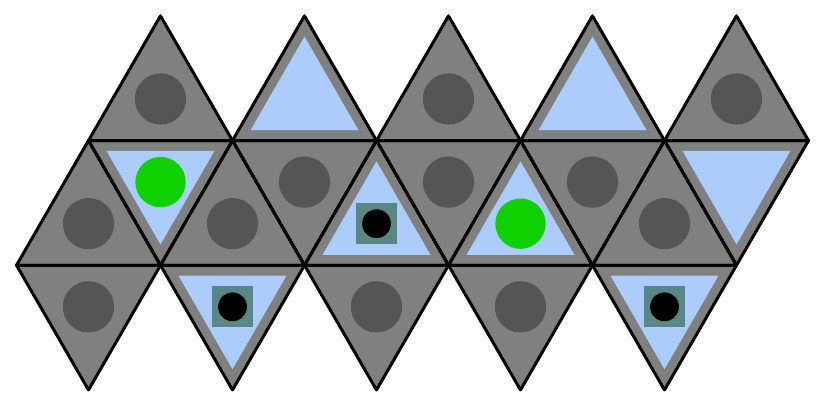}}
\caption{Left: Schematic of the LEM with 12 motors (dark gray) attached. The length of an edge is \SI{400}{mm}. The laser is represented by a green arrow and the three high-speed cameras are in gray-blue (two on the bottom and one on the opposite window). Right: Net of the (inner) faces of the polyhedron showing the impellers (gray discs), the windows (blue), the laser (green discs) and the cameras/objectives (blue squares with black discs inside).}
\label{fig:lem_setup}
\end{figure}

To generate a homogeneous isotropic turbulent flow, we use the LEM \cite{zimmermann2010lagrangian} at the \'Ecole Normale Supérieure de Lyon. The set-up is a convex regular icosahedral (20-faced polyhedron) vessel shown in figure~\ref{fig:lem_setup}, with an edge length of \SI{400}{mm} which results in a volume of around \SI{140}{L}. The vessel is filled with deionized and degassed water and the flow is stirred by impellers with a diameter of \SI{142.5}{mm} featuring 8 equally-spaced straight blades of \SI{5}{mm} width and \SI{7}{mm} height. The motors are mounted to the centers of 12 of the 20 faces (in the most symmetrical way possible), see figure~\ref{fig:lem_setup}. This set-up is similar to the one developed in G\"ottingen, Germany \cite{zimmermann2010lagrangian}, though the G\"ottingen version uses 12 propellers mounted to the corners of the vessel. The present set-up was previously used to study clustering of finite-size particles \cite{fiabane2012clustering}.

We operate the set-up by rotating all the 12 impellers in the same direction (such that impellers which face each other counter-rotate) and we vary their rotation frequency from \SI{1.8}{Hz} to \SI{11.9}{Hz}, resulting in Taylor-based Reynolds numbers of $205 \leq Re_\lambda \leq 602$ (determined in subsection~\ref{subsec:lem_eulerian_stats}). The flow is known to have excellent isotropy and homogeneity around the center of the volume \cite{zimmermann2010lagrangian}, and thus is capable of generating HIT. We will confirm these findings for the current LEM in subsection~\ref{subsec:homogeneity_isotropy}. Water is initially at ambient temperature around \SI{20}{\degree C} then warms up due to the rotation of the impellers during the experimental recordings, and stabilizes at an operating value of the order of \SI{25}{\degree C} or which the associated water kinematic viscosity is $\nu = \SI{0.9e-6}{m^2/s}$.

The flow is seeded with neutrally-buoyant spherical polyethylene tracers (Cospheric \texttt{UVPMS-BR-0.995}) with a density $\SI{0.985}{kg/m^3} \leq \rho_p \leq \SI{1.005}{kg/m^3}$ and a diameter $\SI{212}{\micro\meter} \leq d_p \leq \SI{250}{\micro\meter}$. These particles fluoresce in red under green (laser) illumination. The typical requirement to avoid finite size effects, especially for the accurate measurement of acceleration, is that $d_p \leq 5 \ell_K$ with $d_p$ the particle diameter and $\ell_K$ the Kolmogorov length scale \cite{voth2002measurement, qureshi2007turbulent, brown2009acceleration, calzavarini2009acceleration, volk2011dynamics}. This holds for all our cases, except for the most turbulent one ($\text{Re}_\lambda=602$) with $d_p \approx 6 \ell_K$ (see table~\ref{tab:lem_eulerian_parameters}). Thus, the particles should faithfully follow the flow and we expect to measure the relevant velocities and accelerations, even for small-scale Lagrangian dynamics.

\begin{figure}
\centerline{\includegraphics[width=\textwidth]{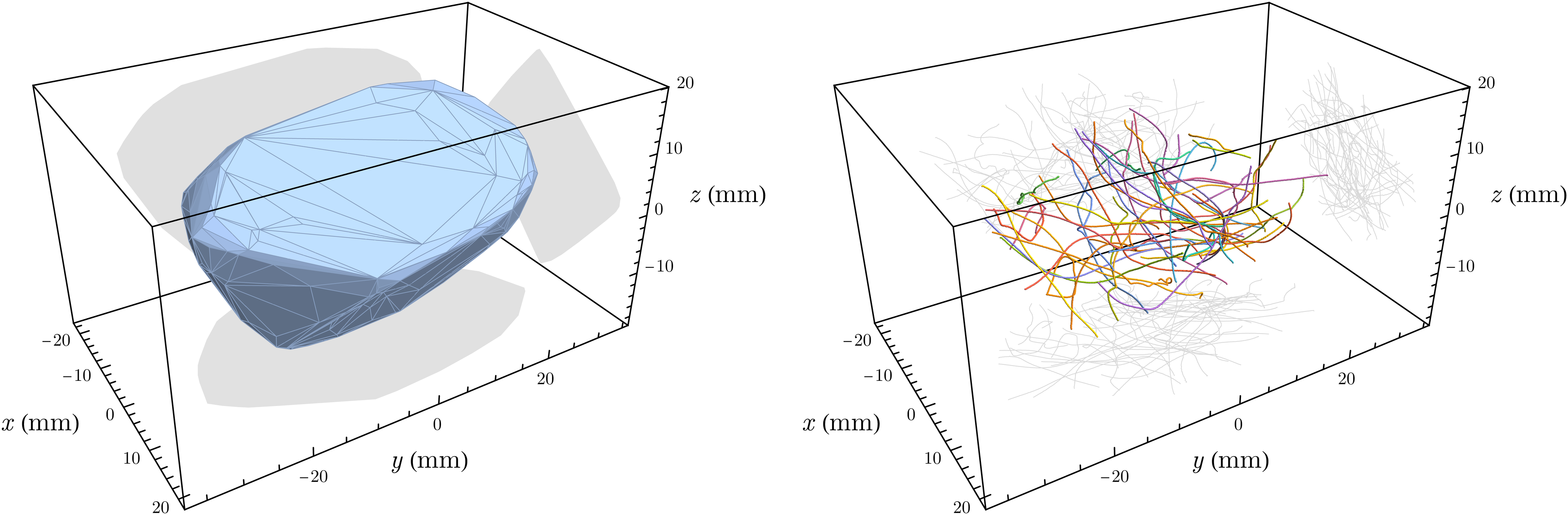}}
\caption{Left: Convex hull of the measurement volume in the LEM captured by the three-camera set-up. Right: Rendering of the 50 longest tracks of one recording (case 600). A total of $25 \times 10^3$ points are shown, where each point is drawn as a sphere with the diameter of the particle. `Shadows' of the convex hull and the trajectories are projected in gray to the bottom and back planes.}
\label{fig:hull_and_tracks}
\end{figure}

We illuminate the particles in the flow with a high-power pulsed laser (Quantronix Condor Dual, Nd:YAG, power \SI{180}{W}, wavelength \SI{532}{nm} (green), pulse width \SI{120}{ns}). The laser beam is expanded and collimated to approximately \SI{15}{cm} in diameter and aligned such that it enters the set-up perpendicular to one of the windows and passes through the center of the tank, as shown in figure~\ref{fig:lem_setup}. The particles are imaged by three high-speed cameras (Phantom V12, Vision Research Inc., $1280 \times 800$ pixels) mounted with \SI{100}{mm} macro objectives (Zeiss Milvus). From the icosahedral geometry we know that the cameras cannot be mounted perpendicular to each other, thus we choose a configuration to have the angles between the cameras around \SI{90}{\degree} (ideal for 3D-matching). The arrangement of the cameras and the laser beam gives a measurement volume of around $33 \times 57 \times \SI{37}{mm^3}$, as seen in figure~\ref{fig:hull_and_tracks}. Hence one pixel corresponds to roughly \SI{40}{\micro\meter}. We synchronize the three cameras and the laser to operate at \SI{3125}{Hz} for the lowest four rotation frequencies and \SI{6250}{Hz} for the highest three such that the sampling of the Kolmogorov time scale $\tau_\eta$ ranges from 97 down to 11 frames. For the seven rotation frequencies considered (detailed in the following), 40 independent recordings of 8000 frames are recorded (one case with 100 recordings).

\subsection{Processing}\label{subsec:processing}
We perform Lagrangian particle tracking (LPT) to obtain a Lagrangian description of the flow. The LPT algorithm utilized can be divided in three steps: particle detection, 3D reconstruction, and temporal tracking. The particle tracking source codes are based on our open source homemade toolbox \cite{githubptvlyon}. A brief description of the method is:

\begin{enumerate}
\itemsep0em 
\item Two-dimensional images are first analyzed to measure the positions of the centers of the particles. The particle detection procedure is an \textit{ad hoc} process in which local maxima are fitted in both the horizontal and vertical directions to obtain the centers of the particles with sub-pixel accuracy \cite{adrian2011particle, raffel2018piv}.
\item After the particle centers for all images and all cameras have been determined, the actual three-dimensional positions of the particles can be reconstructed, knowing that each camera image is a two-dimensional projection of the measurement volume. More typically, methods based on optical models are used to achieve real particle positions, but for this study a geometric method detailed in~\cite{machicoane2019simplified} is used due to its increased precision and ease of implementation. Then the matching algorithm, recently developed by Ref.~\cite{bourgoin2020using}, and optimized for fast computation, is employed to obtain the 3D positions of the seeded particles.
\item The 3D reconstruction gives a cloud of points for every time step. The goal of the tracking is to transform this cloud into trajectories by following particles through time. A classical nearest neighbor approach with predictive tracking based on a linear fit over the five previous positions is used \cite{maas1993particle, malik1993particle, ohmi2000particle, veenman2001resolving, veenman2003establishing, ouellette2006quantitative, viggiano2021lagrangian, basset2022entrainment}.
\end{enumerate}

\subsection{Post-processing of the tracks}\label{subsec:postprocessing}
The tracking of particles results in a set of trajectories with a minimum length of 10 frames (\textit{ad hoc} length to remove presumably false trajectories) for the seven rotation frequencies of the impellers, i.e. for the seven Reynolds numbers considered. A visualization of the tracks is shown in figure~\ref{fig:hull_and_tracks}. See table~\ref{tab:lem_data_sets} for an overview of this data set.
The trajectories reconstructed by the tracking algorithm always exhibit some level of noise due to errors eventually accumulated from particle detection, 3D-matching, and tracking. It is important to properly handle noise, in particular when evaluating statistics associated with differentiated quantities, i.e.\ particle velocity and acceleration. Thus, one of the main questions of the post-processing is how to deal with the noise and separate its contribution from the true signal in order to obtain denoised statistics. A simple and usual method is the convolution of trajectories with a discrete Gaussian kernel \cite{mordant2004experimental2, ouellette2006phd}: the positions are smoothed by convolution with a discrete Gaussian kernel, whereas the velocities and accelerations are computed by convolving tracks with a first- and second-order derivative Gaussian kernel, respectively. The Gaussian kernel is characterized by its width $\sigma$ (in frames) and its length $N_f$: the smoothing is stronger with increasing $\sigma$ and the tracks are shortened by $N_f$ frames (divided between the start and the end of the trajectories).

For some statistics (in particular for single time, single particle statistics, such as velocity and acceleration probability density functions), we will use this method and consider slightly filtered data sets with positions and velocities. The optimal choice of filter width is determined based on the best estimate of filtered acceleration variance as in~\cite{voth2002measurement}, leading to values of $\sigma$ from 30 frames for the lowest Reynolds number case to 6 frames for the highest one (filter length is fixed at $N_f=3\sigma$).

Alternatively, one can use the $\mathrm{d}t$-method \cite{machicoane2017estimating, machicoane2017multi} to account for the impact of the noise on 2-times statistics (such as correlations and structure functions) of velocity and acceleration, without requiring to explicitly differentiate and filter the trajectories. We will therefore also implement this method to obtain statistics on velocity and acceleration without requiring explicit calculation of individual trajectory derivatives. This method, recently developed, enables us to obtain denoised statistics based on an estimation from discrete temporal increments of unfiltered positions.

These seven Lagrangian data sets, used for the following statistics, were previously used to study topological invariants and linking numbers related to helicity \cite{angriman2021broken}.

\begin{table}[H]
\addtolength{\tabcolsep}{5pt}
\centering{
\begin{tabular}{c|c|c|c|c|c|c}
case & $f_{\mathrm{impeller}}$ & $f_{\mathrm{camera}}$ & number of recordings & number of tracks & mean track length & total number of points \\[1pt]
& (\unit{Hz}) & (\unit{Hz}) & & ($10^4$) & (frames) & ($10^6$) \\[3pt]
\hline & & & & & \\[-1.5ex]
200 & 1.8 & 3125 & 40 & 2.4 & 286 & 6.7\\
270 & 2.5 & 3125 & 40 & 4.4 & 176 & 7.8\\
330 & 3.5 & 3125 & 40 & 6.6 & 140 & 9.1\\
380 & 5.0 & 3125 & 100 & 27.8 & 110 & 30.5\\
460 & 7.1 & 6250 & 40  & 8.4  & 136 & 11.5\\
550 & 10.0 & 6250 & 40 & 11.5 & 110 & 12.7\\
600 & 11.9 & 6250 & 42 & 11.2 & 124 & 13.9
\end{tabular}
}
\addtolength{\tabcolsep}{-5pt}
\caption{Recording parameters of the seven experimental data sets. Each recording is 8000 frames long, given a total of $2.7 \times 10^6$ frames per camera.}
\label{tab:lem_data_sets}
\end{table}

\section{Characterization of the flow}\label{sec:characterisation_flow}
Based on the tracks from the seven data sets, we provide a general characterization of the flow. We first check that the flow is homogeneous and isotropic. Then, based on classical Eulerian velocity statistics, we determine, for the seven Reynolds numbers considered, the classical quantities of turbulence, such as the energy dissipation rate $\varepsilon$ and the Eulerian integral scale $L_E$. We finally show how velocity and acceleration variances are affected by statistical bias. Unless otherwise specified, we use in this section slightly filtered data as previously described.

\subsection{Homogeneity and isotropy}\label{subsec:homogeneity_isotropy}

\subsubsection{Isotropy}
\begin{figure}
\hspace{-0.8cm}{\includegraphics[height=0.3\textheight]{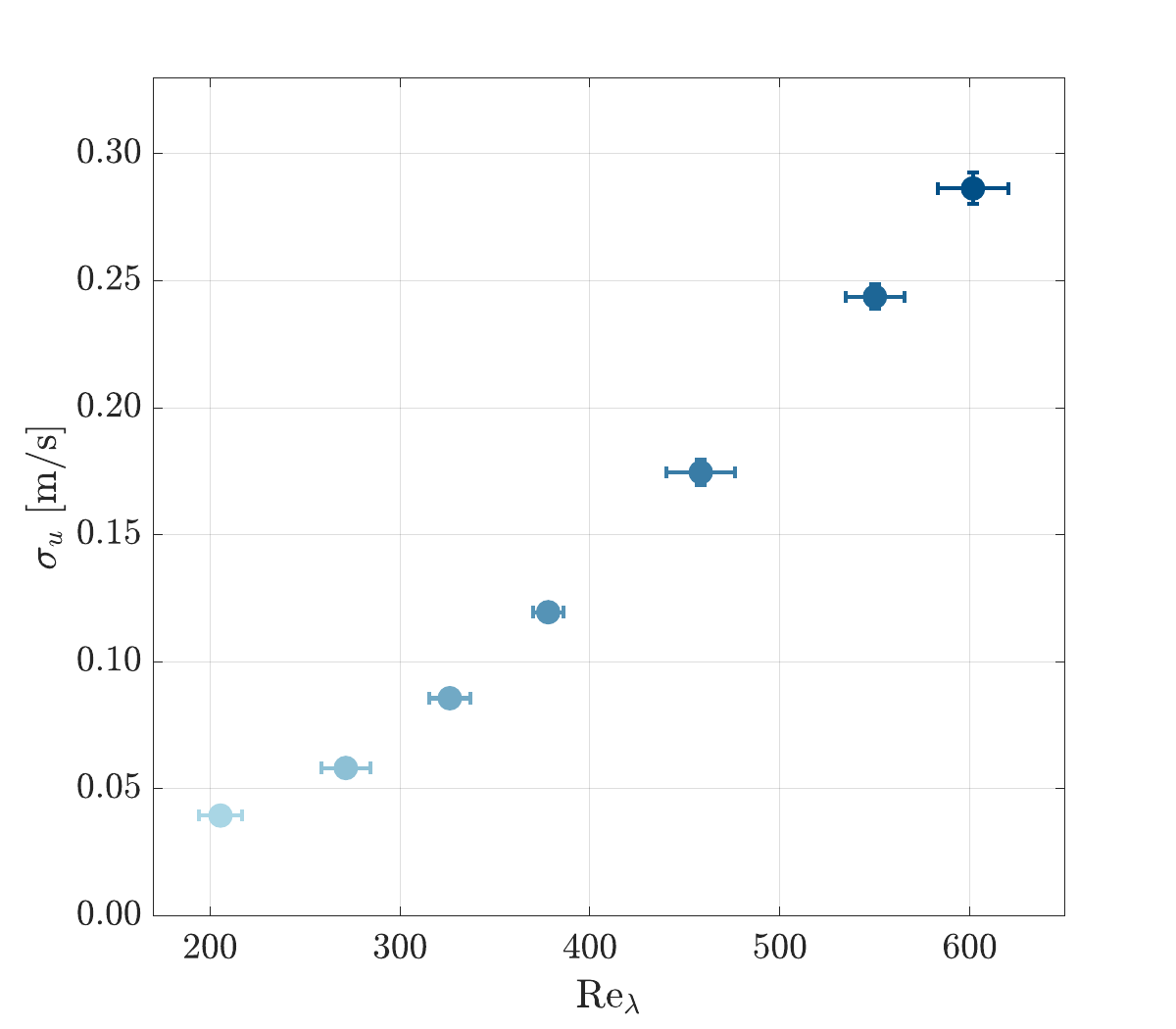}}
\hspace{0.5cm}\includegraphics[height=0.3\textheight]{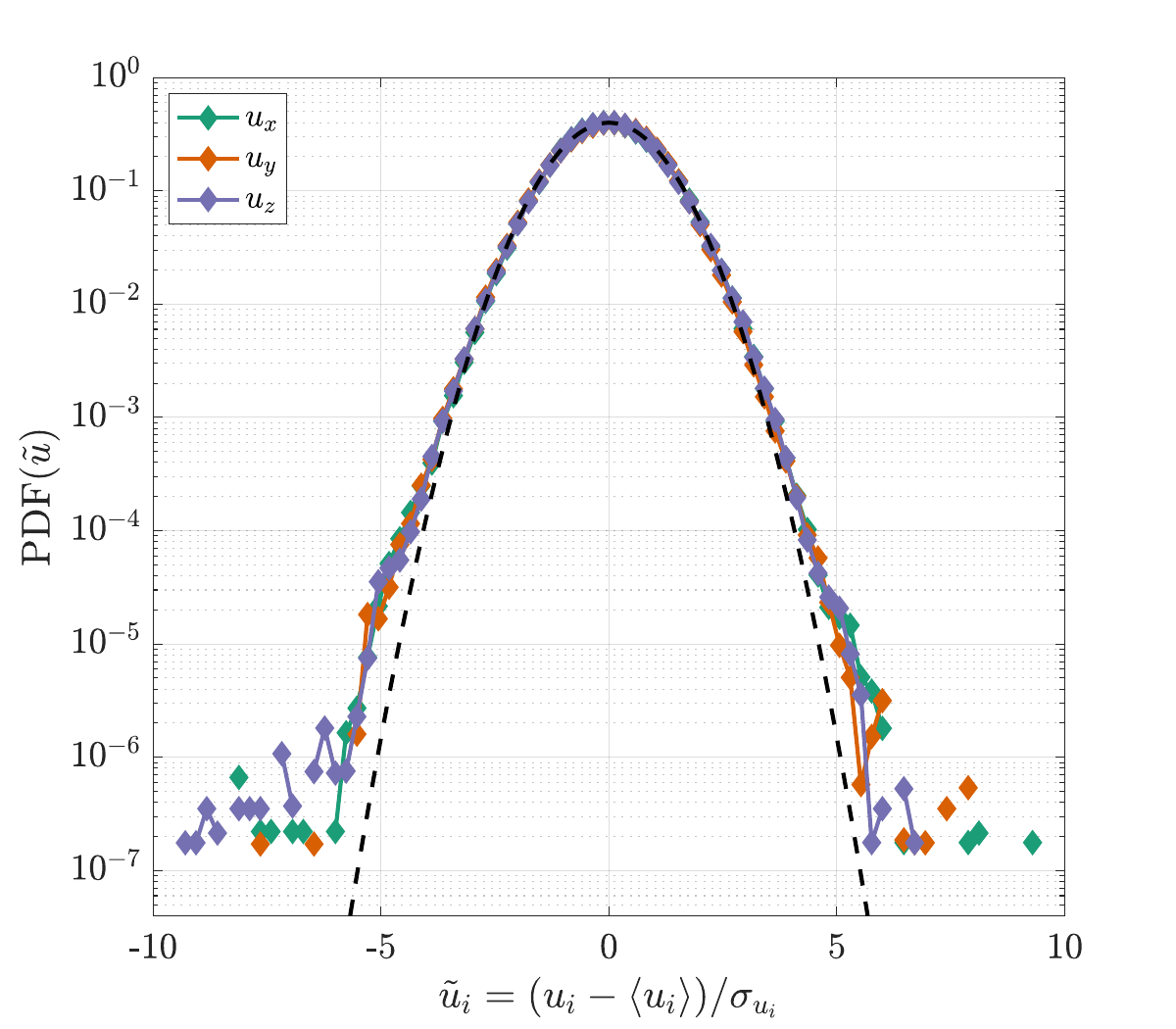}
\caption{Left: Velocity standard deviation $\sigma_u$ (based on calculating the standard deviation of the velocity in $10^4$ uniformly-chosen directions) as a function of the Taylor-based Reynolds number $\text{Re}_\lambda$. The vertical error bars indicate the anisotropy of the fluctuations (standard deviation of the resulting distribution). Right: Standardized PDF of the three velocity components $u_i$ (case 380) (dashed line: Gaussian distribution with zero mean and unit variance).}
\label{fig:lem_isotropy}
\end{figure}

To characterize the isotropy of the flow, we start by considering $u_p$, the projection of velocity along a given direction $p$, for $10^4$ different directions $p$ (uniformly distributed over a sphere). For each direction, we use the whole dataset of trajectories and calculate the associated standard deviation of velocity $\sigma_{u_p}$. We present in figure~\ref{fig:lem_isotropy}(Left) the mean $\langle \sigma_{u_p} \rangle_p$ of these standard deviations, with the standard deviation of the distribution (representative of the anisotropy of the velocity fluctuations across the different directions $p$) as the vertical error bar. For simplicity, we will call this $\sigma_u$ and report those values in  table~\ref{tab:lem_eulerian_parameters}. We find that the anisotropy is maximum for the lowest Reynolds number but does not exceed 5\%. For increasing $\text{Re}_\lambda$ (the values of $\text{Re}_\lambda$ are computed in the next subsection~\ref{subsec:lem_eulerian_stats}), the anisotropy decreases. 

\begin{figure}
\centerline{\hspace{0cm}\includegraphics[width=0.5\textwidth]{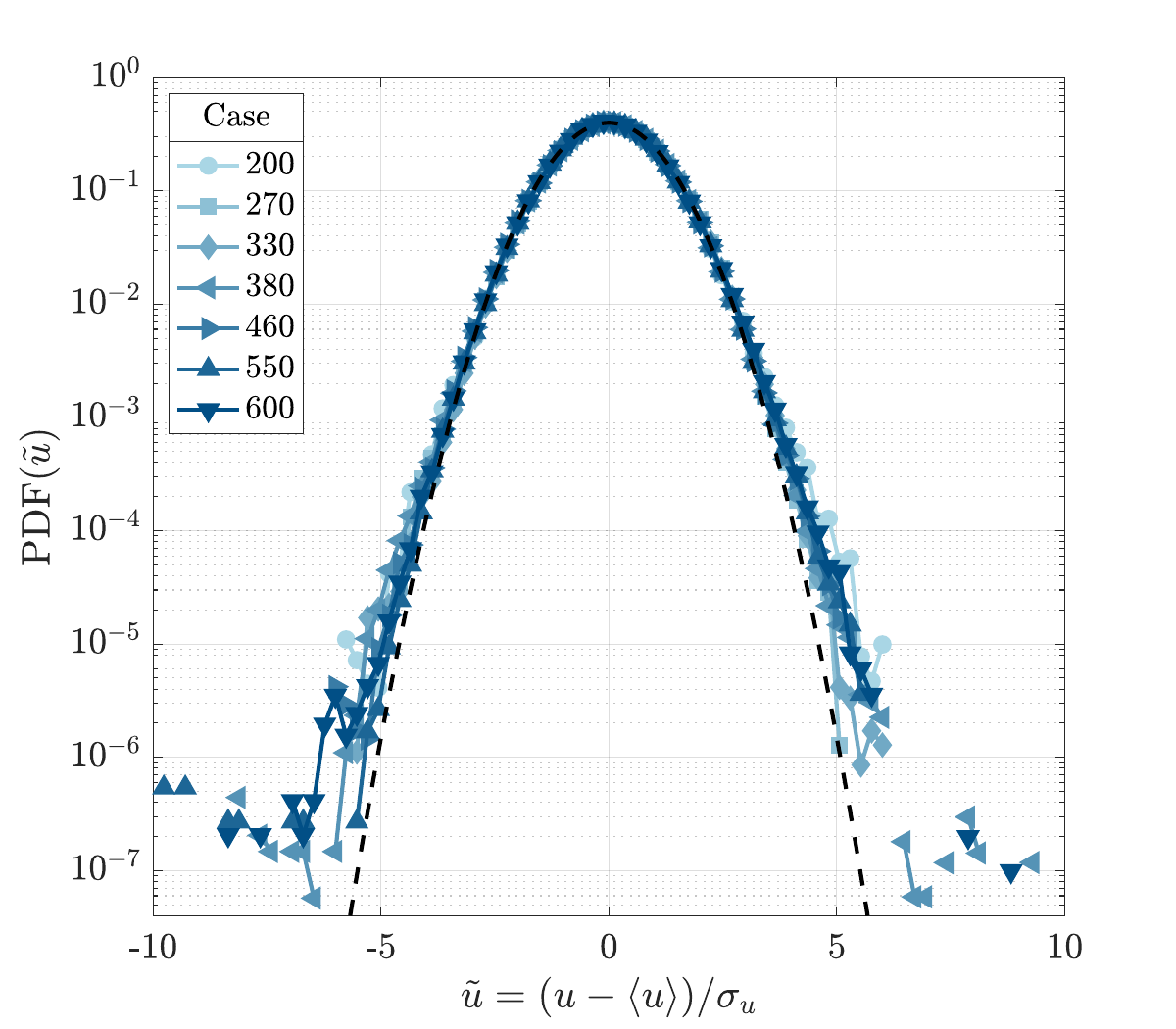}
\hspace{0.0cm}\includegraphics[width=0.5\textwidth]{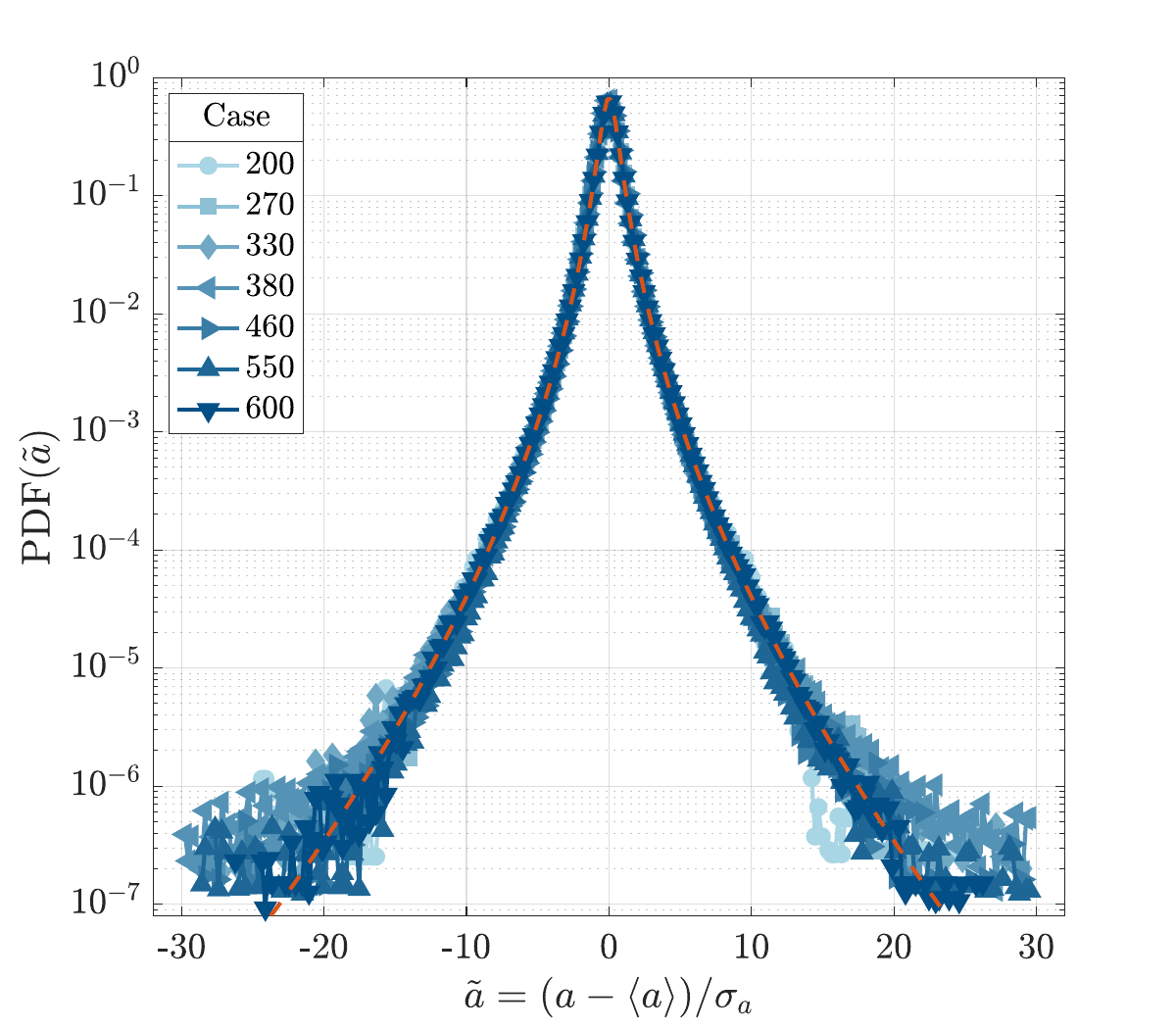}}
\caption{Standardized PDFs (averaged of the three components) for the seven Reynolds numbers. Left: velocity $u$ (dashed line: Gaussian distribution with zero mean and unit variance). Right: acceleration $a$ (dashed line: stretched exponential fit \cite{mordant2004experimental2} $P(a) = A\exp(-a^2/[2(1+|\beta a /\sigma|^\gamma)\sigma^2])$ with $A=0.684$, $\beta = 2.809$, $\sigma = 2.558$, and $\gamma = 1.454$).}
\label{fig:lem_pdfs}
\end{figure}

We also check the isotropy of the velocity probability density function (PDF) in figure~\ref{fig:lem_isotropy}(Right). We represent it as a standardized centered-reduced PDF, i.e.~the PDF of the dimensionless velocity $(u_i - \langle u_i \rangle)/\sigma_{u_i}$ where $\langle u_i \rangle$ is the mean velocity ($\langle \cdot \rangle$ denotes ensemble and time average) and $\sigma_{u_i}$ the associated standard deviation of the velocity. The velocity PDFs appear to be isotropic, close to Gaussian and with no particular difference between the three components. In figure~\ref{fig:lem_pdfs}(Left), we present the centered-reduced velocity PDFs for the seven Reynolds numbers. No large difference between the various Reynolds numbers is observed and all PDFs follow the same slightly super-Gaussian shape (leptokurtic with a kurtosis between 3.04 and 3.26). The origin of the super-Gaussian behavior of the velocity PDF is not totally clear. Deviations in Gaussianity are not uncommon and are usually mild (either sub or super-Gaussian) in situations close to homogeneous isotropic turbulence~\citep{bib:wilczek2011} while they can be more pronounced in markedly inhomogeneous flows~\citep{bib:huisman2013,bib:sy2021}. In the present case, super-Gaussianity may be caused by a mild intermittency arising from fluctuations of the energy input into the measurement volume from the forcing by the twelve impellers surrounding the flow. We also compute the PDFs of the acceleration, see figure~\ref{fig:lem_pdfs}(Right). The anisotropy of the acceleration is negligible and we only present the average of the three components. As usually observed in von K\'arm\'an experiments and direct numerical simulations of HIT, the acceleration PDFs are highly non-Gaussian \cite{toschi2009lagrangian}.

\begin{figure}
\centerline{\includegraphics[width=0.6\textwidth]{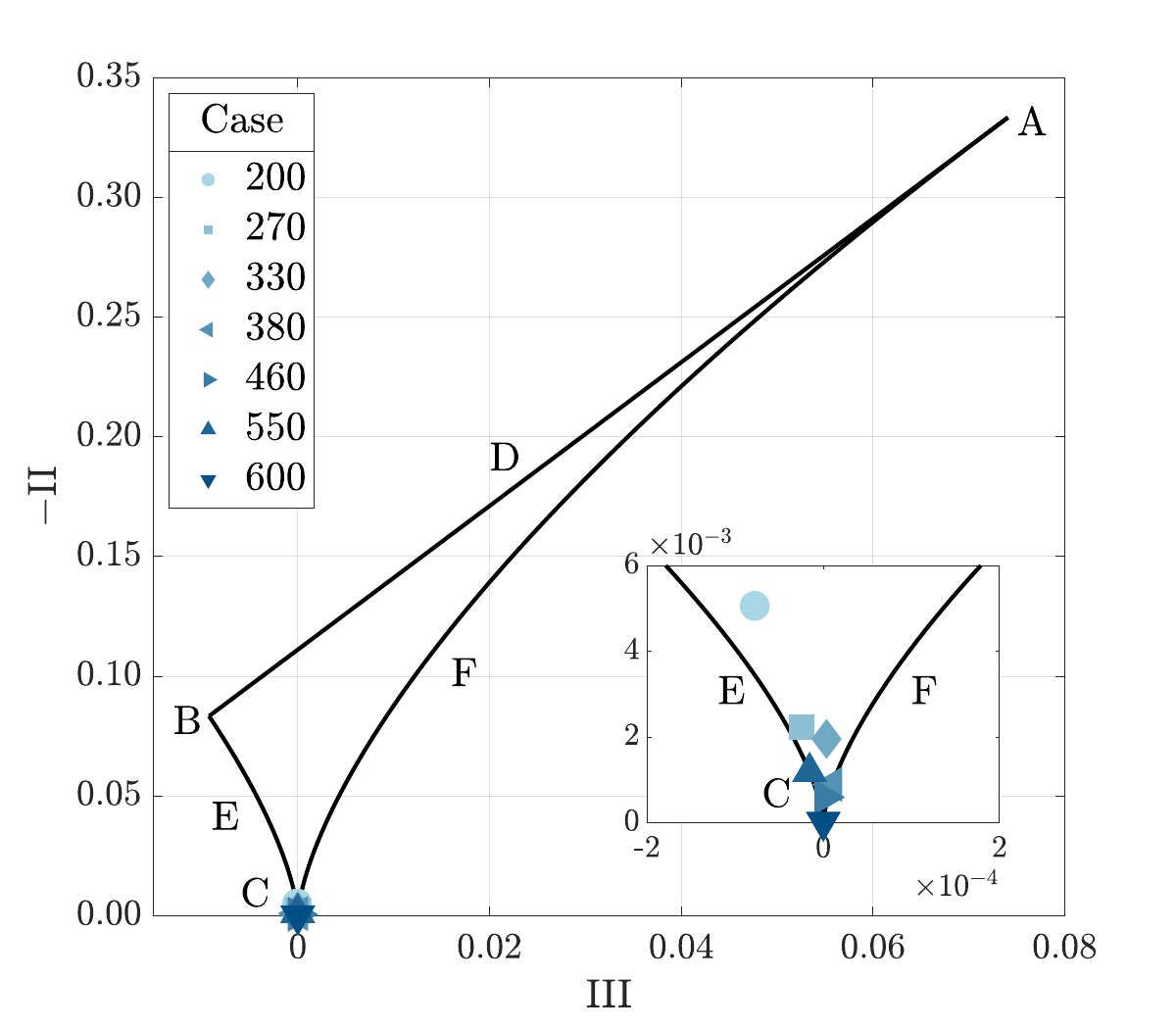}}
\caption{Lumley triangle: all possibles flows are within the triangle. Points A, B, and C refer to 1D, isotropic 2D, and isotropic 3D turbulence, respectively. Sides D, E, and F refer to 2D, axisymmetric disk-like, and axisymmetric rod-like turbulence, respectively. Inset: zoom around the origin.}
\label{fig:lem_lumley_triangle}
\end{figure}

A more sophisticated method to characterize isotropy is based on the Lumley triangle \cite{lumley1977return, lumley1979computational}. This analysis uses the Reynolds stresses by defining the three invariants of the anisotropy tensor:
\begin{equation}
\begin{gathered}
b_{ij} = \dfrac{\langle u_i' u_j' \rangle}{\langle u_k'u_k' \rangle} - \dfrac{\delta_{ij}}3,\\\vspace{0.3cm}
\mathrm{I} = b_{ii}, \;\; \mathrm{II} = \dfrac{b_{ij} b_{ji}}2, \;\; \textrm{and} \;\; \mathrm{III} = \dfrac{b_{ij} b_{jk} b_{ki}}3,
\label{eq:invariants_anisotropy}
\end{gathered}
\end{equation}
where $u'_i$ is the fluctuating velocity based on Reynolds decomposition ($u_i = \langle u_i \rangle + u'_i$) and $\delta_{ij}$ is the Kronecker delta, and where we use the Einstein summation convention and $i=x$, $y$, or $z$. The invariant I is zero by construction (following mass conservation and incompressibility) and the invariants II and III are computed in figure~\ref{fig:lem_lumley_triangle}. As we can see in the inset, all the points are very close to the point C (corresponding to 3D isotropic flow) and the general trend is that it gets more isotropic for increasing Reynolds number (consistent with figure~\ref{fig:lem_isotropy}).

\subsubsection{Homogeneity}
\begin{figure}
\centerline{\hspace{-0.8cm}\includegraphics[height=0.3\textheight]{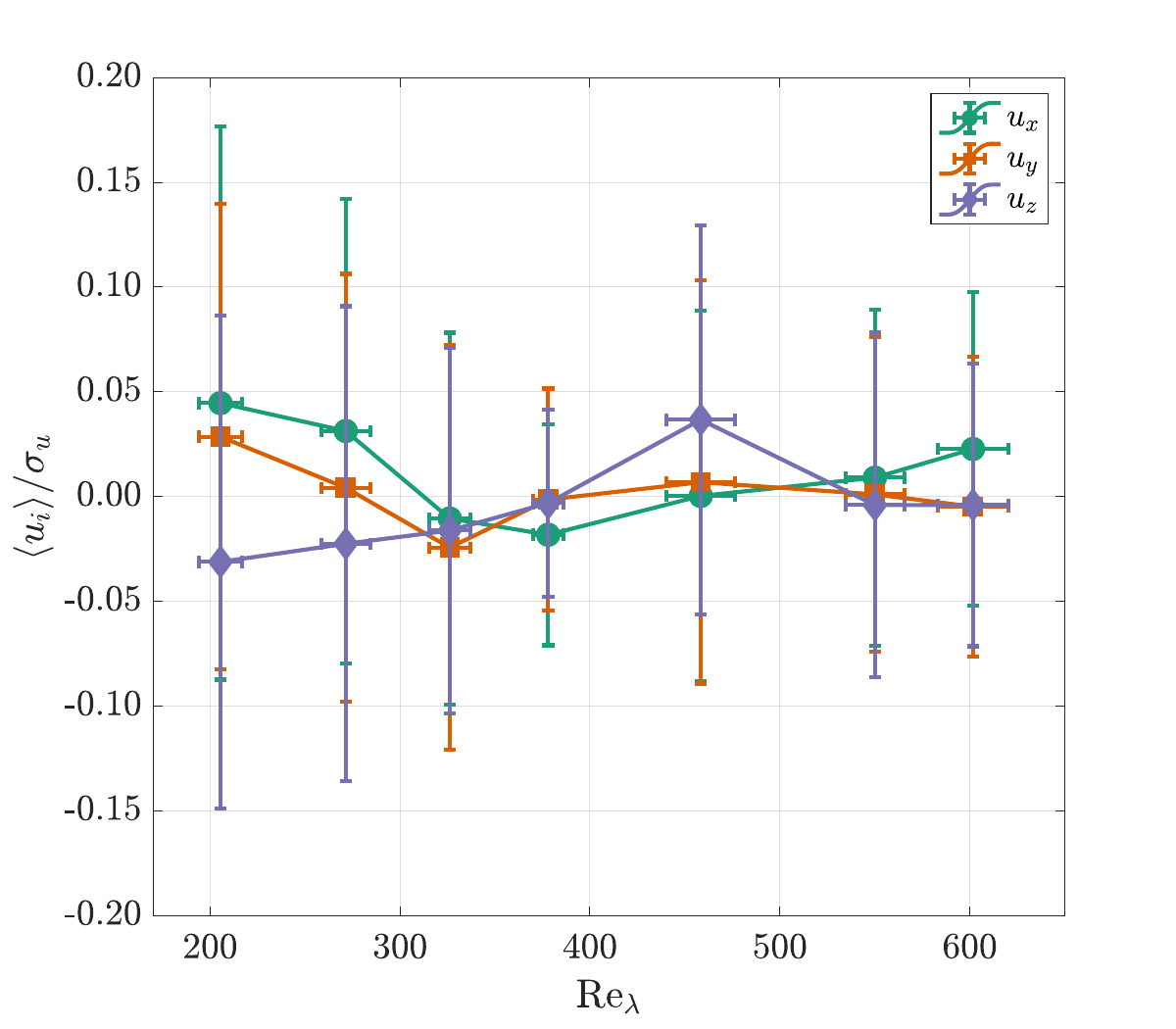}
\hspace{0.5cm}\includegraphics[height=0.3\textheight]{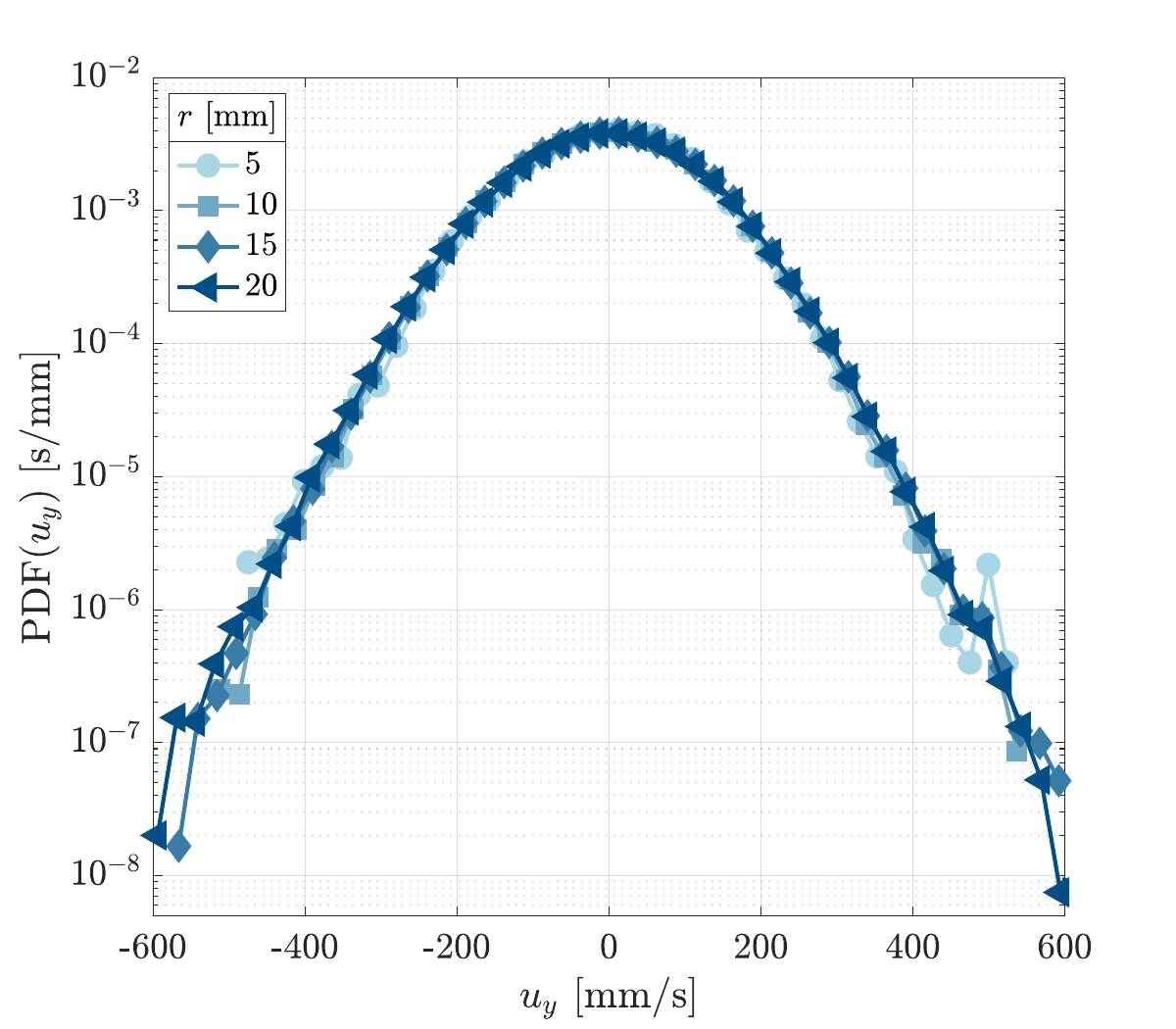}}
\caption{Left: three components of the global mean velocity $\langle u_i \rangle$ normalized by the velocity standard deviation $\sigma_u$ as a function of the Taylor-based Reynolds number $\text{Re}_\lambda$. The error bar is estimated as the standard deviation of the quantity over the different movies used to build the entire dataset. Right: raw PDF of the velocity $u_y$ restricted to spheres with a radius $r=5,~10,~15,$ and \SI{20}{mm} centered in the measurement volume (case 380).}
\label{fig:lem_homogeneity}
\end{figure}

In the present situation, the computation of a mean flow field do not lead to any well organized structure but rather to a small global motion over the entire volume. To characterize such global mean motion, we compute the mean velocity for all three components using all trajectories in the measurement volume. As observed in figure~\ref{fig:lem_homogeneity}(Left), which displays the mean value normalized by the standard deviation (determined in figure~\ref{fig:lem_isotropy}), the contribution of this large scale motion is less than 5\% of the fluctuations and may be considered as negligible in the present situation.

We check now whether the velocity fluctuations are homogeneous. Though the longest side of the measurement volume ($\approx \SI{60}{mm}$) is relatively small as compared to the size of the set-up ($\approx \SI{600}{mm}$), we want to check if the flow fluctuations have a spatial dependence. We therefore compute the velocity PDF for restricted portions of the measurement volume (spheres of increasing radius $r=5,~10,~15,$ and \SI{20}{mm} centered in the measurement volume), as presented in figure~\ref{fig:lem_homogeneity}(Right). As can be observed, the raw velocity PDFs computed for the various spheres are nearly identical (within statistical convergence). Similar plots can be generated for the other components and Reynolds numbers, all showing similar collapse. Thus no spatial dependence is observed for the velocity fluctuations. The same behavior is observed for acceleration.

\subsection{Eulerian velocity statistics}\label{subsec:lem_eulerian_stats}
We switch now to two-point--one-time statistics to characterize the multi-scale spatial correlations of the turbulence, i.e.~properties in a classical Eulerian framework. The determination of Eulerian velocity statistics shown hereafter is based on structure and autocorrelation functions, from which important quantities such as the mean energy dissipation rate $\varepsilon$ and the Eulerian integral length scale $L_E$ can be extracted.

\subsubsection{Measurement of $\varepsilon$}
The mean energy dissipation rate $\varepsilon$ is a fundamental quantity to characterize a turbulent flow, but non-trivial to measure. In this study, we classically use the K41 Eulerian phenomenology \cite{kolmogorov1941local} associated with structure functions to determine it. Structure functions are commonly used to describe multi-scale properties of turbulence through a statistical representation of a flow quantity with a given spatial or temporal separation. In an Eulerian perspective, longitudinal velocity structure functions of order $n$ are defined as
\begin{equation}
S^E_{n,\parallel}(\boldsymbol{x},\boldsymbol{\Delta x}) = \langle [\Delta u_\parallel(\boldsymbol{x},\boldsymbol{\Delta x})]^n \rangle = \langle [u_\parallel(\boldsymbol{x}+\boldsymbol{\Delta x}) - u_\parallel(\boldsymbol{x})]^n \rangle,
\label{eq:Sn}
\end{equation}
where the longitudinal increment $\Delta u_\parallel$ is computed over two points, one at $\boldsymbol{x}$, the other at $\boldsymbol{x}+\boldsymbol{\Delta x}$, with $u_\parallel$ defined as the single longitudinal component of Eulerian velocity along the separation vector $\boldsymbol{\Delta x}$. 

The spatial average $\langle \cdot \rangle$ is taken over the pairs of particles in the entire volume. In HIT, statistics do not depend on the position $\boldsymbol{x}$ and on the direction of the separation vector $\boldsymbol{\Delta x}$, hence $S^E_{n,\parallel}(\boldsymbol{x},\boldsymbol{\Delta x}) = S^E_{n,\parallel}(r)$ with $r = |\boldsymbol{\Delta x}|$. In the same way, a transverse structure function $S^E_{n,\perp}(\Delta x)$ and a total structure function $S^E_{n,\text{tot}}(\Delta x)$ can be defined based on transverse velocity components $u_\perp$ and magnitude of velocity increment $u_{\text{tot}}$, respectively.

Following K41 phenomenology in HIT, for separations within the inertial range ($\ell_K \ll r \ll L_E$, with $\ell_K$ the Kolmogorov length scale and $L_E$ the Eulerian integral length scale) \cite{pope2000turbulent}
\begin{equation}
S_{2,\parallel}^E(r) = C_2 (\varepsilon r)^{2/3},
\label{eq:S2_parallel}
\end{equation}
with $C_2 \approx 2.1$ for flows with $\text{Re}_\lambda > \mathcal{O}(100)$~\cite{bib:sreenivasan1995_PoF}, and
\begin{equation}
S_{2,\perp}^E(r) = \dfrac{4}{3} S_{2,\parallel}^E(r).
\label{eq:S2_perp}
\end{equation}

The longitudinal third-order structure function is also of particular interest, as it is analytically related to the mean energy dissipation rate $\epsilon$, such that for fully developed turbulence and separations within the inertial range ($\ell_K \ll r \ll L_E$) \cite{kolmogorov1941local, kolmogorov1941dissiptation, frisch1995turbulence, pope2000turbulent}
\begin{equation}
S^E_{3,\parallel}(r) = -\dfrac{4}{5} \varepsilon r.
\label{eq:S3}
\end{equation}

\begin{figure}
\centerline{\hspace{-0.8cm}\includegraphics[height=0.3\textheight]{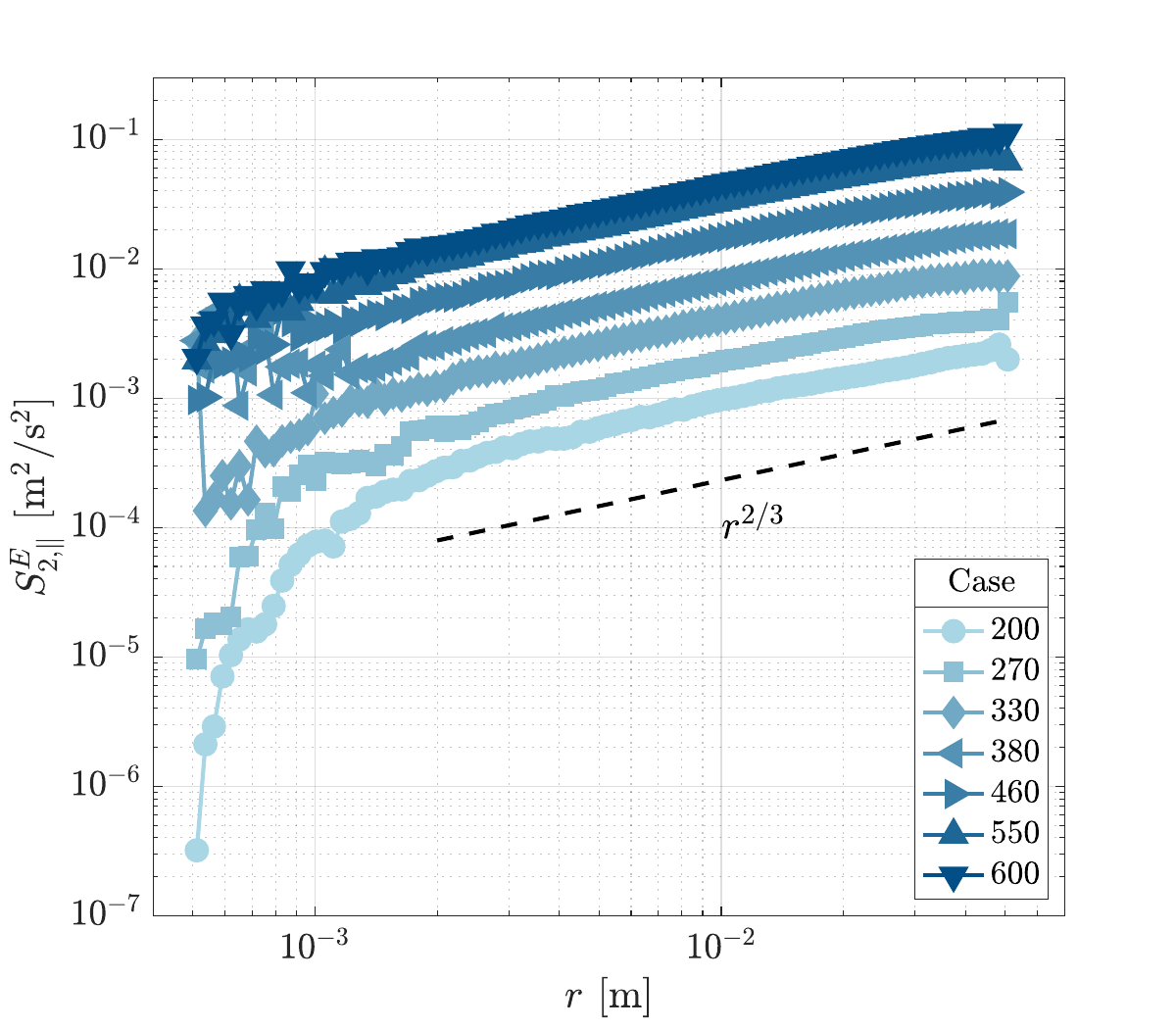}
\hspace{0.5cm}\includegraphics[height=0.3\textheight]{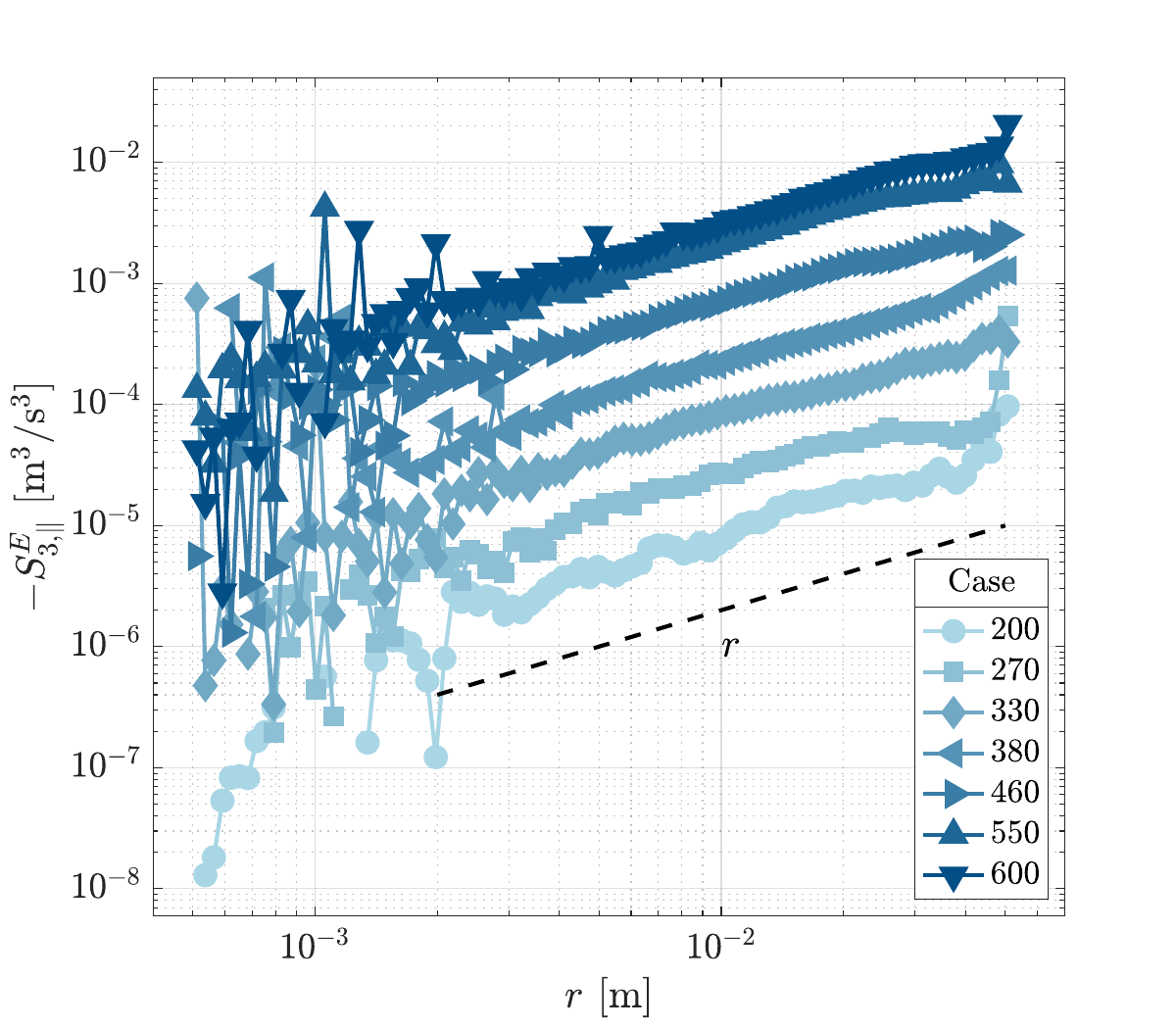}}
\caption{Eulerian second-order (Left) and third-order (right) structure functions for the seven Reynolds numbers}
\label{fig:S2_S3_long}
\end{figure}

Previous relations for $S_{2,\parallel}^E$ (see figure~\ref{fig:S2_S3_long}), $S_{2,\perp}^E$ and for $S^E_{3,\parallel}$ (see figure~\ref{fig:S2_S3_long}) are well verified in the present situation. The inertial regime is well observed for separations from \SI{5}{mm} to \SI{5}{cm}, whereas the injection and dissipative regimes are not clearly resolved. Actually the measurement volume is too small to fully observe the injection scale (we will see in the following that $L_E \approx \SI{5}{cm}$) while the particle seeding is too low to achieve statistical convergence for pairs of particles with separations comparable to the dissipative range. Indeed, the seeding density in our experiment is of the order of 5 tracers per cubic centimeter so that the probability of finding pairs of particles with separations below a few millimeters is scarce. This choice is motivated by the fact that the focus of the present work is primarily on Lagrangian statistics, while we only aim at reasonably accessing Eulerian inertial range statistics to characterize the main turbulence properties (in particular $\varepsilon$). The scarcity of small separations statictics is particularly visible for the estimate of third-order structure function for scales below 5~mm, which appears to be noisier. We therefore may only use in the sequel the second-order structure function extract estimates of $\varepsilon$.

\begin{figure}
\centerline{\includegraphics[width=0.5\textwidth]{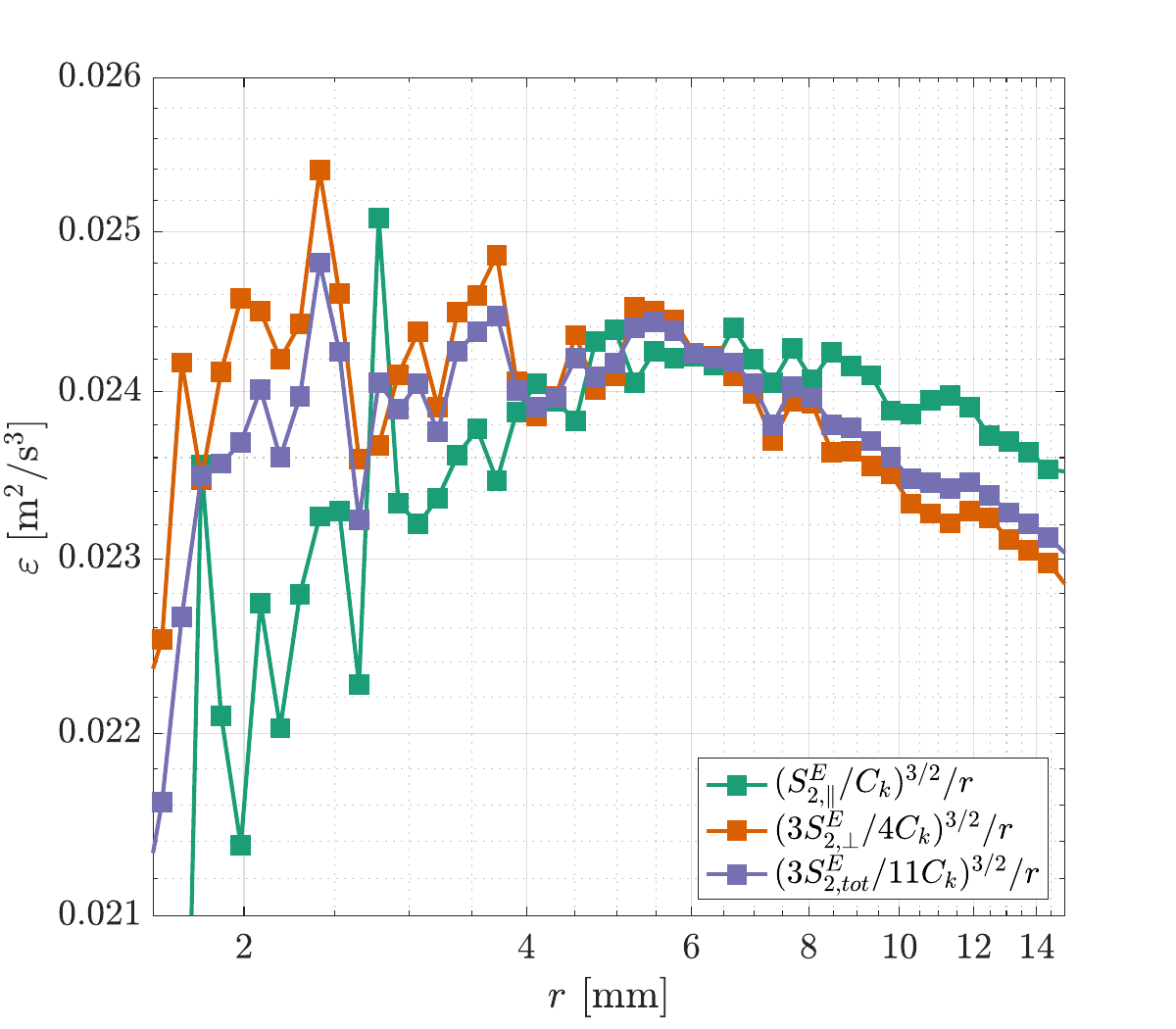}}
\caption{Estimation of the mean energy dissipation rate, $\varepsilon$, with the compensated second-order structure functions for case 380.}
\label{fig:lem_epsilon_f1.000_3125fps}
\end{figure}

Figure~\ref{fig:lem_epsilon_f1.000_3125fps} shows (for case 380) both transverse and longitudinal second-order structure functions compensated such that a plateau at $\varepsilon$ shall appear within inertial scales. It can be seen that for this case, a consistent estimate of $\varepsilon \approx \SI{0.024}{m^2/s^3}$ is obtained. One may also note that the three estimates are consistent, which is again a sign that the properties of the flow are close to being homogeneous and isotropic.

\begin{figure}
\centerline{\hspace{0cm}\includegraphics[width=0.5\textwidth]{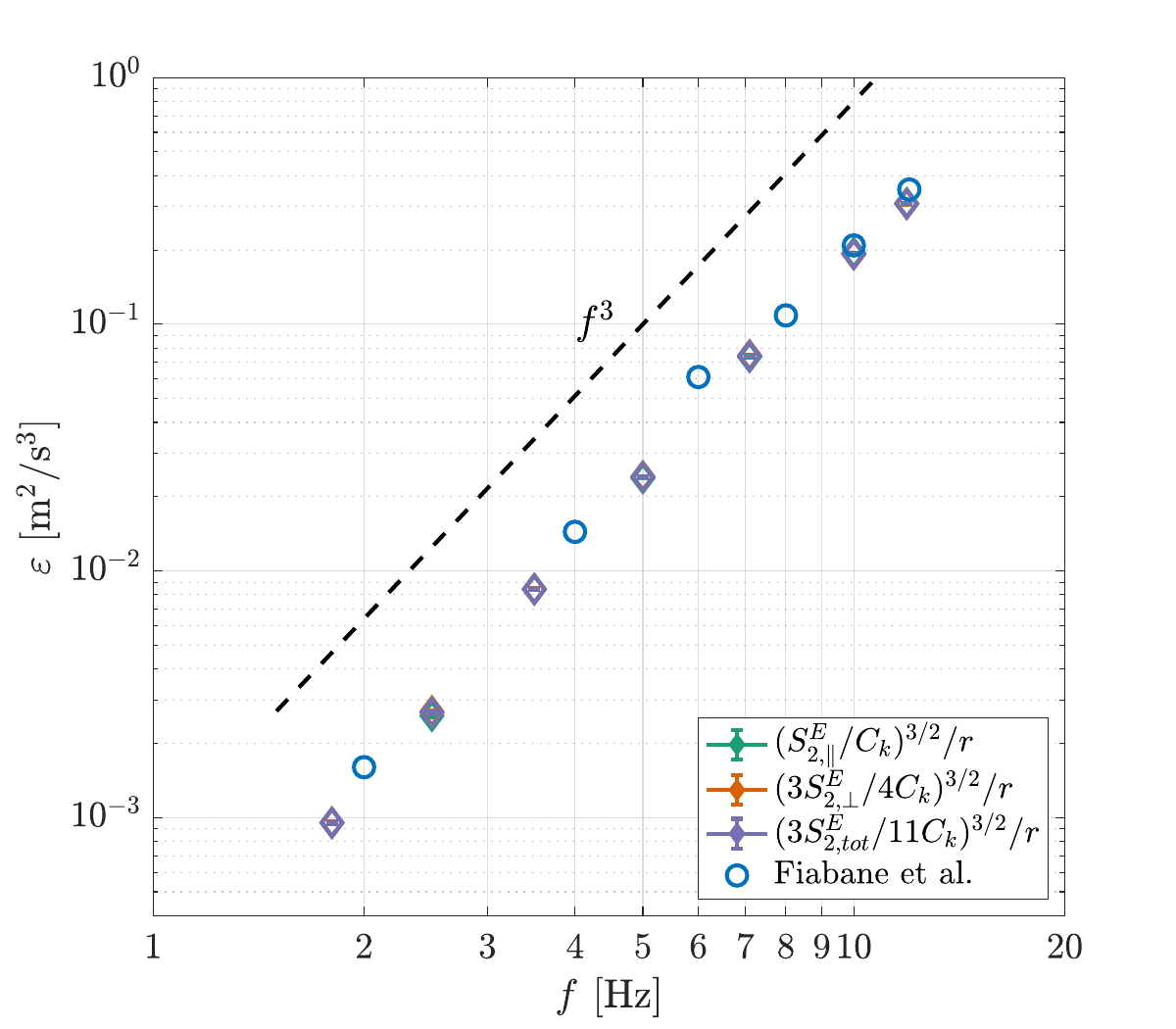}
\hspace{0.0cm}\includegraphics[width=0.5\textwidth]{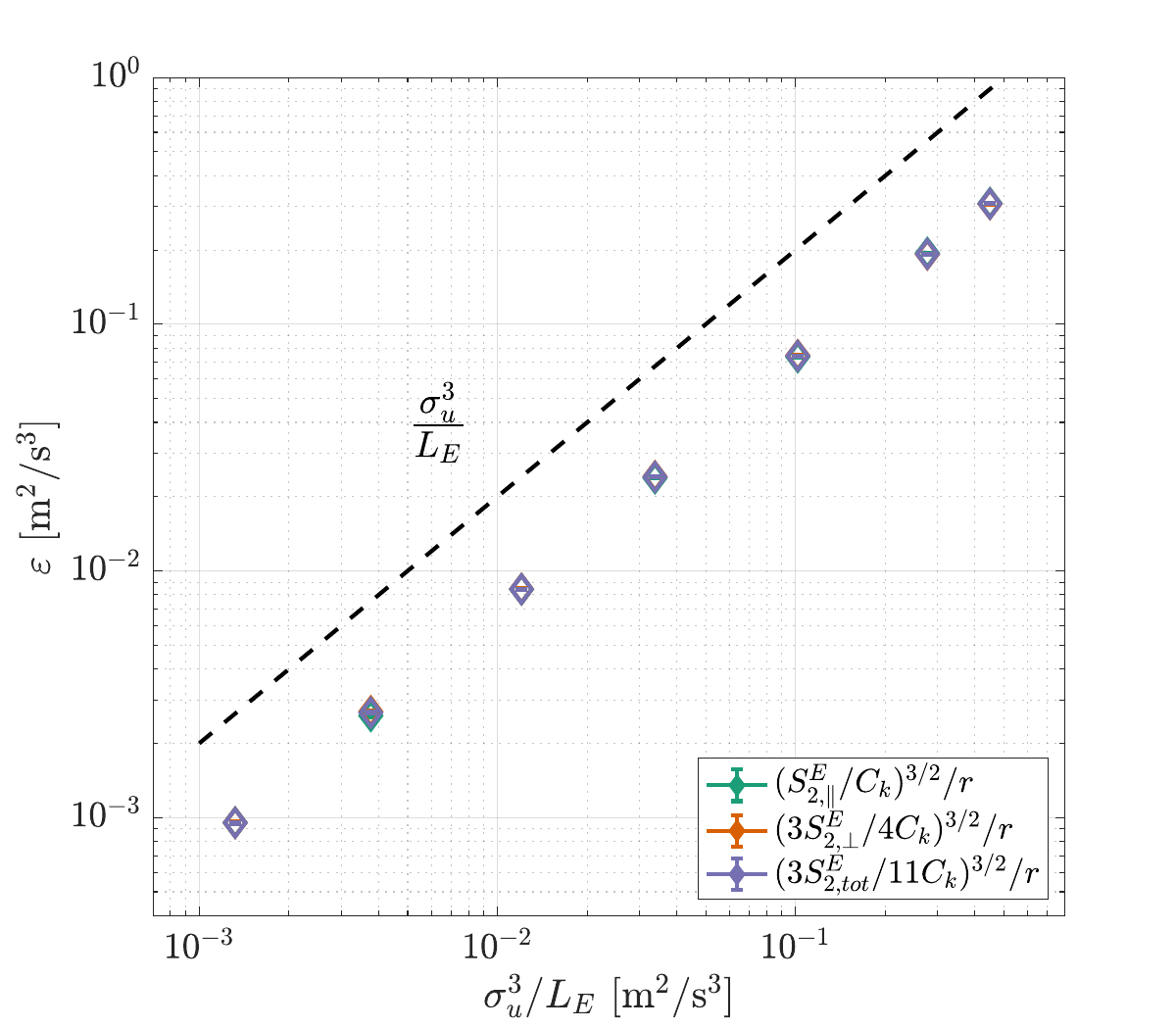}}
\caption{Mean energy dissipation rate, $\varepsilon$ (obtained from compensated second-order structure functions) as a function of the impellers rotation frequency $f$ (Left) and as a function of $\sigma_u^3/L_E$ (Right).}
\label{fig:lem_epsilon}
\end{figure}

Figure~\ref{fig:lem_epsilon} displays the dissipation rate $\varepsilon$ estimated with the different methods as a function of the impeller rotation frequency $f$ and as a function of the dimensional argument $\sigma_u^3/L_E$. Ref.~\cite{fiabane2012clustering} uses the same experimental set-up and the values of $\varepsilon$ they find (obtained with second-order structure functions from particle image velocimetry) are also included, and are found to be close to our values. The evolution goes as $f^3$ as dimensionally expected for turbulent flows generated with such a forcing. According to the Richardson-Kolmogorov cascade theory the dissipation rate $\varepsilon$ of a turbulence at equilibrium state scales as $\varepsilon = C_\varepsilon\sigma_u^3/L_E$ independent of the fluid viscosity \cite{vassilicos2015dissipation}, where $\sigma_u$ is the r.m.s velocity of turbulence, $L_E$ the Eulerian integral length scale and $C_\varepsilon$ is a constant depending on the forcing procedures. This is verified in figure~\ref{fig:lem_epsilon} (right). In our experiments $C_\varepsilon$ are found to be around 0.70, consistent with the value previously reported with similar forcing \cite{vassilicos2015dissipation}. Determining $C_\epsilon$ requires to know the value of the Eulerian integral scale $L_E$, which is described in next sub-section. The corresponding measured values of $\varepsilon$ and $C_\varepsilon$ are reported in table~\ref{tab:lem_eulerian_parameters}.

\subsubsection{Measurement of $L_E$}
We determine the Eulerian integral length scale $L_E$ based on the computation of the normalized Eulerian longitudinal velocity autocorrelation function
\begin{equation}
\mathcal{R}_{uu,\parallel}^E(r) = \dfrac{\langle u_\parallel(\boldsymbol{x}+\boldsymbol{r}) u_\parallel(\boldsymbol{x}) \rangle}{\sigma_u^2}.
\label{eq:Ru}
\end{equation}

\begin{figure}
\centerline{\includegraphics[width=0.7\textwidth]{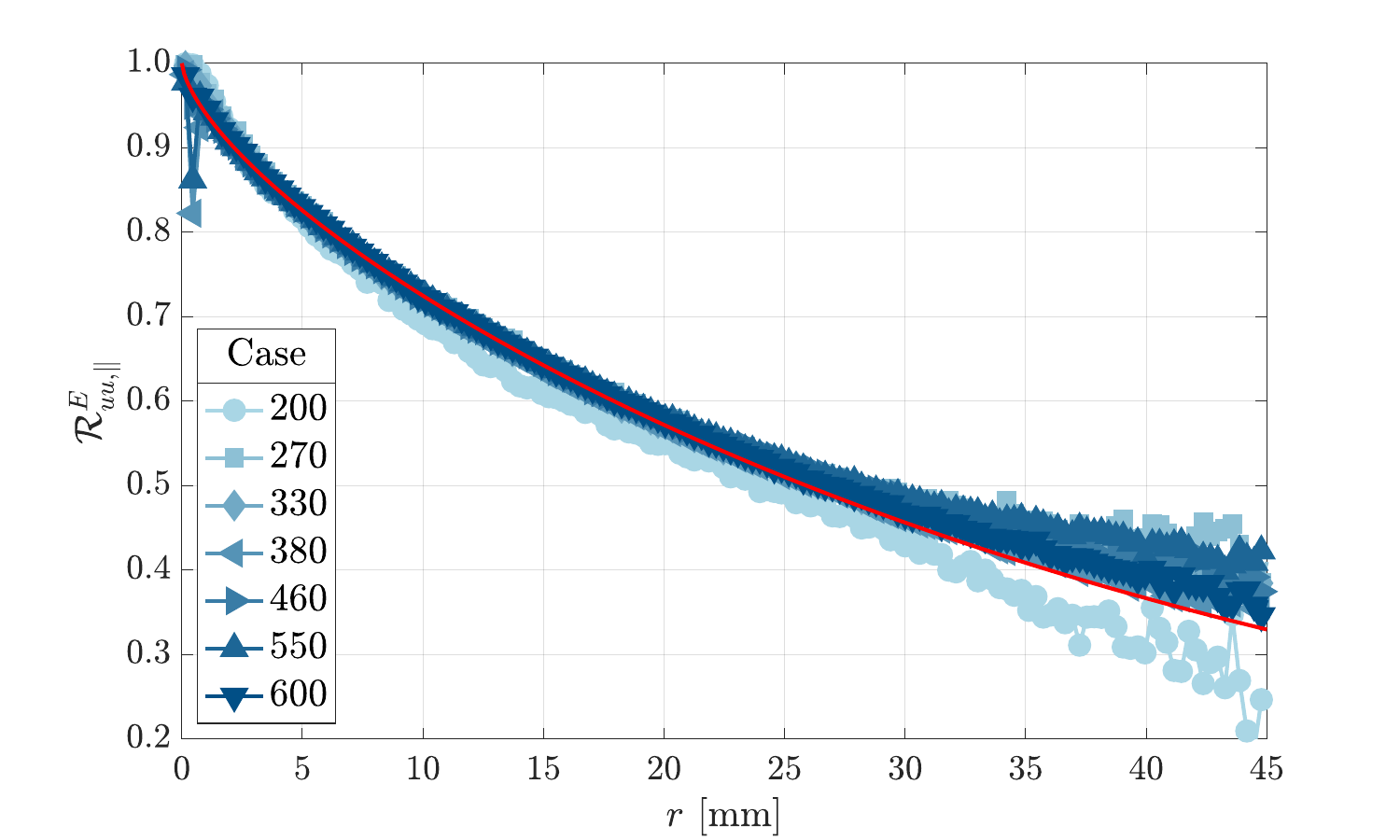}}
\caption{Eulerian longitudinal velocity autocorrelation functions $\mathcal{R}_{uu,\parallel}^E$ for the seven Reynolds numbers (red line: fit~\eqref{eq:fit_Ru} for case 380).}
\label{fig:lem_Ru_pairdisp}
\end{figure}

Assuming homogeneity, $\mathcal{R}_{uu,\parallel}^E(r)$ can be simply computed based on $S_{2,\parallel}^E(r)$ with $\mathcal{R}_{uu,\parallel}^E(r) = 1 - S_{2,\parallel}^E(r)/\left(2\sigma_u^2 \right)$.
Figure~\ref{fig:lem_Ru_pairdisp} displays the corresponding autocorrelation functions for the seven Reynolds numbers using the results of figure~\ref{fig:S2_S3_long}.

Following the classical definition of the Eulerian integral scale $L_E = \int_0^\infty R_{uu,\parallel}^E(r)\textrm{d}r$, one would ideally need to integrate the Eulerian correlation function up to infinity. However, the experimental functions are reasonably converged for a limited range of separations $\mathcal{O}(\SI{1}{mm}) \leq r \leq \mathcal{O}(\SI{4}{cm})$. Thus, we use the following function 
\begin{equation}
\mathcal{R}_{uu,\parallel}^E(r) = 1 - \dfrac{ \left( \dfrac{r}{d} \right) ^{2/3} }{ \left[ 1 + \left( \dfrac{r}{d} \right)^2 \right] ^{1/3} },
\label{eq:fit_Ru}
\end{equation}
based on a Batchelor parameterization \cite{batchelor1951pressure, grossmann1997application} to fit the inertial and injection regimes, whose contribution dominates the integral (besides, as already pointed the Eulerian dissipative range is not resolved in the present data). Figure~\ref{fig:lem_Ru_pairdisp} shows that this parameterization accurately fits the experimental autocorrelation functions up to the largest accessible scales ($r \approx \SI{4}{cm}$). From the parameterization in Eq.~\eqref{eq:fit_Ru} it follows analytically that the Eulerian integral length scale is as follows:

\begin{equation}
L_E = \int_0^\infty \mathcal{R}_{uu,\parallel}^E(r) \:\mathrm{d}r = \sqrt{\pi} \dfrac{ \Gamma \left(\frac{5}{6}\right)}{\Gamma \left(\frac{1}{3}\right)} d \approx 0.747 d,
\label{eq:LE}
\end{equation}
where $\Gamma$ is the (complete) gamma function. The corresponding values of $L_E$ for the seven cases investigated here are reported in table~\ref{tab:lem_eulerian_parameters}. The integral scale $L_E$ is found to not strongly depend on the forcing frequency of the impellers and is of the order of \SI{5}{cm}. This is consistent with the qualitative interpretation that the forcing must be commensurate with the impeller dimensions, which are \SI{7}{cm} in radius.

\subsubsection{Eulerian parameters}
The values of the velocity standard deviation $\sigma_u$, the mean energy dissipation rate $\varepsilon$, and the Eulerian integral length scale $L_E$ are reported in table~\ref{tab:lem_eulerian_parameters} along with various other important quantities: the Kolmogorov length scale $\ell_K = \left( \tfrac{\nu^3}{\varepsilon} \right) ^{1/4}$ and time scale $\tau_\eta = \left( \tfrac{\nu}{\varepsilon} \right) ^{1/2}$, the Taylor microscale $\lambda = \left( 15 \nu \tfrac{\sigma_u^2}{\varepsilon} \right)^{1/2}$ and the Taylor-based Reynolds number $\text{Re}_\lambda = \tfrac{\sigma_u \lambda}{\nu}$, the Eulerian integral time scale $T_E = \tfrac{L_E}{\sigma_u}$, and the constant $C_\varepsilon$ defined as $C_\varepsilon = \varepsilon \frac{L_E}{\sigma_u^3}$.

\begin{table}[H]
\addtolength{\tabcolsep}{2pt}
\centerline{\begin{tabular}{c|c|c|c|c|c|c|c|c|c}
case & $\sigma_u$ & $\varepsilon$ & $\ell_K$ & $\tau_\eta$ & $\lambda$ & $\text{Re}_\lambda$ & $L_E$ & $T_E$ & $C_\varepsilon$\\[1pt]
& (\unit{mm/s}) & (\unit{m^2/s^3}) & (\unit{\micro\meter}) & (\unit{ms}) & (\unit{mm}) & & (\unit{cm}) & (\unit{s}) & \\[3pt]
\hline & & & & & & & & \\[-1.5ex]
200 & $39.4 \pm 1.6$ & $(9.5 \pm 0.07) \times 10^{-4}$ & $166.3 \pm 0.29$ & $30.7 \pm 0.11$ & $4.69 \pm 0.19$ & $205 \pm 11$ & $4.64 \pm 0.04$ & $1.18 \pm 0.05$ & $0.72 \pm 0.09$\\
270 & $58.2 \pm 1.9$ & $(2.6 \pm 0.04) \times 10^{-3}$ & $129.6 \pm 0.54$ & $18.6 \pm 0.16$ & $4.20 \pm 0.14$ & $271 \pm 13$ & $5.23 \pm 0.04$ & $0.90 \pm 0.03$ & $0.69 \pm 0.07$\\
330 & $85.6 \pm 2.0$ & $(8.4 \pm 0.04) \times 10^{-3}$ & $ 96.5 \pm 0.13$ & $10.3 \pm 0.02$ & $3.43 \pm 0.08$ & $326 \pm 11$ & $5.21 \pm 0.02$ & $0.61 \pm 0.01$ & $0.70 \pm 0.05$\\
380 & $119.5 \pm 1.8$ & $(2.4 \pm 0.01) \times 10^{-2}$ & $ 74.4 \pm 0.09$ & $6.1 \pm 0.02$ & $2.85 \pm 0.04$ & $378 \pm 8$ & $5.05 \pm 0.06$ & $0.42 \pm 0.01$ & $0.70 \pm 0.03$\\
460 & $174.6 \pm 4.9$ & $(7.4 \pm 0.03) \times 10^{-2}$ & $ 56.1 \pm 0.06$ & $3.5 \pm 0.01$ & $2.36 \pm 0.06$ & $458 \pm 18$ & $5.23 \pm 0.03$ & $0.30 \pm 0.01$ & $0.72 \pm 0.06$\\
550 & $243.7 \pm 4.8$ & $(1.9 \pm 0.01) \times 10^{-1}$ & $ 44.0 \pm 0.03$ & $2.2 \pm 0.01$ & $2.03 \pm 0.04$ & $550 \pm 15$ & $5.23 \pm 0.05$ & $0.21 \pm 0.01$ & $0.70 \pm 0.04$\\
600 & $286.3 \pm 6.2$ & $(3.1 \pm 0.01) \times 10^{-1}$ & $ 39.2 \pm 0.03$ & $1.7 \pm 0.01$ & $1.89 \pm 0.04$ & $602 \pm 18$ & $5.23 \pm 0.01$ & $0.18 \pm 0.01$ & $0.69 \pm 0.04$\\
\end{tabular}}
\addtolength{\tabcolsep}{-2pt}
\caption{Eulerian parameters of the LEM flow for the seven cases considered.}
\label{tab:lem_eulerian_parameters}
\end{table}

\section{Lagrangian statistics}\label{sec:lem_lagrangian_stats}
Through one-point--one-time and two-point--one-time statistics presented in the previous section, we have characterized the turbulence generated in the LEM using a classical Eulerian approach, showing that this flow is a good experimental model of HIT. We propose now to compute one-particle--two-time statistics, i.e. a Lagrangian characterization of the flow. We are first interested in velocity statistics which are marginally affected by experimental noise but strongly altered by finite volume effects and are difficult to access experimentally. Then we present acceleration statistics which are, on the contrary, highly affected by noise but robust to statistical bias. 

\subsection{Velocity statistics}

\subsubsection{Autocorrelation functions}

We start with the velocity autocorrelation function, defined as $R_{uu}^L(\tau)=\langle u(t) u(t+\tau) \rangle$ or in the normalized form $\mathcal{R}_{uu}^L(\tau) = R_{uu}^L(\tau)/\sigma_u^2$. Here, to account for the bias of the track length on $\sigma_u$, we define the normalized autocorrelation function as
\begin{equation}
\mathcal{R}_{uu}^L(\tau) = \dfrac{\langle u(t) u(t+\tau) \rangle}{\sqrt{\langle u(t+\tau)^2 \rangle \langle u(t)^2 \rangle}}.
\label{eq:Ruu}
\end{equation}
where $u$ corresponds to any filtered velocity component. The corrective term $\sqrt{\langle u(t+\tau)^2 \rangle \langle u(t)^2 \rangle}$ and the time averaging $\langle\cdot\rangle$ are performed with trajectories longer than a given time lag $\tau$. 

\begin{figure}
\includegraphics[width=0.75\textwidth]{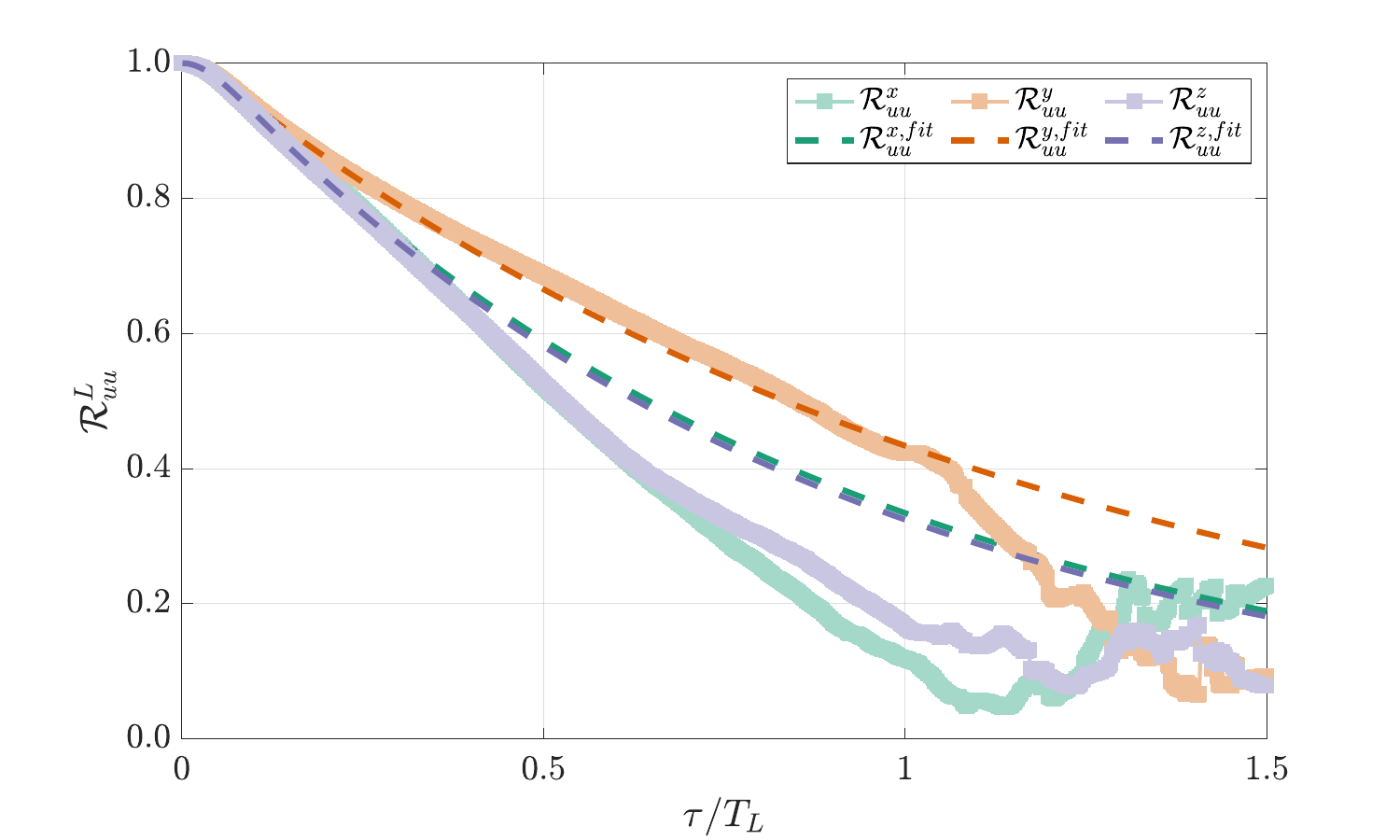}
\includegraphics[width=0.75\textwidth]{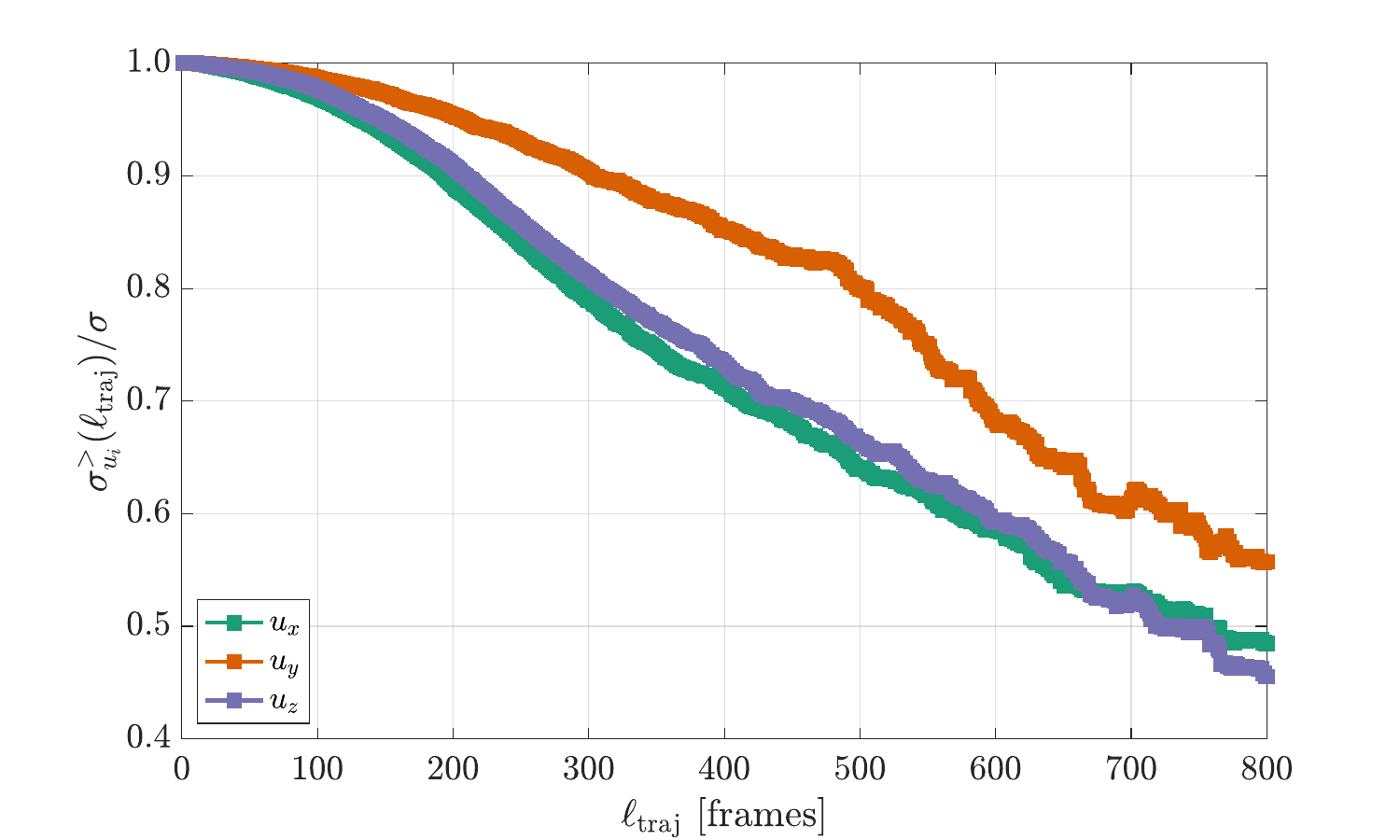}
\caption{Top: Normalized Lagrangian autocorrelation functions $\mathcal{R}_{uu}^L$ for the three velocity components as a function of normalized time $\tau/T_L$ (case 380). The $y$ component is the longest dimension and therefore is less affected by finite volume effect. Dashed line: fit of infinite-layer model. Bottom: Normalized standard deviations conditioned on trajectory length $\ell_{\text{traj}}$ for the three components of velocity $\sigma_{u_i}^>(\ell_{\text{traj}})$ (case 380).}
\label{fig:lem_RuuL}
\end{figure}

Because $\mathcal{R}_{uu}^L(\tau)$ is expected to depend slightly on the width of the filter used to get rid of noise, we implement the dt-method \cite{machicoane2017estimating, machicoane2017multi} to compute $\mathcal{R}_{uu}^L(\tau)$ (and $\mathcal{R}_{aa}^L(\tau)$ in section~\ref{sec:lem_lagrangian_stats} B) directly from the unfiltered trajectories for an optimal handling of experimental noise. Figure~\ref{fig:lem_RuuL} (top) displays the corresponding autocorrelation functions for the three velocity components of case $380$. Although the flow shows good isotropy and homogeneity, it can be observed that the $y$-component of the velocity, $\mathcal{R}_{uu}^y(\tau)$, exhibits a slower decay than the other components $\mathcal{R}_{uu}^x(\tau)$ and $\mathcal{R}_{uu}^z(\tau)$. Accordingly, the shape of $\mathcal{R}_{uu}^y(\tau)$ is found to be much closer to an exponential shape at large time lags as observed in numerical simulations \cite{yeung1989lagrangian}.

Such effect can be understood as a finite measurement volume effect already mentioned in Ref.~\cite{ouellette2006small}, which leads to a rapid decrease of the Lagrangian velocity variance when increasing the track length. This is visible in figure~\ref{fig:lem_RuuL} (bottom) which displays the rms velocity of trajectories longer than a threshold $\ell_{\text{traj}}$, noted $\sigma_{u_i}^>(\ell_{\text{traj}})$. This evolution is due to the restricted measurement volume: the fastest trajectories rapidly leave the measurement volume whereas the slowest ones stay longer, leading to an over-representation of the slow tracks among the longest ones and a decrease of the velocity standard deviation. The measurement volume is longer along $y$, thus the evolution is slower for the $y$ component. In the same way the measurement volume is the shortest along $x$, thus the finite measurement volume effect is stronger. We shall therefore use the $y$-component of the velocity to estimate the Lagrangian integral time-scale defined as $T_L =\int_0^\infty \mathcal{R}_{uu}^L(\tau)\mathrm{d}\tau$. However, computing directly this time-scale would require trajectories much longer than $T_L$ which is not possible in the present situation. To overcome this problem, we choose to fit the autocorrelation functions $\mathcal{R}_{uu}^L(\tau)$. For this purpose, two different stochastic models may be used: the two-layer model from Ref.~\cite{sawford1991reynolds} and the infinite-layer model recently proposed by Ref.~\cite{viggiano2020modelling}. Those two models give really close results for velocity autocorrelation functions, but for acceleration autocorrelation function the infinite-layer model is much more accurate. Here we use the infinite-layer for velocity and acceleration to be consistent. For velocity autocorrelation function the infinite-layer model reads 
\begin{equation}
\mathcal{R}_{uu}^L(\tau) = \dfrac{1}{2\erfc{\left(\dfrac{\tau_2}{\tau_1}\right)}\exp\left(\dfrac{|\tau|}{\tau_1}\right)} \left[1+\erf{\left(\dfrac{|\tau|}{2\tau_2}-\dfrac{\tau_2}{\tau_1}\right)}+\exp{\left(\dfrac{2|\tau|}{\tau_1}\right)}\erfc{\left(\dfrac{|\tau|}{2\tau_2}+\dfrac{\tau_2}{\tau_1}\right)}\right],
\label{eq:corrv_fit}
\end{equation}
where $\tau_1$ and $\tau_2$ are two fit parameters and correspond to large and short time scales, respectively. They are related to the Lagrangian integral time scale and the dissipative time scale. The error function $\erf(x) = \tfrac{2}{\sqrt{\pi}} \int_0^x e^{-t^2} \:\mathrm{d}t$ is used, with the associated complementary error function $\erfc(x) = 1 - \erf(x)$. 

As shown in figure~\ref{fig:lem_RuuL} (top), the fit close follows the Lagrangian autocorrelation function up to the estimated integral-scale for the $y$-component, $\mathcal{R}_{uu}^y(\tau)$, which is less affected by finite volume effects. We therefore use fits of $\mathcal{R}_{uu}^y(\tau)$ for the various runs to discuss the evolution of $T_L$ with $\text{Re}_\lambda$. The values obtained from the integration of the fits are given in table \ref{tab:lem_lagrangian_parameters} together with the corresponding values of $\tau_1$ and $\tau_2$. We find that $T_L$ follows the same scaling as $T_E$, the ratio being $T_E/T_L \approx 2.4$, which is consistent with estimates obtained from constants derived from the acceleration autocorrelation function $\mathcal{R}_{aa}^L(\tau)$ and to the second-order structure function $S_2^L(\tau)$. We shall therefore first explain how these quantities are computed from the data before discussing how the various Lagrangian time-scales scale with the Reynolds number.

\subsubsection{Second-order structure functions}

\begin{figure}
\centerline{\hspace{0cm}\includegraphics[width=0.6\textwidth]{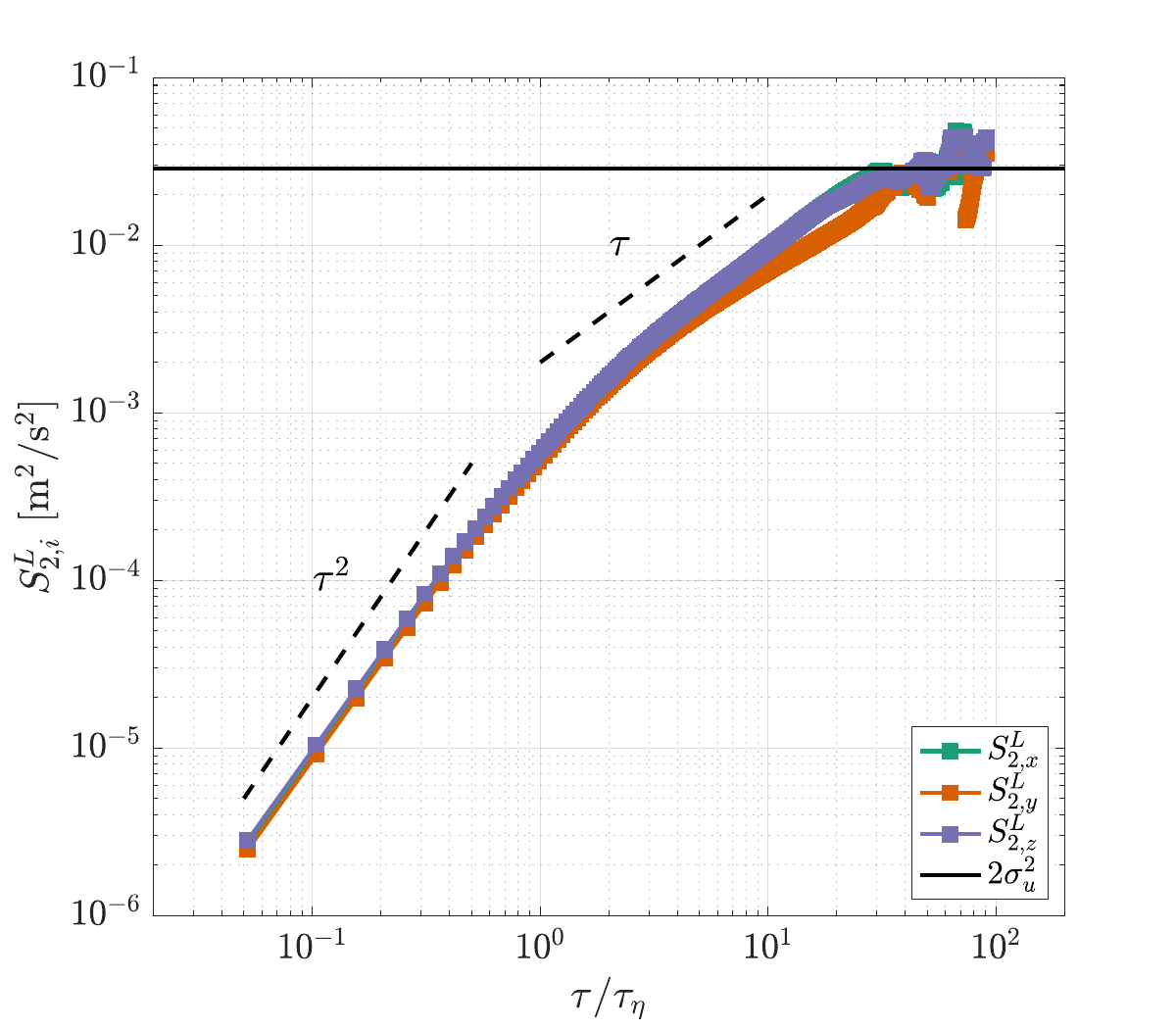}
\hspace{0.0cm}}
\caption{Lagrangian second order structure functions $S_2^L(\tau)$ for the three velocity components as a function of the normalized time $\tau/\tau_\eta$. Linear ($\propto\tau$) and quadratic ($\propto\tau^2$) power-law scalings are shown for reference.}
\label{fig:lem_S2L_corrected}
\end{figure}

Similar to what was done to $\mathcal{R}_{uu}^L(\tau)$, we adapt a corrective term $2\sigma_u^2/(\langle u(t+\tau)^2 \rangle + \langle u(t)^2 \rangle)$ \footnote{The present definition uses an arithmetic mean instead of a geometric mean, which proved to be efficient for computing $S_2^L(\tau)$ \cite{dumont2021phd}. However the two estimates of the velocity variance give here close results and the second order structure function does not change much when changing the way the average is computed} to compensate for the bias of finite length of the trajectories on $\sigma_u^2$
, and thus the Lagrangian second-order structure function is defined as \cite{dumont2021phd}
\begin{equation}
S_2^L(\tau) = \dfrac{2 \sigma_u^2}{\langle u(t+\tau)^2 \rangle + \langle u(t)^2 \rangle} \langle | u(t+\tau) - u(t) |^2 \rangle,
\label{eq:S2L}
\end{equation}
where $u$ corresponds to any filtered velocity component. We also note that the Lagrangian second-order structure function can be analytically derived directly from the velocity autocorrelation function as $S_2^L(\tau)=2\sigma_u^2(1-\mathcal{R}_{uu}^L(\tau))$. The results for an intermediate Reynolds are presented here and the same analysis is performed for the six others. In figure~\ref{fig:lem_S2L_corrected} we show $S_2^L(\tau)$ computed with Eq.~\eqref{eq:S2L} for case 380. For large $\tau$, $S_2^L$ approaches the value $2\sigma_u^2$. For small $\tau$ the flow shows good isotropy between the 3 components and follows the quadratic power-law scaling. We note that in our experiments, small time scales are not fully resolved in favor of having more statistically significant data for all other time scales. The quadratic power-law scaling is enforced by the Gaussian filter we used, which holds true for time scales smaller than the width of the Gaussian filter. For the case 380, this corresponds to $\tau < 0.4 \tau_\eta$ (or 8 frames). In between the large and small time scales, $S_2^L(\tau)$ follows a power-law scaling with an exponent close to 1 but exhibits some anisotropy between the three components. This is probably due to the large scale anisotropy of the shape of the measurement volume which are reflected in the anisotropy in velocity components.

In analogy to Kolmogorov phenomenology for Eulerian statistics, at inertial scales ($\tau_\eta \ll \tau \ll T_L$) the Lagrangian second-order structure functions is expected to scale as
\begin{equation}
S_2^L(\tau) = C_0\varepsilon\tau,
\label{eq:S2L_C0}
\end{equation}
 \cite{yeung2002lagrangian, toschi2009lagrangian} where $C_0$ is an universal constant. The constant $C_0$ depends on the Reynolds number and tends to $C_0 \approx 7$ for $\text{Re}_\lambda \rightarrow \infty$ from DNS results \cite{sawford1991reynolds, yeung2006reynolds, biferale2008lagrangian, sawford2011kolmogorov} and was observed to depend on the direction in anisotropic von K\'arm\'an flows \cite{ouellette2006small}. The usual method to determine $C_0$ is to consider the compensated structure function such that
\begin{equation}
C_0 = \max \left( \dfrac{S_2^L(\tau)}{\varepsilon \tau} \right).
\label{eq:C0}
\end{equation}
If the inertial range is long enough (i.e. at very high Reynolds numbers), a plateau can be observed, but usually the value of $C_0$ is based on the peak value rather than the values of a clear plateau \cite{yeung2006reynolds, sawford2011kolmogorov}. An alternate definition of this constant, discussed in particular by Ref.~\cite{huck2019lagrangian}, is
\begin{equation}
C_0^* = \max \left( \dfrac{1}{\varepsilon} \dfrac{\mathrm{d}S_2^L(\tau)}{\mathrm{d}\tau}\right),
\label{eq:C0*}
\end{equation}

which is reached at the zero-crossing time of the acceleration autocorrelation function, $\tau_0$, such that $R_{aa}^L(\tau = \tau_0) = 0$ (because for a statistically stationary signal $R_{aa}^L(\tau) = -\ddot R_{uu}^L(\tau)$). Thus we also have
\begin{equation}
C_0^* = \dfrac{1}{\varepsilon} \left. \dfrac{\mathrm{d}S_2^L(\tau)}{\mathrm{d}\tau}\right|_{\tau=\tau_0}.
\label{eq:C0*_tau0}
\end{equation}
This alternative definition presents several advantages:
\begin{itemize}
\item From measurements of $\tau_0$ (presented in the following), we exactly know where to consider the derivative of $S_2^L$ to estimate $C_0^*$.
\item The peak for $C_0$ appears in the inertial range, whereas the peak for $C_0^*$ is at the transition between the dissipative and inertial ranges such that the estimation of this quantity is less affected by finite volume effects.
\item Based on simple models of $S_2^L$ \cite{sawford1991reynolds, viggiano2020modelling}, we can show that
\begin{equation}
C_0^* \approx \dfrac{2 \sigma_u^2}{\varepsilon T_L}.
\label{eq:C0*_TL}
\end{equation}
Based on these models, which simply give an analytical formula for $S_2^L$, it can be shown that this value is an upper bound (reached for infinite Reynolds number) and both $C_0$ and $C_0^*$ are close, though $C_0^*$ is closer.
\end{itemize}

\begin{figure}
\centerline{\includegraphics[width=0.8\textwidth]{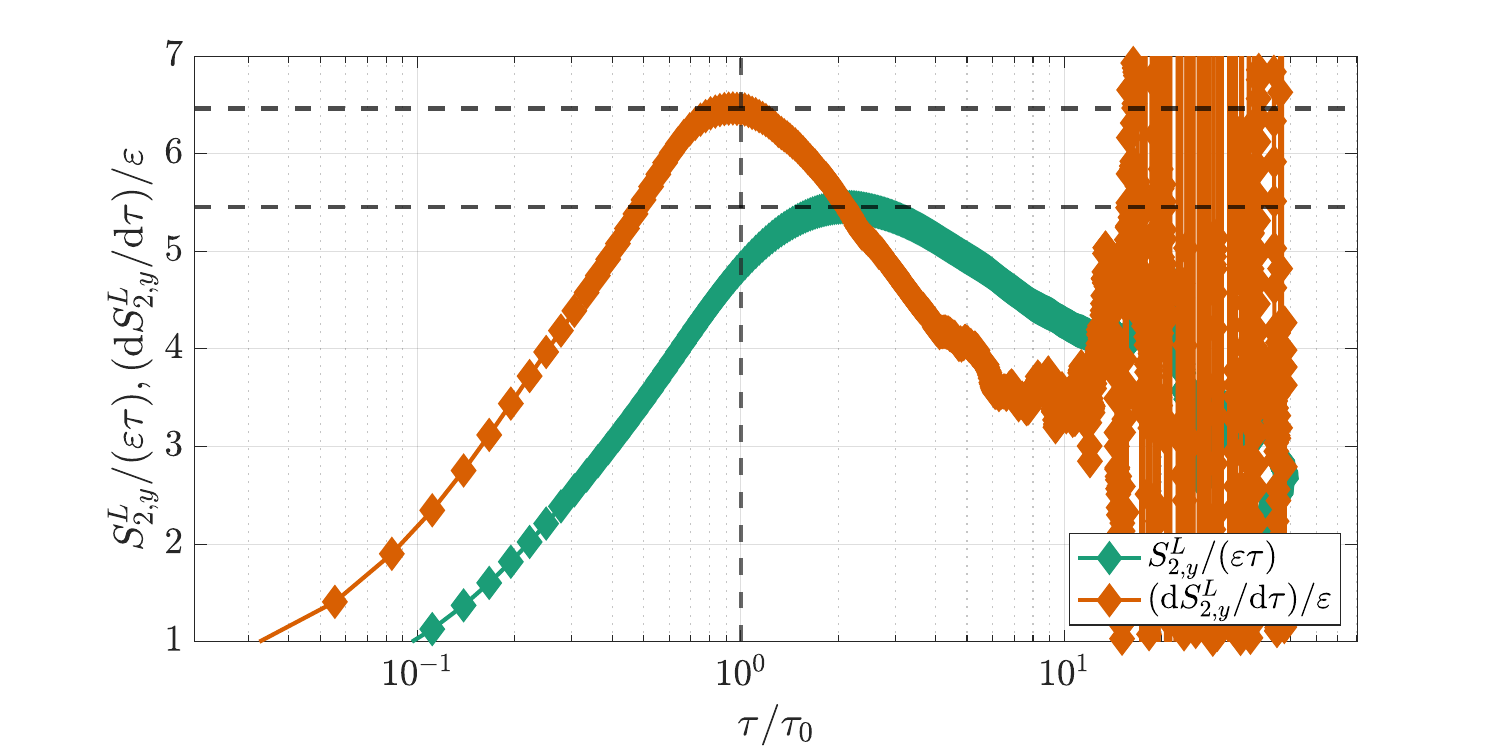}}
\caption{$S_{2,y}^L/(\varepsilon \tau)$ and $\displaystyle \frac{dS_{2,y}^L}{d\tau}\frac{1}{\varepsilon}$ as a function of $\tau/\tau_0$ where $\tau_0$ is the zero crossing time of the acceleration autocorrelation function.
Only the $y$ component is shown for simplicity (case 380). The horizontal dashed lines indicate the value of $C_0$ and $C_0^*$.}
\label{fig:lem_C0_corrected}
\end{figure}

\begin{figure}
\centerline{\includegraphics[width=0.8\textwidth]{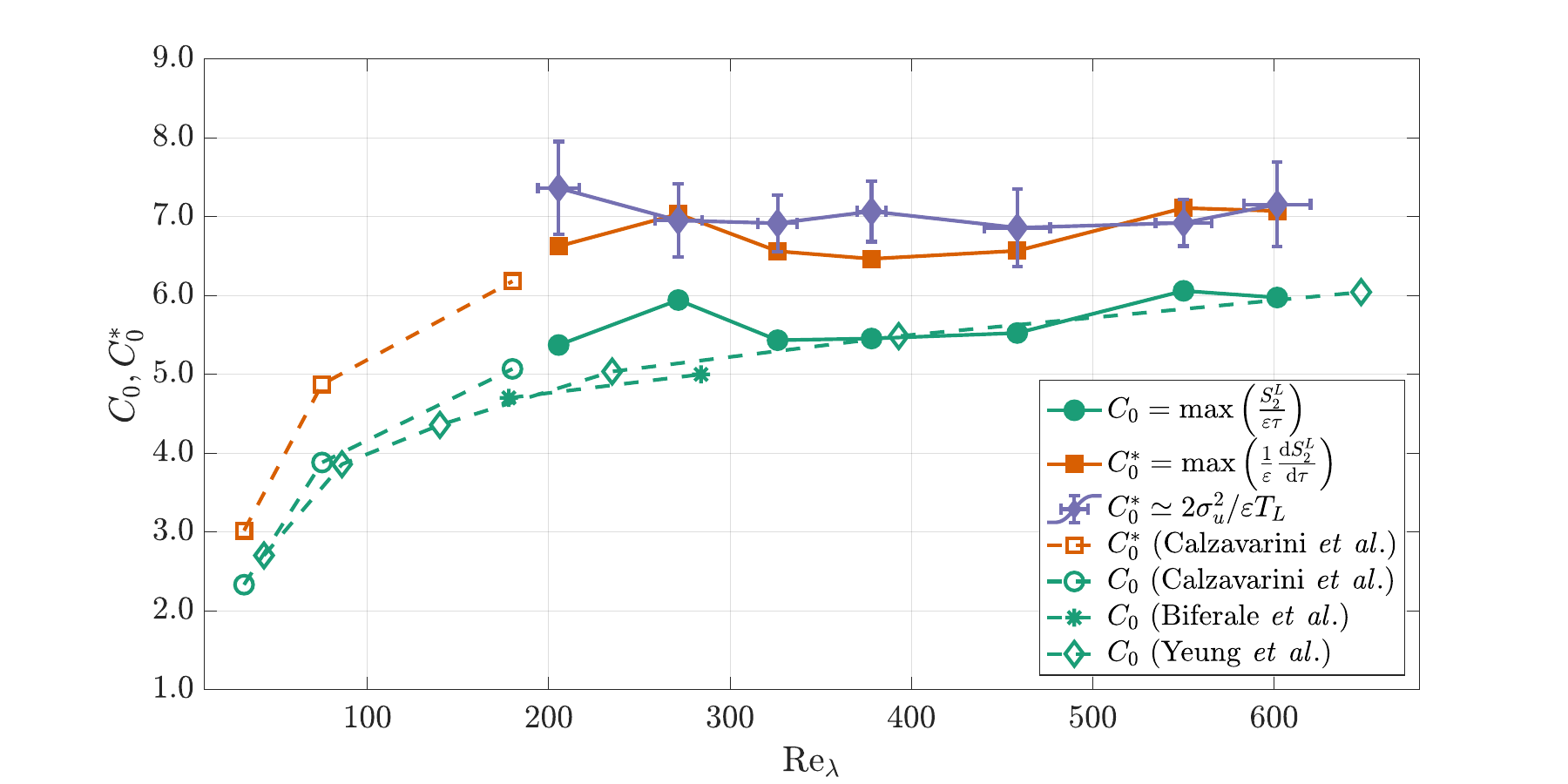}}
\caption{Constants $C_0$ and $C_0^*$ from different estimations as a function of the Taylor-based Reynolds number $\text{Re}_\lambda$. Only the $y$ component of the velocity was used. DNS results are shown for comparision: open orange squares ($C_0^*$) and open green circles ($C_0$) are processed with the data from \citet{calzavarini2009acceleration,CALZAVARINI2012237}, green stars are from \cite{biferale2008lagrangian} and green diamonds are from \cite{yeung2006reynolds}.}
\label{fig:lem_C0_Re}
\end{figure}

Figure~\ref{fig:lem_C0_corrected} displays, for the case $\text{Re}_\lambda=380$, $S_{2}^L/(\varepsilon \tau)$ and $\displaystyle \frac{1}{\varepsilon}\frac{dS_{2}^L}{d\tau}$ as functions of the reduced time $\tau/\tau_0$, where $\tau_0$ is the zero crossing of the acceleration autocorrelation function. As explained, the constant $C_0$ determined from Eq.~\ref{eq:C0} appears at a time lag larger than $\tau_0$, and is smaller than $C_0^\star$ computed from Eq.~\ref{eq:C0*}. We also note that the value of $C_0^\star$ determined by Eq.~\eqref{eq:C0*_tau0} is the same as given by Eq.~\eqref{eq:C0*} because the maximum is found to occur at $\tau_0$ up to experimental precision, which helps to determine the value of $C_0^\star$ despite the noise occurring at large time lags amplified by the numerical derivation. For simplicity, we present only the results for the $y$ component; however, the same analysis has been conducted for the other two components and for all cases. In figure~\ref{fig:lem_C0_corrected} we have the constant $C_0 \approx 5.5$ and $C_0^\star \approx 6.5$, slightly smaller than the typical value ($C_0 \approx 7$ when $\text{Re}_\lambda\rightarrow \infty$) reported in previous work \cite{sawford1991reynolds, yeung2006reynolds, biferale2008lagrangian, sawford2011kolmogorov}. Figure~\ref{fig:lem_C0_Re} shows the values of $C_0$ and $C_0^*$ obtained experimentally for the different values of $Re_\lambda$ we explored. It can be seen that the agreement between $C_0^\star$ and the upper-bound $2 \sigma_u^2/\varepsilon T_L$ improves for increasing $\text{Re}_\lambda$. The value of $C_0^*$ is approximately 6.8 and appears to be independent of $\text{Re}_\lambda$ within the measurement range. We also report in figure~\ref{fig:lem_C0_Re} values for $C_0$ reported in previous numerical studies from DNS of HIT~\citep{biferale2008lagrangian, sawford2013lagrangian, yeung2006reynolds}. To the best of our knowledge, no such values have been reported for $C_0^*$. We therefore show estimates of $C_0^*$ from past numerical simulations carried in our group~\cite{CALZAVARINI2012237,calzavarini2009acceleration} at lower values of $Re_\lambda$. The agreement with our experiments is very good for both $C_0$ and $C_0^*$. We will use the value of $C_0^*$ in the following sections (see table~\ref{tab:lem_lagrangian_parameters}).

\subsection{Acceleration statistics}

\begin{figure}
\includegraphics[width=1\textwidth]{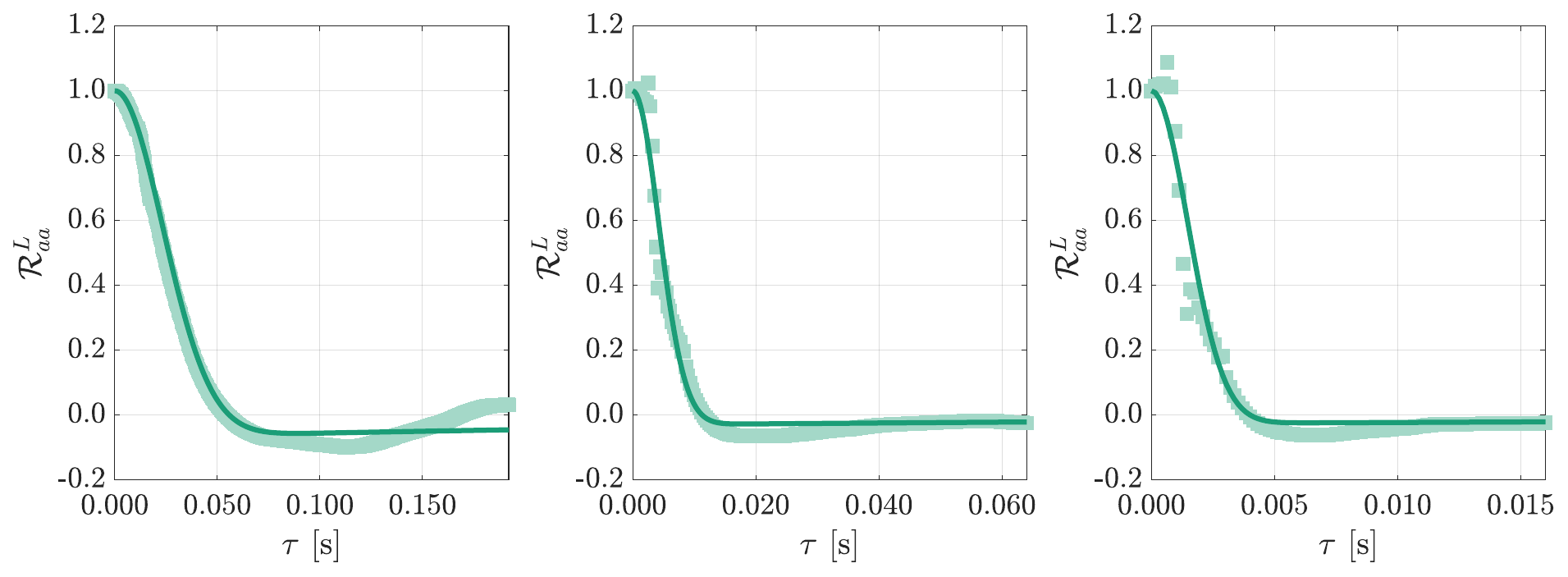}
\caption{Normalized Lagrangian autocorrelation functions $\mathcal{R}_{aa}^L$ as a function of the time lag $\tau$ (case 380). Solid line: fits of infinite-layer model. From left to right: case 200, case 380, case 600.}
\label{fig:corray_tau}
\end{figure}

\begin{figure}
\includegraphics[width=0.8\textwidth]{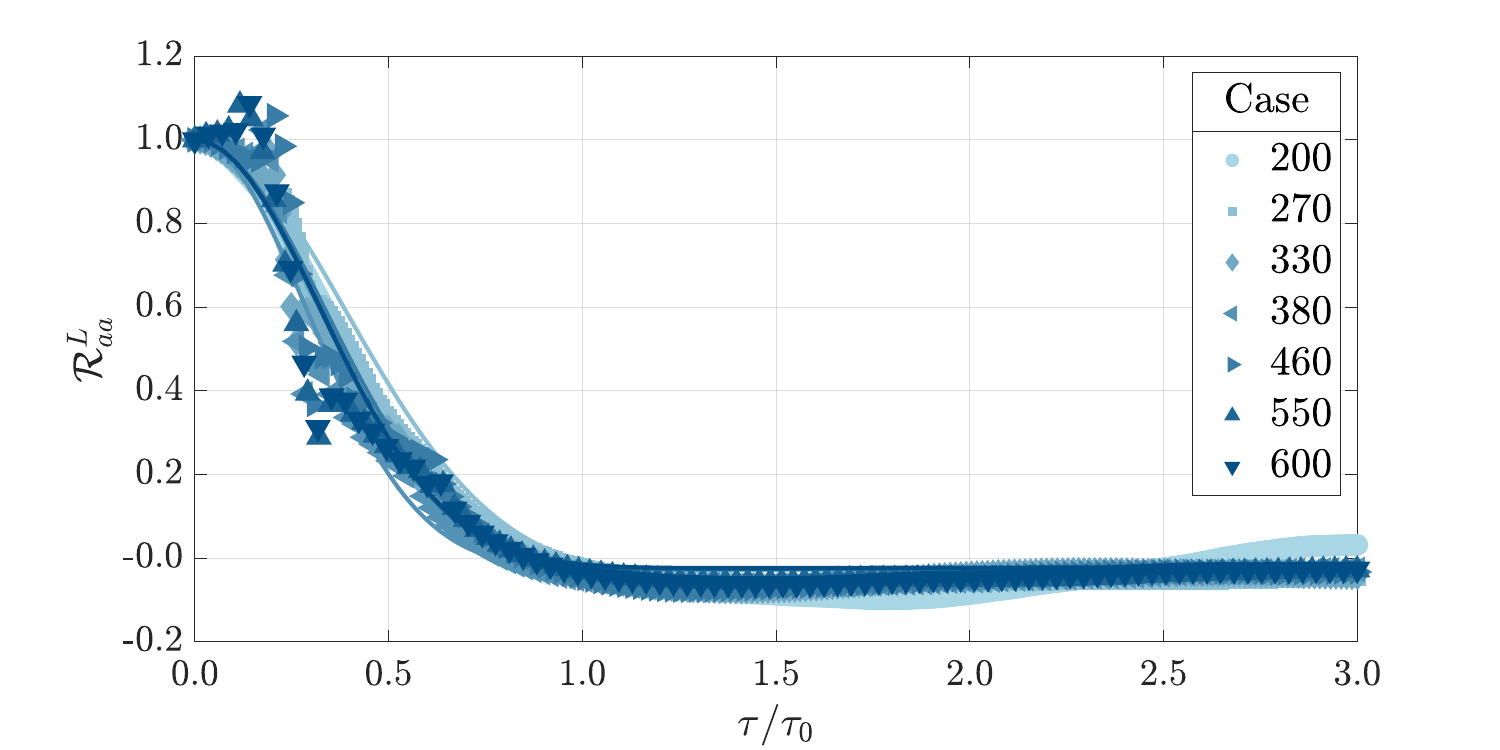}
\caption{Normalized Lagrangian autocorrelation functions $\mathcal{R}_{aa}^L$ for the $y$ components as a function of the normalized time $\tau/\tau_0$. Solid line: fits of infinite-layer model. The plot is zoomed in at short time scales for clarity.}
\label{fig:lem_RaaL}
\end{figure}

We now turn to the analysis of the Lagrangian autocorrelation function of acceleration $R_{aa}^L(\tau) = \langle a(t+\tau) a(t) \rangle$ (normalized $\mathcal{R}_{aa}^L = R_{aa}^L/\sigma_a^2$). Unlike the velocity autocorrelation function, the evolution of the acceleration is dominated by short time scales and is almost insensitive to any finite volume bias. Thus, we adopt the classical definition of $R_{aa}^L(\tau)$, as the inclusion of a corrective term does not affect the results. As we already explained, the fitting of infinite-layer model \cite{viggiano2020modelling} is more accurate for accelerations statistics.
Using this model the Lagrangian autocorrelation function for acceleration is given by
\begin{equation}
\begin{split}
R_{aa}^L(\tau) = & \dfrac{\sigma_a^2}{2\left[ \dfrac{\tau_1}{\sqrt{\pi}\tau_2}\exp\left(-\dfrac{\tau_2^2}{\tau_1^2}\right) - \erfc\left(\dfrac{\tau_2}{\tau_1}\right) \right]} \left[ \dfrac{2\tau_1}{\sqrt{\pi}\tau_2}\exp\left(-\dfrac{\tau^2}{4\tau_2^2} - \dfrac{\tau_2^2}{\tau_1^2}\right) \right.\\\\
&\left. - \exp\left(-\dfrac{|\tau|}{\tau_1}\right)\left(1+\erf\left(\dfrac{|\tau|}{2\tau_2}-\dfrac{\tau_2}{\tau_1}\right)\right) - \exp\left(\dfrac{|\tau|}{\tau_1}\right)\erfc\left(\dfrac{|\tau|}{2\tau_2}+\dfrac{\tau_2}{\tau_1}\right) \right],
\label{eq:corra_fit}
\end{split}
\end{equation}

where the time scales $\tau_1$ and $\tau_2$ are, respectively, representative of the large and dissipative Lagrangian scales as in equation~\eqref{eq:corrv_fit}. The autocorrelation functions of acceleration for case 200, case 380, case 600 are presented in figure~\ref{fig:corray_tau} with the fits of equation~\eqref{eq:corra_fit}. For clarity only the component at the longest dimension $\mathcal{R}^y_{aa}$ is shown. For the case 200 ($\tau_\eta = \SI{31}{ms}$), we do not observe any marked signature of noise at small scale and the fit is accurate from the smallest to the largest considered time lags. For the case 380 ($\tau_\eta = \SI{6.4}{ms}$), the part before the zero-crossing time appears to be slightly affected by noise but without compromising the determination of $\sigma_a^2$. For the case 600 ($\tau_\eta = \SI{1.7}{ms}$), the noise affects the behavior of $\mathcal{R}^y_{aa}$ at small time with large oscillations. Higher time resolution is required to improve it. Nevertheless, the fit~\eqref{eq:corra_fit} surprisingly gives satisfying evolutions up to $\tau = \tau_0$, even though with a strong signature of noise as for the case 600 at short time, and in particular the zero-crossing time $\tau_0$ appears to be quite robust. We also present in figure~\ref{fig:lem_RaaL} that the seven normalized Lagrangian autocorrelation functions nearly collapse into a single curve by normalizing the time $\tau$ with the zero-crossing time $\tau_0$. This is because the shape of $\mathcal{R}^y_{aa}(\tau/\tau_0)$ is dominated by its evolution at lags $\tau \leq \tau_0$, but the evolution of these curves at large lags is determined by the ratio $T_L/\tau_\eta$ which changes with $\text{Re}_\lambda$ such that it is impossible that all these curves superpose onto a master curve.

\begin{figure}
\centerline{\includegraphics[width=0.8\textwidth]{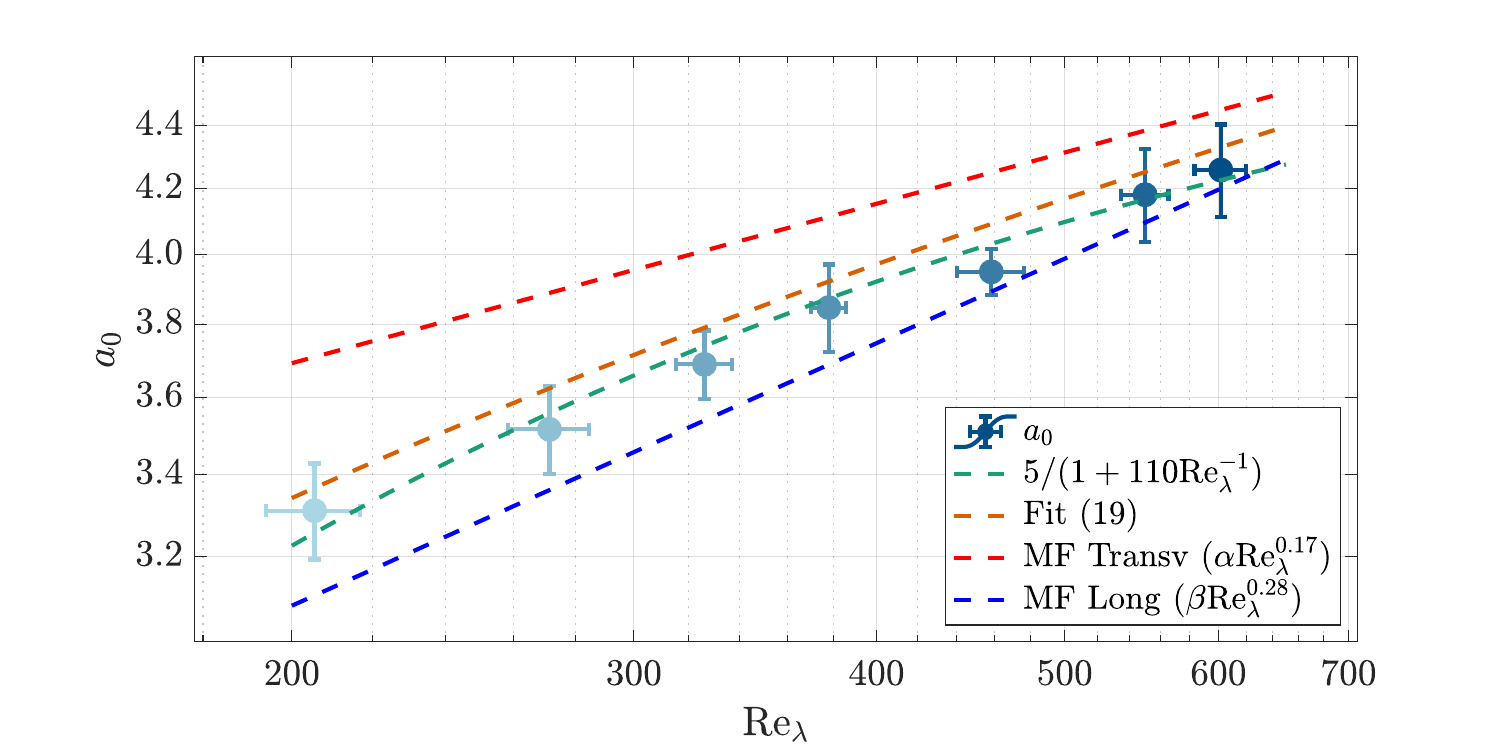}}
\caption{Acceleration constant $a_0$ as a function of the Taylor-based Reynolds number $\text{Re}_\lambda$. The green dashed line refers to the fit $a_0 = 5/(1+110\text{Re}_\lambda^{-1})$ given in \cite{sawford2013lagrangian}. The orange dashed line is the fit (\ref{eq:a0}) adapted from \cite{Bec_Vallee_2024}. The red curve is obtained from multifractal prediction \citep{lanotte2013new} using bridge relation for transverse increments (MF Tranv) leading to $\gamma = 0.17$, and the blue curve is obtained using the bridge relation for longitudinal increments (MF long) leading to $\gamma = 0.28$. The prefactor $\alpha$ and $\beta$ are arbitrary constants as the multifractal prediction valid scaling-wise.}
\label{fig:lem_a0}
\end{figure}

The fitting of $\mathcal{R}_{aa}^L$ gives estimates of three parameters: $\sigma_a^2$, $\tau_1$, and $\tau_2$. The values of $\tau_1$ and $\tau_2$ will be discussed in the next section. The acceleration variance $\sigma_a^2$ is traditionally characterized by the scaling constant $a_0$ through the Heisenberg--Yaglom relation \cite{monin1975statistical, laporta2001fluid}:
\begin{equation}
\sigma_a^2 = a_0 \dfrac{\varepsilon^{3/2}}{\nu^{1/2}}. 
\end{equation}

The dependence of the constant $a_0$ on $\text{Re}_\lambda$ can be fitted by $a_0 = 5/(1+110\text{Re}_\lambda^{-1})$ \cite{sawford2013lagrangian} and is found to be around 4 in DNS \cite{sawford1991reynolds, vedula1999similarity, calzavarini2009acceleration, sawford2013lagrangian} and other experiments \cite{voth2002measurement, qureshi2007turbulent, brown2009acceleration}. Another formula for $a_0$ was recently proposed \cite{Bec_Vallee_2024} which can be adapted to 
\begin{equation}
a_0 \approx \frac{a\text{Re}_\lambda^{2\gamma}}{[1+R^*/\text{Re}_\lambda]^{1-2\gamma}}, \label{eq:a0}
\end{equation}
where $\gamma=0.078$ is the exponent of asymptotic scaling $\sigma_a^2 \propto Re_\lambda ^{2\gamma}$ for moderate and large Reynolds numbers, $a = 1.67$ and $R^*= 34.64$ are two fit parameters. 

Figure~\ref{fig:lem_a0} shows the present estimations of $a_0(\text{Re}_\lambda)$ together with both fits. While we can notice that both fits provide reasonable estimation for the constant $a_0$, the present data are found to be very close to $a_0 = 5/(1+110\text{Re}_\lambda^{-1})$, which was derived from a fit of DNS data, thus smaller than values classically reported in von K\'arm\'an flows \cite{voth2002measurement,volk2011dynamics} for which turbulence is dominated by the presence of a stagnation point near the center \cite{huck2017production}, leading to an increase of acceleration variance \cite{lee_2015}. We also show in figure~\ref{fig:lem_a0} two scalings proposed for the Reynolds dependence of $a_0$ in the framework the multifractal approaches\citep{lanotte2013new}, which are also in very good agreement with the present experimental data..

\subsection{Discussion on Lagrangian time scales}
\begin{figure}
\centerline{\hspace{0cm}\includegraphics[width=0.5\textwidth]{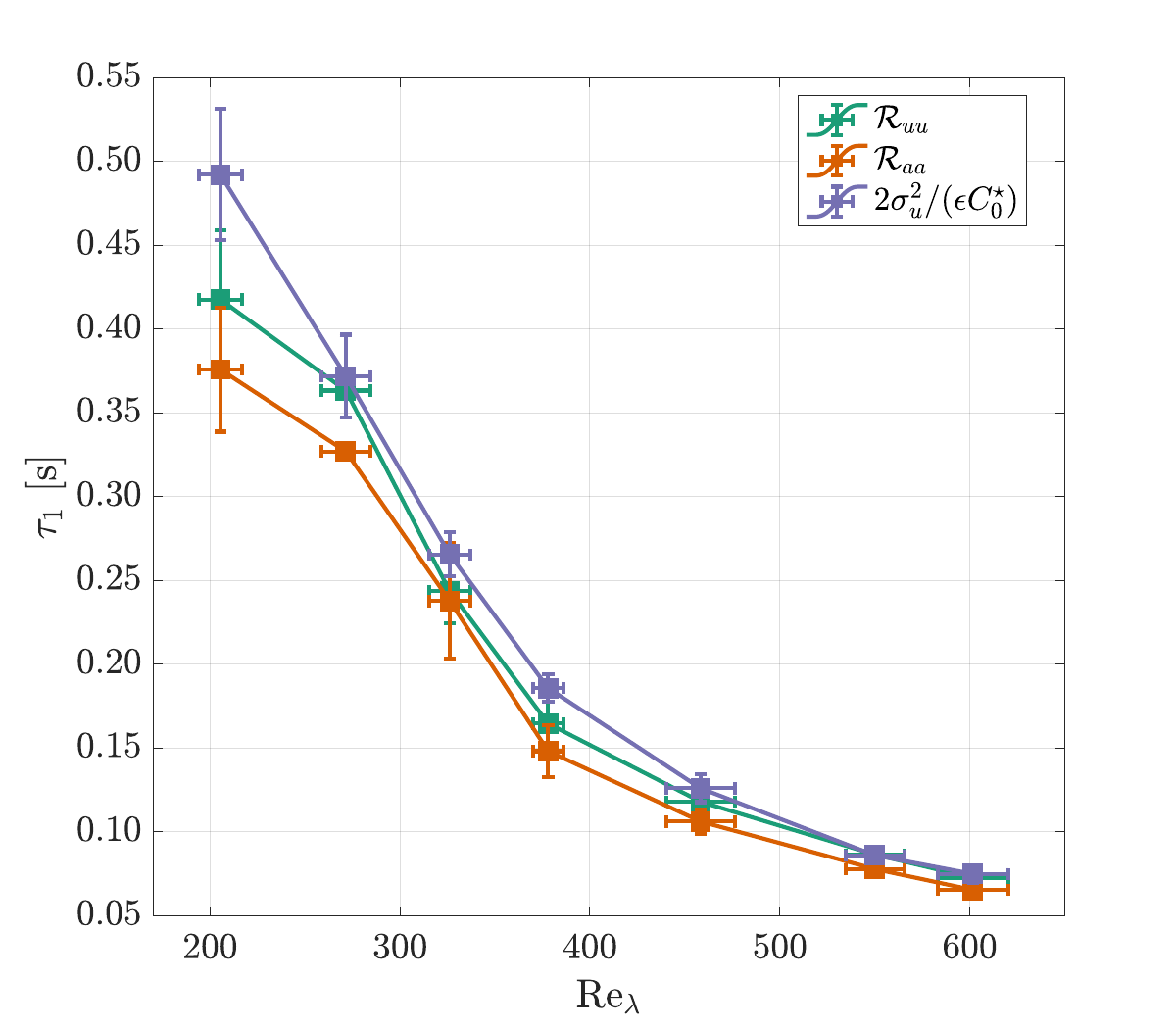}
\hspace{0cm}\includegraphics[width=0.5\textwidth]{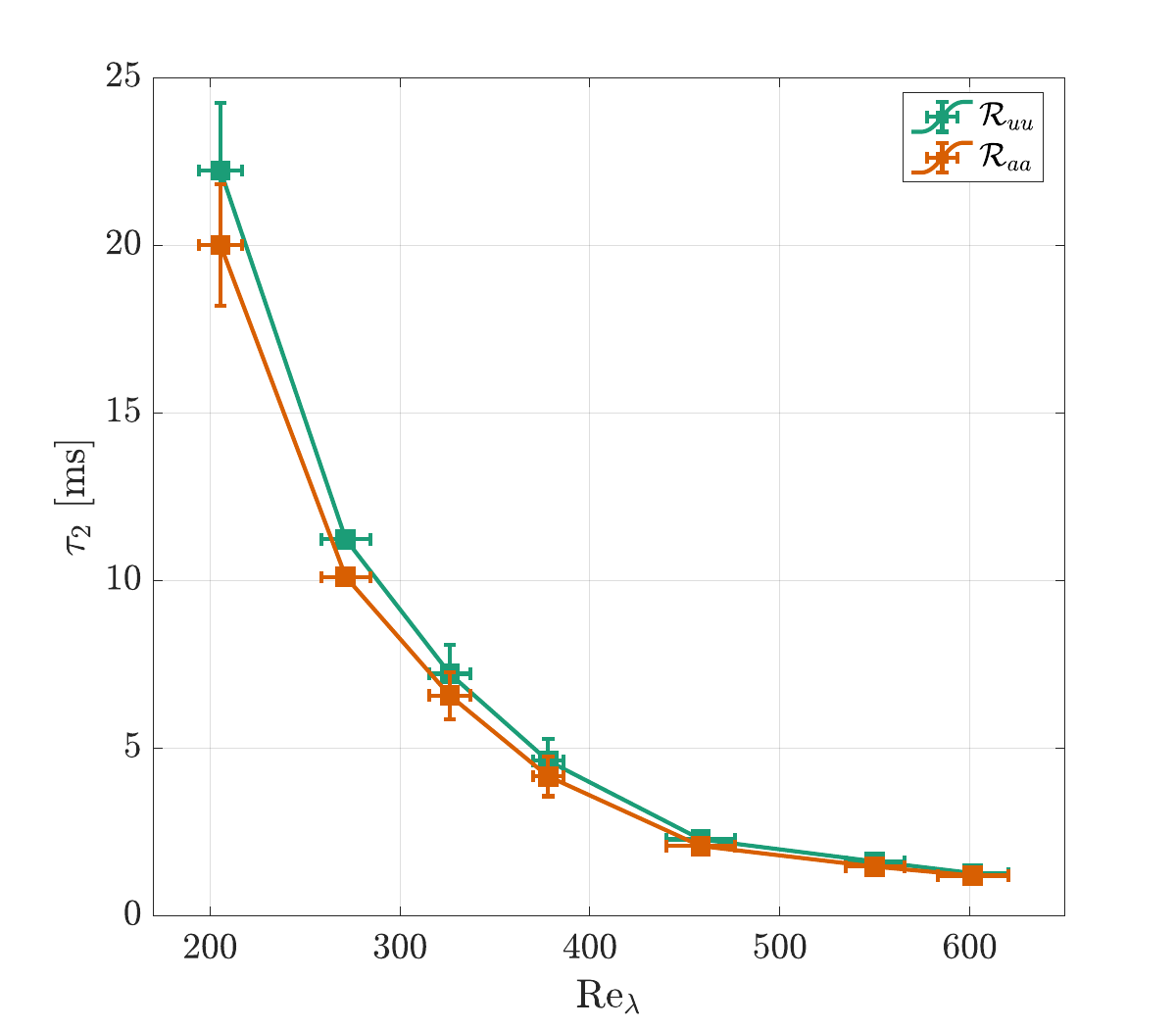}}
\caption{Lagrangian time scales from different estimations as a function of the Taylor-based Reynolds number $\text{Re}_\lambda$. Left: Large time scale $\tau_1$. Right: Short time scale $\tau_2$.}
\label{fig:lem_tau}
\end{figure}

With the fits \eqref{eq:corrv_fit} and \eqref{eq:corra_fit} for velocity and accelerations, two time scales $\tau_1$ and $\tau_2$, corresponding to large and small time scales, can be extracted. As presented in figure~\ref{fig:lem_tau}, $\tau_1$ and $\tau_2$ from two fits agree reasonably as $\mathcal{R}_{aa}^L = -\ddot{\mathcal{R}}^L_{uu}$. In addition, the infinite-layer model gives $T_L = \tau_1 \exp(-(\tau_2/\tau_1)^2)/\erfc(\tau_2/\tau_1) \approx \tau_1$ with $\tau_2 \ll \tau_1$. This is also examined in figure~\ref{fig:lem_tau} where the estimation of Eq.~\eqref{eq:C0*_TL} provides an upper bounds for $\tau_1$. The observed discrepancy between the estimations from $\mathcal{R}_{aa}^L$ and $\mathcal{R}_{uu}^L$ probably occur because fitting the acceleration autocorrelation function put more weight on the small scales while fitting the velocity autocorrelation function put more weight on the large scales. We thus keep the values of $\tau_1$ from $\mathcal{R}_{uu}^L$ and the values of $\tau_2$ from $\mathcal{R}_{aa}^L$, which are presented in table~\ref{tab:lem_lagrangian_parameters}.

From these two characteristic time scales $\tau_1$ and $\tau_2$, which so far are essentially fitting parameters of consistent parameterizations of different measured quantities (velocity and acceleration autocorrelation functions), different physically relevant time scales can be extracted (the values obtained are presented in table~\ref{tab:lem_lagrangian_parameters}):

\begin{itemize}
\item The zero-crossing time scale $\tau_0$, defined as $\mathcal{R}_{aa}^L (\tau = \tau_0) = 0$ and previously used to obtain $C_0^*$, can be numerically extracted from the fitting function~\eqref{eq:corra_fit}. As shown in figure~\ref{fig:lem_tauatauK} (left) we have $\tau_0 \approx 2.4 \tau_\eta$. We can notice that the ratio of $\tau_0/\tau_\eta$ increases a bit as the Reynolds number increases. It could be due to the presence of noise at high Reynolds number, as already observed in figure~\ref{fig:corray_tau}. The fitting of Eq.~\eqref{eq:corra_fit} is sensitive to noise part of $\mathcal{R}_{aa}^L (\tau = \tau_0)$, particularly at short time scale that $\tau<\tau_0$. Slightly smaller values are usually reported with $\tau_0 \approx 2 \tau_\eta$ from DNS \cite{calzavarini2009acceleration} or experiments \cite{mordant2004threedimensional, volk2008acceleration, huck2019lagrangian}. Here, $\tau_0$ and $\tau_a$ are proportional to $\tau_\eta$, making them both representative of dissipative Lagrangian time scales.

\item The acceleration integral time scale $\tau_a$ is defined as
\begin{equation}
\tau_a = \int_0^{\tau_0} \mathcal{R}^L_{aa}(\tau) \:\mathrm{d}\tau,
\label{eq:taua}
\end{equation}
which can also be extracted from the fitting function~\eqref{eq:corra_fit}. In figure~\ref{fig:lem_tauatauK} (left) the ratio $\tau_a/\tau_\eta$ is rough around $1$ and also weakly depends on $\text{Re}_\lambda$, which are also slightly larger than the value previously reported ($\tau_a \approx 0.9 \tau_\eta$ from Ref.~\cite{calzavarini2009acceleration}). The great interest of this time scale is that, through the kinematic relation $R_{aa}^L(\tau) = -\ddot R_{uu}^L(\tau)$, we have the exact relation
\begin{equation}
\dfrac{\tau_a}{\tau_\eta} = \dfrac{C_0^*}{2a_0},
\label{eq:tauatauK}
\end{equation}
connecting the Lagrangian small time scale to the Eulerian equivalent. This relation~\eqref{eq:tauatauK} is verified in figure~\ref{fig:lem_tauatauK} (right),
especially at small $\text{Re}_\lambda$. The discrepancy between these two ratios increases as $\text{Re}_\lambda$ increases, which again is probably due to the sensitivity of acceleration autocorrelation to the noisy signal at short time scale and the fitting parameters in the infinite-layer model Eq.~\eqref{eq:corra_fit}.

\begin{figure}
\includegraphics[width=0.495\textwidth]{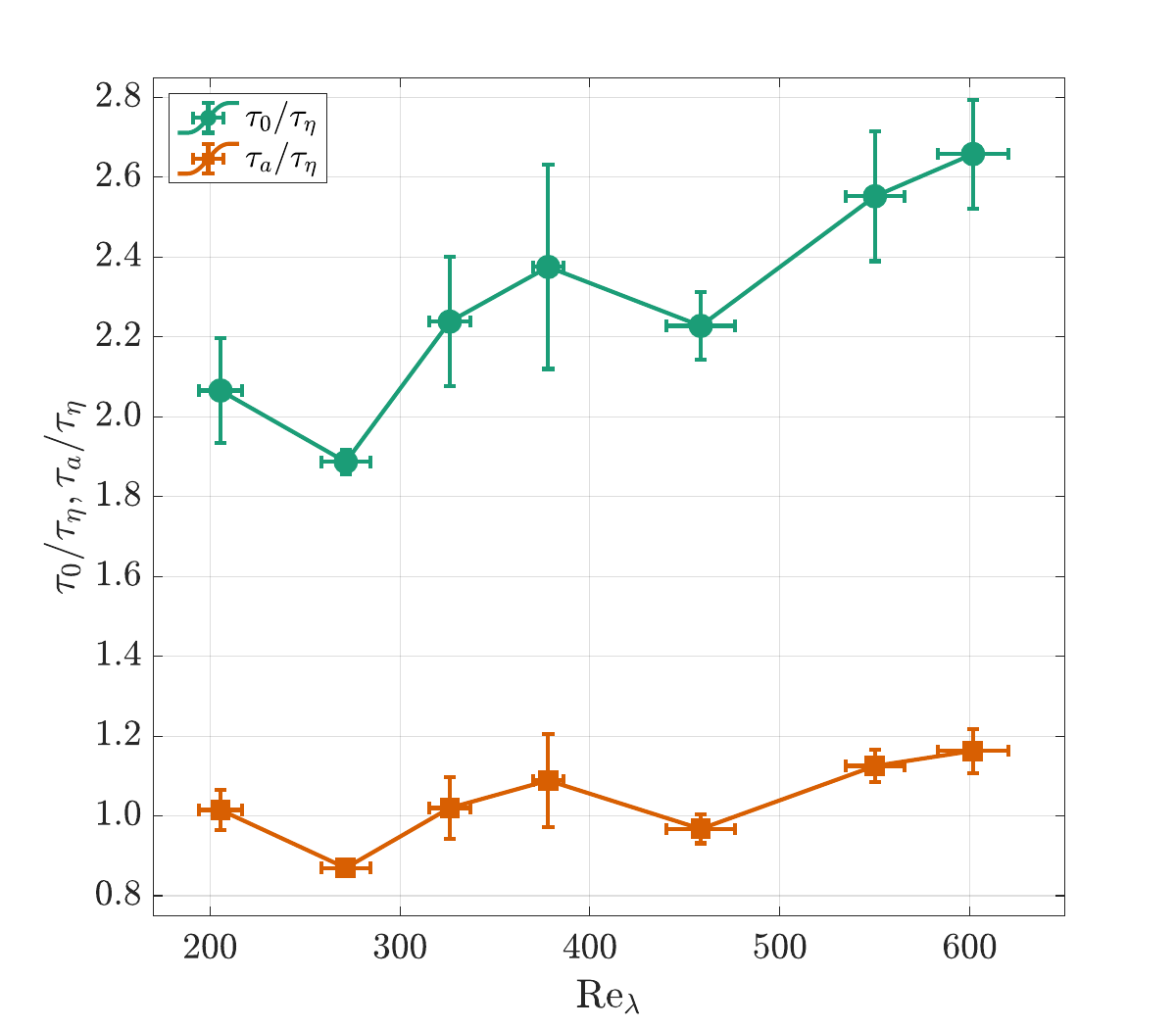}
\includegraphics[width=0.495\textwidth]{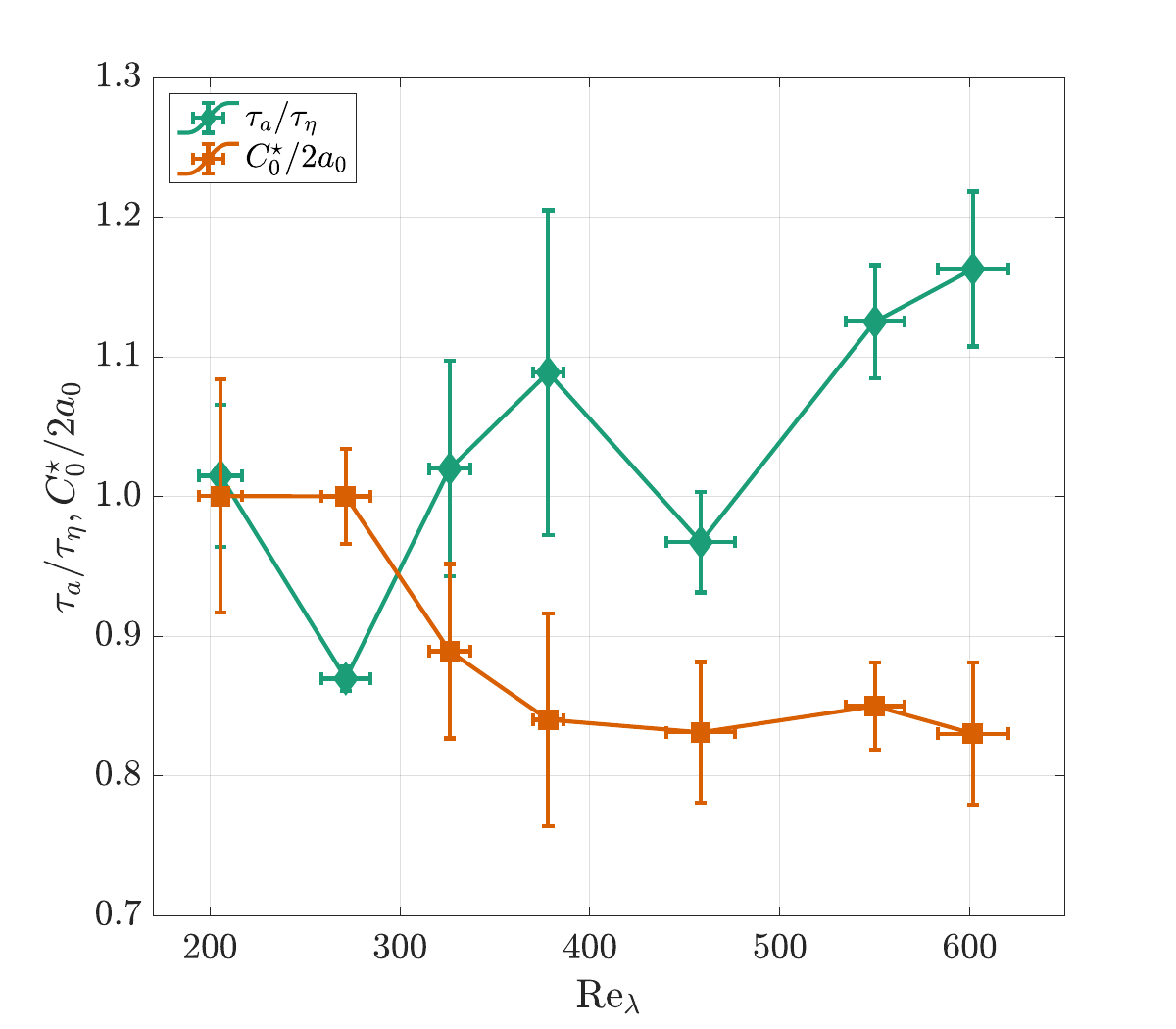}
\caption{Left: Comparison of $\tau_0/\tau_\eta$ and $\tau_a/\tau_\eta$ as a function of the Taylor-based Reynolds number $\text{Re}_\lambda$. Right: Comparison of $\tau_a/\tau_\eta$ and $C_0^*/2a_0$ as a function of the Taylor-based Reynolds number $\text{Re}_\lambda$.}
\label{fig:lem_tauatauK}
\end{figure}

\begin{figure}
\centerline{\includegraphics[width=0.75\textwidth]{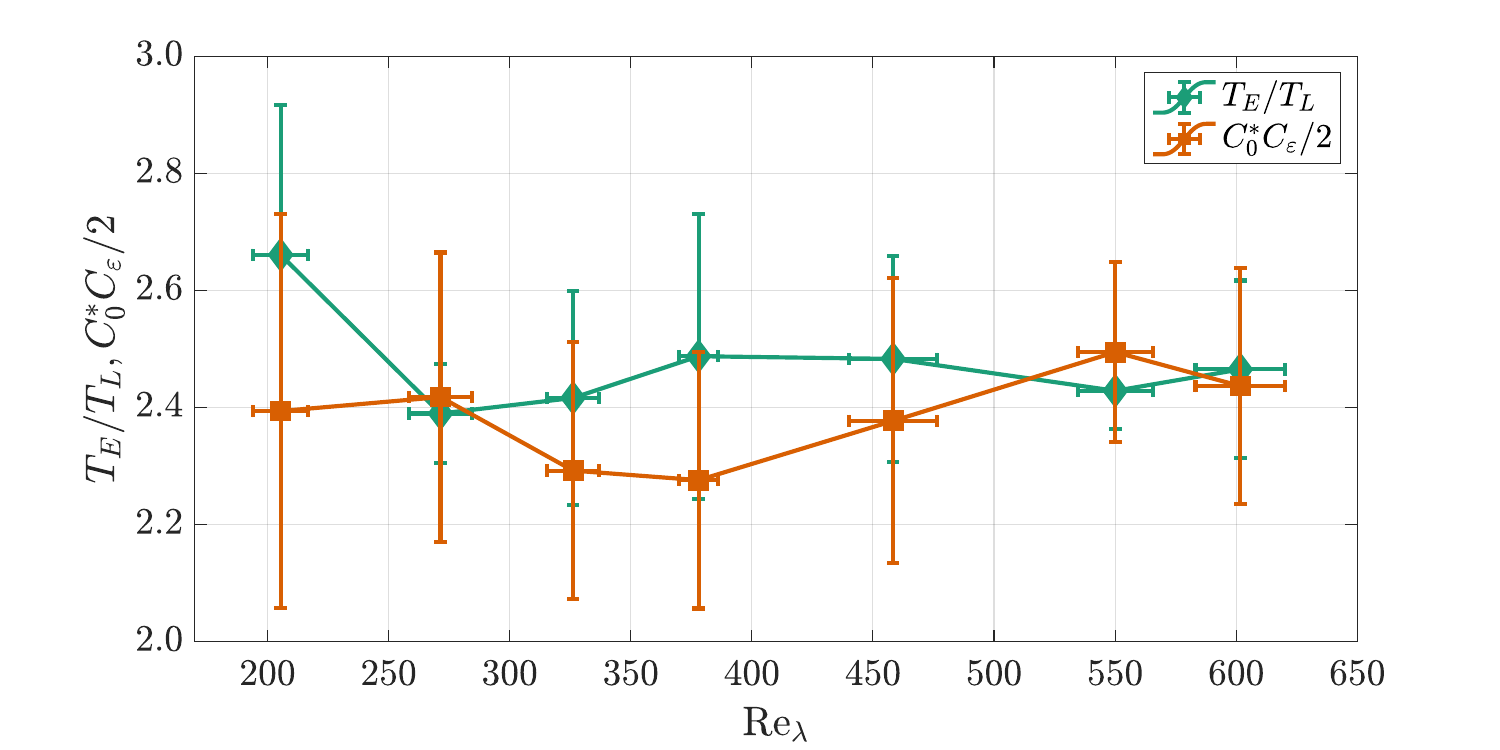}}
\caption{Comparison of $T_E/T_L$ and $C_0^* C_\varepsilon /2$ as a function of the Taylor-based Reynolds number $\text{Re}_\lambda$.
\label{fig:lem_TETL}}
\end{figure}

\item The Lagrangian integral time scale $T_L$ is simply defined as
\begin{equation}
T_L = \int_0^{\infty} \mathcal{R}^L_{uu}(\tau) \:\mathrm{d}\tau.
\label{eq:TL}
\end{equation}
While for the dissipative dynamics we verified that $\tau_a$ can be kinematically related to $\tau_\eta$ through the universal Lagrangian constants $C_0^*$ and $a_0$, the connection between the Eulerian and Lagrangian time scales is a challenging question of turbulence that is hitherto still unclear. A connection can be made through considering the relation~\eqref{eq:C0*_TL} $C_0^* = 2 \sigma_u^2 / (\varepsilon T_L)$ for $T_L$ and the classical relation previously introduced $\varepsilon = C_\varepsilon \sigma_u^2 / T_E$ for $T_E$. We then obtain the following relation
\begin{equation}
\dfrac{T_E}{T_L} = \dfrac{C_0^* C_\varepsilon}{2}.
\label{eq:C0*_TETL}
\end{equation}
This relation~\eqref{eq:C0*_TETL} is well verified as shown in figure~\ref{fig:lem_TETL} with $T_E/T_L \approx 2.4$.
\end{itemize}

\begin{table}[t]
\addtolength{\tabcolsep}{5pt}
\centerline{\begin{tabular}{c|c|c|c|c|c|c|c|c}
case & $C_0^*$ & $\sigma_a$ & $a_0$ & $\tau_1$ & $\tau_2$ & $\tau_0$ & $\tau_a$ & $T_L$ \\[1pt]
& & (\unit{m/s^2}) & & (\unit{ms}) & (\unit{ms}) & (\unit{ms}) & (\unit{ms}) & (\unit{ms}) \\[3pt]
\hline & & & & & & & \\[-1.5ex]
200 & $6.62 $ & $0.321 \pm 0.005$ & $3.31 \pm 0.12$ & $418 \pm 41$ & $20.00 \pm 1.82$ & $63.4 \pm 4.0$ & $31.2 \pm 1.56$ & $443 \pm 39$\\
270 & $7.03 $ & $0.698 \pm 0.007$ & $3.52 \pm 0.12$ & $363 \pm 4$ & $10.10 \pm 0.17$ & $35.2 \pm 0.5$ & $16.2 \pm 0.09$ & $376 \pm 4$\\
330 & $6.56 $ & $1.723 \pm 0.021$ & $3.69 \pm 0.09$ & $244 \pm 19$ & $6.57 \pm 0.70$ & $23.1 \pm 1.7$ & $10.5 \pm 0.80$ & $252 \pm 18$\\
380 & $6.47 $ & $3.856 \pm 0.060$ & $3.85 \pm 0.12$ & $165 \pm 17$ & $4.16 \pm 0.59$ & $14.6\pm 1.6$ & $6.70 \pm 0.72$ & $170 \pm 16$\\
460 & $6.57 $ & $9.124 \pm 0.072$ & $3.95 \pm 0.07$ & $118 \pm 8$ & $2.08 \pm 0.12$ & $7.8 \pm 0.3$ & $3.38 \pm 0.13$ & $121 \pm 8$\\
550 & $7.11 $ & $19.43 \pm 0.327$ & $4.18 \pm 0.14$ & $ 86 \pm 2$ & $1.46 \pm 0.10$ & $5.5 \pm 0.3$ & $2.42 \pm 0.09$ & $88 \pm 1$\\
600 & $7.07 $ & $27.79 \pm 0.467$ & $4.26 \pm 0.15$ & $ 73 \pm 4$ & $1.20 \pm 0.09$ & $4.5 \pm 0.2$ & $1.99 \pm 0.09$ & $74 \pm 4$\\
\end{tabular}}
\addtolength{\tabcolsep}{-5pt}
\caption{Lagrangian parameters of the LEM flow for the seven Reynolds numbers. The values of $\tau_1$ are estimated from velocity correlation functions, and the values of $\tau_2$ are estimated from acceleration correlation functions. $C_0^*$ is estimated based on the $y$ component, while other quantities are the average mean of three components and the uncertainties indicate the standard derivation of three components.}
\label{tab:lem_lagrangian_parameters}
\end{table}

\begin{figure}
\centerline{\includegraphics[width=0.7\textwidth]{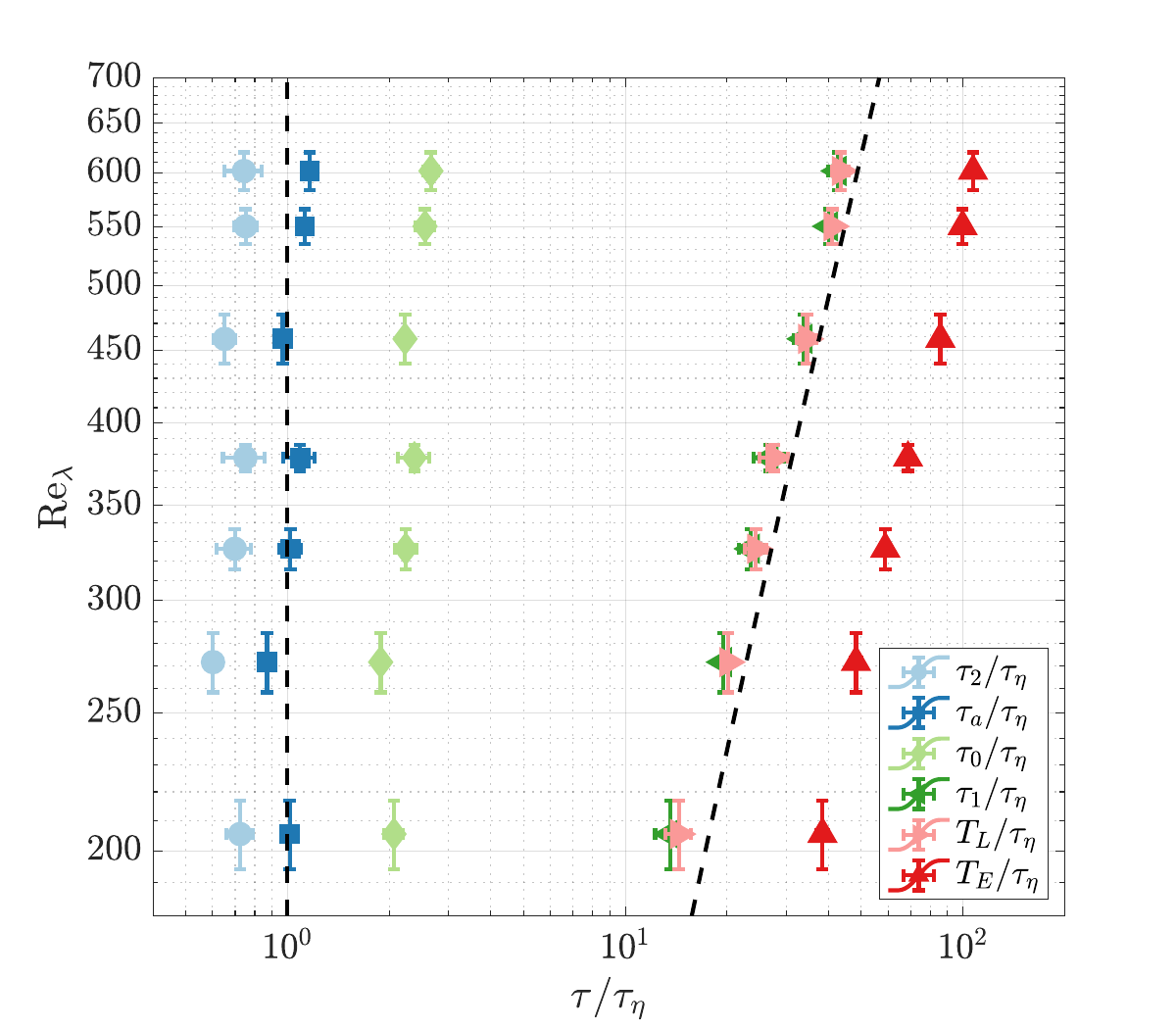}}
\caption{Visualization of all the time scales as a function of the Taylor-based Reynolds number $\text{Re}_\lambda$. Two dashed lines are included to separate different regimes of time scales. left: $\tau=\tau_\eta$, right: $\tau/\tau_\eta=(4.77+(\text{Re}_\lambda/12.6)^{4/3})^{3/4}$.}
\label{fig:all_time_scales}
\end{figure}

To summarize our findings we have plotted in figure~\ref{fig:all_time_scales} all the Lagrangian time scales derived from our datasets, as a function of the Taylor-based Reynolds number in one large diagram, where two dashed lines are included to separate different regimes of time scales. Small time scales, including $\tau_2$, $\tau_a$ and $\tau_0$, are of the order of $\tau_\eta$, characterizing the transition from dissipative to inertial scale. For large time scales, the Lagrangian integral time scales found in our flow are close to the empirical fitting $T_L/\tau_\eta=(4.77+(\text{Re}_\lambda/12.6)^{4/3})^{3/4}$ \cite{sawford2013lagrangian}. As shown in figure~\ref{fig:all_time_scales} large time scales, $\tau_1$, $T_L$ and $T_E$, are of the order estimated from this fitting relation (the right dashed line), and thus can be used to separate the inertial and energy injection regime.

\section{Conclusion}\label{sec:conclusion}

In this paper, we have shown a detailed Eulerian and Lagrangian characterization of an experimental homogeneous and isotropic turbulent flow, mainly using second-order statistics. All the Eulerian and Lagrangian parameters are measured with specific attention to minimize any experimental noise and statistical bias.

The flow presents good homogeneity and isotropy, as shown with classical statistical tools such as PDFs, Lumley triangle, and negligible mean velocity. Slightly leptokurtic velocity PDFs and highly non-Gaussian acceleration PDFs are reported. Through second-order structure functions and autocorrelation functions, Eulerian velocity statistics are computed to determine the mean energy dissipation rate $\varepsilon$, the Eulerian integral length scale $L_E$ and all classical associated turbulent quantities. The isotropy at the inertial scales is shown through the relation between the longitudinal and transverse structure functions, which are used to obtain the energy dissipation rate. All Eulerian parameters are summarized in Table~\ref{tab:lem_eulerian_parameters}. The statistical bias of trajectory length is also highlighted to be correctly taken into account.

The main part of this study is dedicated to Lagrangian statistics on velocity and acceleration. Velocity statistics, which present large time dynamics, are highly affected by statistical bias and adapted corrections are proposed. The difference between the constants $C_0$ and $C_0^*$ is discussed and $C_0^*$ is preferred both because of its clear definition and because it is more robust to estimate. The constant $C_0^*$ is found to be almost independent of $\text{Re}_\lambda$ with a value around 6.8, which is consistent with the highest Reynolds number limit predicted from DNS in HIT. The Lagrangian autocorrelation functions of velocity and acceleration are also computed, corrected, and fitted with parameterization from infinitely differentiable stochastic models \cite{viggiano2020modelling}. From the velocity autocorrelation, the large time scale $\tau_1$ is determined mainly associated with the Lagrangian integral time scale $T_L$. From the acceleration autocorrelation, we obtain the acceleration variance and the acceleration constant $a_0$. The constant $a_0$ is found to be weakly dependent on $\text{Re}_\lambda$ with a value of around 4, consistent with the recently-proposed \cite{Bec_Vallee_2024} fitting relations \eqref{eq:a0}. We also have access to short time dynamics with the time scales $\tau_2$, $\tau_0$, and $\tau_a$ which successfully satisfy the exact relation~\eqref{eq:tauatauK}. These time scales $\tau_0$ and $\tau_a$ can be considered as Lagrangian dissipative time scales and an associated Lagrangian Reynolds number $\text{Re}^L_\lambda = T_L/\tau_0$ or $T_L/\tau_a$ can be build. The connection between the Eulerian and Lagrangian time scales $T_E$ and $T_L$ with $T_E/T_L \approx 2.4$ is also consistent with relation~\eqref{eq:C0*_TETL}.

We hope this study will be used as a reference for experiments in HIT, with a full second-order characterization of a single flow at seven Reynolds numbers. The attention dedicated to the precise determination of physical quantities could be used as strong protocols to measure them, while the measurement of turbulent quantities is often really diverse between various studies. We restricted our discussions on Lagrangian second-order statistics and let the discussion on intermittency for further studies as they require much better statistical convergence.

\section*{Acknowledgment}
We thank the organization of the 26th International Conference of the Theoretical and Applied Mechanics (ICTAM 2024) of the The International Union of Theoretical and Applied Mechanics (IUTAM) in Daegu, Korea, held over August 25-30, 2024, for creating the platform at which, and for bringing together the audience to which, this work was first presented. This work was supported by state funding managed by the National Research Agency under the France 2030 program, with reference ANR-24-RRII-0001 (projetc ``ALEAS'') and by the National Research Agency under the PRCI program, with the reference ANR-23-CE30-0050 (project ``FSPINT''). We also thank Enrico Calzavarini for granting us permission to process his DNS data for the purpose of comparison with our measurements.

\section*{AUTHOR DECLARATIONS}
\subsection*{Conflict of Interest}
The authors have no conflicts to disclose.

\subsection*{Author Contributions}
\textbf{Cheng Wang}: Data curation (equal), Formal analysis (equal), Writing - original draft (equal), writing - review \& editing (equal), Validation (equal). \textbf{Sander Huisman}: Conceptualisation (equal), Data curation (equal), Formal analysis (equal), Investigation (equal), Writing - original draft (equal), Writing - review \& editing (equal). \textbf{Thomas Basset}: Data curation (equal), Formal analysis (equal), Writing - original draft (equal), Writing - review \& editing (equal). \textbf{Romain Volk}: Conceptualization (equal), Formal analysis (equal), Investigation (equal), Validaiton (equal), Writing - review \& editing (equal). \textbf{Mickael Bourgoin}: Conceptualization (equal), Data curation (equal), Formal analysis (equal), Formal analysis (equal), Investigation (equal), Validation (equal), Writing - original draft (equal), Writing - review \& editing (equal).

\section*{Data availability}
The data that support the findings of this study are available from the corresponding author upon reasonable request.

%\bibliography{mybib}

\begin{thebibliography}{80}%
\makeatletter
\providecommand \@ifxundefined [1]{%
 \@ifx{#1\undefined}
}%
\providecommand \@ifnum [1]{%
 \ifnum #1\expandafter \@firstoftwo
 \else \expandafter \@secondoftwo
 \fi
}%
\providecommand \@ifx [1]{%
 \ifx #1\expandafter \@firstoftwo
 \else \expandafter \@secondoftwo
 \fi
}%
\providecommand \natexlab [1]{#1}%
\providecommand \enquote  [1]{``#1''}%
\providecommand \bibnamefont  [1]{#1}%
\providecommand \bibfnamefont [1]{#1}%
\providecommand \citenamefont [1]{#1}%
\providecommand \href@noop [0]{\@secondoftwo}%
\providecommand \href [0]{\begingroup \@sanitize@url \@href}%
\providecommand \@href[1]{\@@startlink{#1}\@@href}%
\providecommand \@@href[1]{\endgroup#1\@@endlink}%
\providecommand \@sanitize@url [0]{\catcode `\\12\catcode `\$12\catcode
  `\&12\catcode `\#12\catcode `\^12\catcode `\_12\catcode `\%12\relax}%
\providecommand \@@startlink[1]{}%
\providecommand \@@endlink[0]{}%
\providecommand \url  [0]{\begingroup\@sanitize@url \@url }%
\providecommand \@url [1]{\endgroup\@href {#1}{\urlprefix }}%
\providecommand \urlprefix  [0]{URL }%
\providecommand \Eprint [0]{\href }%
\providecommand \doibase [0]{https://doi.org/}%
\providecommand \selectlanguage [0]{\@gobble}%
\providecommand \bibinfo  [0]{\@secondoftwo}%
\providecommand \bibfield  [0]{\@secondoftwo}%
\providecommand \translation [1]{[#1]}%
\providecommand \BibitemOpen [0]{}%
\providecommand \bibitemStop [0]{}%
\providecommand \bibitemNoStop [0]{.\EOS\space}%
\providecommand \EOS [0]{\spacefactor3000\relax}%
\providecommand \BibitemShut  [1]{\csname bibitem#1\endcsname}%
\let\auto@bib@innerbib\@empty
%</preamble>
\bibitem [{\citenamefont {Richarson}(1922)}]{richardson1922weather}%
  \BibitemOpen
  \bibfield  {author} {\bibinfo {author} {\bibfnamefont {L.~F.}\ \bibnamefont
  {Richarson}},\ }\href@noop {} {\emph {\bibinfo {title} {Weather Prediction by
  Numerical Process}}}\ (\bibinfo  {publisher} {Cambridge University Press},\
  \bibinfo {year} {1922})\BibitemShut {NoStop}%
\bibitem [{\citenamefont
  {Kolmogorov}(1941{\natexlab{a}})}]{kolmogorov1941local}%
  \BibitemOpen
  \bibfield  {author} {\bibinfo {author} {\bibfnamefont {A.~N.}\ \bibnamefont
  {Kolmogorov}},\ }\bibfield  {title} {\bibinfo {title} {Local structure of
  turbulence in an incompressible viscous fluid at very large {Reynolds}
  numbers},\ }\href@noop {} {\bibfield  {journal} {\bibinfo  {journal} {Dokl.
  Akad. Nauk SSSR}\ }\textbf {\bibinfo {volume} {30}},\ \bibinfo {pages} {299}
  (\bibinfo {year} {1941}{\natexlab{a}})}\BibitemShut {NoStop}%
\bibitem [{\citenamefont {Bourgoin}\ \emph {et~al.}(2018)\citenamefont
  {Bourgoin}, \citenamefont {Baudet}, \citenamefont {Kharche}, \citenamefont
  {Mordant}, \citenamefont {Vandenberghe}, \citenamefont {Sumbekova},
  \citenamefont {Stelzenmuller}, \citenamefont {Aliseda}, \citenamefont
  {Gibert}, \citenamefont {Roche}, \citenamefont {Volk}, \citenamefont
  {Barois}, \citenamefont {Caballero}, \citenamefont {Chevillard},
  \citenamefont {Pinton}, \citenamefont {Fiabane}, \citenamefont {Delville},
  \citenamefont {Fourment}, \citenamefont {Bouha}, \citenamefont {Danaila},
  \citenamefont {Bodenschatz}, \citenamefont {Bewley}, \citenamefont
  {Sinhuber}, \citenamefont {Segalini}, \citenamefont {\"Orl\"u}, \citenamefont
  {Torrano}, \citenamefont {Mantik}, \citenamefont {Guariglia}, \citenamefont
  {Uruba}, \citenamefont {Skala}, \citenamefont {Puczylowski},\ and\
  \citenamefont {Peinke}}]{bourgoin2018investigation}%
  \BibitemOpen
  \bibfield  {author} {\bibinfo {author} {\bibfnamefont {M.}~\bibnamefont
  {Bourgoin}}, \bibinfo {author} {\bibfnamefont {C.}~\bibnamefont {Baudet}},
  \bibinfo {author} {\bibfnamefont {S.}~\bibnamefont {Kharche}}, \bibinfo
  {author} {\bibfnamefont {N.}~\bibnamefont {Mordant}}, \bibinfo {author}
  {\bibfnamefont {T.}~\bibnamefont {Vandenberghe}}, \bibinfo {author}
  {\bibfnamefont {S.}~\bibnamefont {Sumbekova}}, \bibinfo {author}
  {\bibfnamefont {N.}~\bibnamefont {Stelzenmuller}}, \bibinfo {author}
  {\bibfnamefont {A.}~\bibnamefont {Aliseda}}, \bibinfo {author} {\bibfnamefont
  {M.}~\bibnamefont {Gibert}}, \bibinfo {author} {\bibfnamefont {P.-E.}\
  \bibnamefont {Roche}}, \bibinfo {author} {\bibfnamefont {R.}~\bibnamefont
  {Volk}}, \bibinfo {author} {\bibfnamefont {T.}~\bibnamefont {Barois}},
  \bibinfo {author} {\bibfnamefont {M.~L.}\ \bibnamefont {Caballero}}, \bibinfo
  {author} {\bibfnamefont {L.}~\bibnamefont {Chevillard}}, \bibinfo {author}
  {\bibfnamefont {J.-F.}\ \bibnamefont {Pinton}}, \bibinfo {author}
  {\bibfnamefont {L.}~\bibnamefont {Fiabane}}, \bibinfo {author} {\bibfnamefont
  {J.}~\bibnamefont {Delville}}, \bibinfo {author} {\bibfnamefont
  {C.}~\bibnamefont {Fourment}}, \bibinfo {author} {\bibfnamefont
  {A.}~\bibnamefont {Bouha}}, \bibinfo {author} {\bibfnamefont
  {L.}~\bibnamefont {Danaila}}, \bibinfo {author} {\bibfnamefont
  {E.}~\bibnamefont {Bodenschatz}}, \bibinfo {author} {\bibfnamefont
  {G.}~\bibnamefont {Bewley}}, \bibinfo {author} {\bibfnamefont
  {M.}~\bibnamefont {Sinhuber}}, \bibinfo {author} {\bibfnamefont
  {A.}~\bibnamefont {Segalini}}, \bibinfo {author} {\bibfnamefont
  {R.}~\bibnamefont {\"Orl\"u}}, \bibinfo {author} {\bibfnamefont
  {I.}~\bibnamefont {Torrano}}, \bibinfo {author} {\bibfnamefont
  {J.}~\bibnamefont {Mantik}}, \bibinfo {author} {\bibfnamefont
  {D.}~\bibnamefont {Guariglia}}, \bibinfo {author} {\bibfnamefont
  {V.}~\bibnamefont {Uruba}}, \bibinfo {author} {\bibfnamefont
  {V.}~\bibnamefont {Skala}}, \bibinfo {author} {\bibfnamefont
  {J.}~\bibnamefont {Puczylowski}},\ and\ \bibinfo {author} {\bibfnamefont
  {J.}~\bibnamefont {Peinke}},\ }\bibfield  {title} {\bibinfo {title}
  {Investigation of the small-scale statistics of turbulence in the {Modane
  S1MA} wind tunnel},\ }\href@noop {} {\bibfield  {journal} {\bibinfo
  {journal} {CEAS Aeronaut. J.}\ }\textbf {\bibinfo {volume} {9}},\ \bibinfo
  {pages} {269} (\bibinfo {year} {2018})}\BibitemShut {NoStop}%
\bibitem [{\citenamefont {She}\ \emph {et~al.}(1991)\citenamefont {She},
  \citenamefont {Jackson},\ and\ \citenamefont {Orszag}}]{she1991structure}%
  \BibitemOpen
  \bibfield  {author} {\bibinfo {author} {\bibfnamefont {Z.-S.}\ \bibnamefont
  {She}}, \bibinfo {author} {\bibfnamefont {E.}~\bibnamefont {Jackson}},\ and\
  \bibinfo {author} {\bibfnamefont {S.~A.}\ \bibnamefont {Orszag}},\ }\bibfield
   {title} {\bibinfo {title} {Structure and dynamics of homogeneous turbulence:
  models and simulations},\ }\href@noop {} {\bibfield  {journal} {\bibinfo
  {journal} {Proc. Math. Phys. Sci.}\ }\textbf {\bibinfo {volume} {434}},\
  \bibinfo {pages} {101} (\bibinfo {year} {1991})}\BibitemShut {NoStop}%
\bibitem [{\citenamefont {Comte-Bellot}(1976)}]{comtebellot1976hotwire}%
  \BibitemOpen
  \bibfield  {author} {\bibinfo {author} {\bibfnamefont {G.}~\bibnamefont
  {Comte-Bellot}},\ }\bibfield  {title} {\bibinfo {title} {Hot-wire
  anemometry},\ }\href@noop {} {\bibfield  {journal} {\bibinfo  {journal}
  {Annu. Rev. Fluid Mech.}\ }\textbf {\bibinfo {volume} {53}},\ \bibinfo
  {pages} {209} (\bibinfo {year} {1976})}\BibitemShut {NoStop}%
\bibitem [{\citenamefont {Raffel}\ \emph {et~al.}(2018)\citenamefont {Raffel},
  \citenamefont {Willert}, \citenamefont {Scarano}, \citenamefont {K{\"a}hler},
  \citenamefont {Wereley},\ and\ \citenamefont {Kompenhans}}]{raffel2018piv}%
  \BibitemOpen
  \bibfield  {author} {\bibinfo {author} {\bibfnamefont {M.}~\bibnamefont
  {Raffel}}, \bibinfo {author} {\bibfnamefont {C.~E.}\ \bibnamefont {Willert}},
  \bibinfo {author} {\bibfnamefont {F.}~\bibnamefont {Scarano}}, \bibinfo
  {author} {\bibfnamefont {C.~J.}\ \bibnamefont {K{\"a}hler}}, \bibinfo
  {author} {\bibfnamefont {S.~T.}\ \bibnamefont {Wereley}},\ and\ \bibinfo
  {author} {\bibfnamefont {J.}~\bibnamefont {Kompenhans}},\ }\bibinfo {title}
  {{PIV Uncertainty and Measurement Accuracy}},\ in\ \href@noop {} {\emph
  {\bibinfo {booktitle} {Particle Image Velocimetry: A Practical Guide}}}\
  (\bibinfo  {publisher} {Springer},\ \bibinfo {year} {2018})\ pp.\ \bibinfo
  {pages} {203--241}\BibitemShut {NoStop}%
\bibitem [{\citenamefont {Adrian}\ and\ \citenamefont
  {Westerweel}(2011)}]{adrian2011particle}%
  \BibitemOpen
  \bibfield  {author} {\bibinfo {author} {\bibfnamefont {R.~J.}\ \bibnamefont
  {Adrian}}\ and\ \bibinfo {author} {\bibfnamefont {J.}~\bibnamefont
  {Westerweel}},\ }\href@noop {} {\emph {\bibinfo {title} {Particle Image
  Velocimetry}}}\ (\bibinfo  {publisher} {Cambridge University Press},\
  \bibinfo {year} {2011})\BibitemShut {NoStop}%
\bibitem [{\citenamefont {Ishihara}\ \emph {et~al.}(2007)\citenamefont
  {Ishihara}, \citenamefont {Kaneda}, \citenamefont {Yokokawa}, \citenamefont
  {Itakura},\ and\ \citenamefont {Uno}}]{ishihara2007small}%
  \BibitemOpen
  \bibfield  {author} {\bibinfo {author} {\bibfnamefont {T.}~\bibnamefont
  {Ishihara}}, \bibinfo {author} {\bibfnamefont {Y.}~\bibnamefont {Kaneda}},
  \bibinfo {author} {\bibfnamefont {M.}~\bibnamefont {Yokokawa}}, \bibinfo
  {author} {\bibfnamefont {K.}~\bibnamefont {Itakura}},\ and\ \bibinfo {author}
  {\bibfnamefont {A.}~\bibnamefont {Uno}},\ }\bibfield  {title} {\bibinfo
  {title} {Small-scale statistics in high-resolution direct numerical
  simulation of turbulence: {Reynolds} number dependence of one-point velocity
  gradient statistics},\ }\href@noop {} {\bibfield  {journal} {\bibinfo
  {journal} {J. Fluid Mech.}\ }\textbf {\bibinfo {volume} {592}},\ \bibinfo
  {pages} {335–366} (\bibinfo {year} {2007})}\BibitemShut {NoStop}%
\bibitem [{\citenamefont {Khurshid}\ \emph {et~al.}(2023)\citenamefont
  {Khurshid}, \citenamefont {Donzis},\ and\ \citenamefont
  {Sreenivasan}}]{bib:khurshid2023_PRE}%
  \BibitemOpen
  \bibfield  {author} {\bibinfo {author} {\bibfnamefont {S.}~\bibnamefont
  {Khurshid}}, \bibinfo {author} {\bibfnamefont {D.~A.}\ \bibnamefont
  {Donzis}},\ and\ \bibinfo {author} {\bibfnamefont {K.~R.}\ \bibnamefont
  {Sreenivasan}},\ }\bibfield  {title} {\bibinfo {title} {{Emergence of
  universal scaling in isotropic turbulence}},\ }\href
  {https://doi.org/10.1103/PHYSREVE.107.045102/FIGURES/4/MEDIUM} {\bibfield
  {journal} {\bibinfo  {journal} {Phys. Rev. E}\ }\textbf {\bibinfo {volume}
  {107}},\ \bibinfo {pages} {045102} (\bibinfo {year} {2023})}\BibitemShut
  {NoStop}%
\bibitem [{\citenamefont {Taylor}(1922)}]{taylor1922diffusion}%
  \BibitemOpen
  \bibfield  {author} {\bibinfo {author} {\bibfnamefont {G.~I.}\ \bibnamefont
  {Taylor}},\ }\bibfield  {title} {\bibinfo {title} {Diffusion by continuous
  movements},\ }\href@noop {} {\bibfield  {journal} {\bibinfo  {journal} {Proc.
  Lond. Math. Soc.}\ }\textbf {\bibinfo {volume} {20}},\ \bibinfo {pages} {196}
  (\bibinfo {year} {1922})}\BibitemShut {NoStop}%
\bibitem [{\citenamefont {Richardson}(1926)}]{richardson1926diffusion}%
  \BibitemOpen
  \bibfield  {author} {\bibinfo {author} {\bibfnamefont {L.~F.}\ \bibnamefont
  {Richardson}},\ }\bibfield  {title} {\bibinfo {title} {Atmospheric diffusion
  shown on a distance-neighbour graph},\ }\href@noop {} {\bibfield  {journal}
  {\bibinfo  {journal} {Proc. R. Soc. Lond. A}\ }\textbf {\bibinfo {volume}
  {110}},\ \bibinfo {pages} {709} (\bibinfo {year} {1926})}\BibitemShut
  {NoStop}%
\bibitem [{\citenamefont {Sato}\ and\ \citenamefont
  {Yamamoto}(1987)}]{sato1987lagrangian}%
  \BibitemOpen
  \bibfield  {author} {\bibinfo {author} {\bibfnamefont {Y.}~\bibnamefont
  {Sato}}\ and\ \bibinfo {author} {\bibfnamefont {K.}~\bibnamefont
  {Yamamoto}},\ }\bibfield  {title} {\bibinfo {title} {{{Lagrangian}}
  measurement of fluid-particle motion in an isotropic turbulent field},\
  }\href@noop {} {\bibfield  {journal} {\bibinfo  {journal} {J. Fluid Mech.}\
  }\textbf {\bibinfo {volume} {175}},\ \bibinfo {pages} {183–199} (\bibinfo
  {year} {1987})}\BibitemShut {NoStop}%
\bibitem [{\citenamefont {Maas}\ \emph {et~al.}(1993)\citenamefont {Maas},
  \citenamefont {Gruen},\ and\ \citenamefont {Papantoniou}}]{maas1993particle}%
  \BibitemOpen
  \bibfield  {author} {\bibinfo {author} {\bibfnamefont {H.~G.}\ \bibnamefont
  {Maas}}, \bibinfo {author} {\bibfnamefont {A.}~\bibnamefont {Gruen}},\ and\
  \bibinfo {author} {\bibfnamefont {D.}~\bibnamefont {Papantoniou}},\
  }\bibfield  {title} {\bibinfo {title} {Particle tracking velocimetry in
  three-dimensional flows. {Part 1}. {Photogrammetric} determination of
  particle coordinates},\ }\href@noop {} {\bibfield  {journal} {\bibinfo
  {journal} {Exp. Fluids}\ }\textbf {\bibinfo {volume} {15}},\ \bibinfo {pages}
  {133} (\bibinfo {year} {1993})}\BibitemShut {NoStop}%
\bibitem [{\citenamefont {Malik}\ \emph {et~al.}(1993)\citenamefont {Malik},
  \citenamefont {Dracos},\ and\ \citenamefont
  {Papantoniou}}]{malik1993particle}%
  \BibitemOpen
  \bibfield  {author} {\bibinfo {author} {\bibfnamefont {N.~A.}\ \bibnamefont
  {Malik}}, \bibinfo {author} {\bibfnamefont {T.}~\bibnamefont {Dracos}},\ and\
  \bibinfo {author} {\bibfnamefont {D.~A.}\ \bibnamefont {Papantoniou}},\
  }\bibfield  {title} {\bibinfo {title} {Particle tracking velocimetry in
  three-dimensional flows. {Part} 2. {Particle} tracking},\ }\href@noop {}
  {\bibfield  {journal} {\bibinfo  {journal} {Exp. Fluids}\ }\textbf {\bibinfo
  {volume} {15}},\ \bibinfo {pages} {279} (\bibinfo {year} {1993})}\BibitemShut
  {NoStop}%
\bibitem [{\citenamefont {Virant}\ and\ \citenamefont
  {Dracos}(1997)}]{virant1997ptv}%
  \BibitemOpen
  \bibfield  {author} {\bibinfo {author} {\bibfnamefont {M.}~\bibnamefont
  {Virant}}\ and\ \bibinfo {author} {\bibfnamefont {T.}~\bibnamefont
  {Dracos}},\ }\bibfield  {title} {\bibinfo {title} {{3D PTV} and its
  application on {{{Lagrangian}}} motion},\ }\href@noop {} {\bibfield
  {journal} {\bibinfo  {journal} {Meas. Sci. Technol.}\ }\textbf {\bibinfo
  {volume} {8}},\ \bibinfo {pages} {1539} (\bibinfo {year} {1997})}\BibitemShut
  {NoStop}%
\bibitem [{\citenamefont {Schr\"oder}\ and\ \citenamefont
  {Schanz}(2023)}]{schroder2023lagrangian}%
  \BibitemOpen
  \bibfield  {author} {\bibinfo {author} {\bibfnamefont {A.}~\bibnamefont
  {Schr\"oder}}\ and\ \bibinfo {author} {\bibfnamefont {D.}~\bibnamefont
  {Schanz}},\ }\bibfield  {title} {\bibinfo {title} {{3D} {{{Lagrangian}}}
  particle tracking in fluid mechanics},\ }\href@noop {} {\bibfield  {journal}
  {\bibinfo  {journal} {Annu. Rev. Fluid Mech.}\ }\textbf {\bibinfo {volume}
  {55}},\ \bibinfo {pages} {511} (\bibinfo {year} {2023})}\BibitemShut
  {NoStop}%
\bibitem [{\citenamefont {Yeung}(2002)}]{yeung2002lagrangian}%
  \BibitemOpen
  \bibfield  {author} {\bibinfo {author} {\bibfnamefont {P.~K.}\ \bibnamefont
  {Yeung}},\ }\bibfield  {title} {\bibinfo {title} {{{Lagrangian}}
  investigations of turbulence},\ }\href@noop {} {\bibfield  {journal}
  {\bibinfo  {journal} {Annu. Rev. Fluid Mech.}\ }\textbf {\bibinfo {volume}
  {34}},\ \bibinfo {pages} {115} (\bibinfo {year} {2002})}\BibitemShut
  {NoStop}%
\bibitem [{\citenamefont {Buaria}\ \emph {et~al.}(2016)\citenamefont {Buaria},
  \citenamefont {Yeung},\ and\ \citenamefont {Sawford}}]{buaria2016lagrangian}%
  \BibitemOpen
  \bibfield  {author} {\bibinfo {author} {\bibfnamefont {D.}~\bibnamefont
  {Buaria}}, \bibinfo {author} {\bibfnamefont {P.~K.}\ \bibnamefont {Yeung}},\
  and\ \bibinfo {author} {\bibfnamefont {B.~L.}\ \bibnamefont {Sawford}},\
  }\bibfield  {title} {\bibinfo {title} {A {{{Lagrangian}}} study of turbulent
  mixing: forward and backward dispersion of molecular trajectories in
  isotropic turbulence},\ }\href@noop {} {\bibfield  {journal} {\bibinfo
  {journal} {J. Fluid Mech.}\ }\textbf {\bibinfo {volume} {799}},\ \bibinfo
  {pages} {352–382} (\bibinfo {year} {2016})}\BibitemShut {NoStop}%
\bibitem [{\citenamefont {Voth}\ \emph {et~al.}(2002)\citenamefont {Voth},
  \citenamefont {{La Porta}}, \citenamefont {Crawford}, \citenamefont
  {Alexander},\ and\ \citenamefont {Bodenschatz}}]{voth2002measurement}%
  \BibitemOpen
  \bibfield  {author} {\bibinfo {author} {\bibfnamefont {G.~A.}\ \bibnamefont
  {Voth}}, \bibinfo {author} {\bibfnamefont {A.}~\bibnamefont {{La Porta}}},
  \bibinfo {author} {\bibfnamefont {A.~M.}\ \bibnamefont {Crawford}}, \bibinfo
  {author} {\bibfnamefont {J.}~\bibnamefont {Alexander}},\ and\ \bibinfo
  {author} {\bibfnamefont {E.}~\bibnamefont {Bodenschatz}},\ }\bibfield
  {title} {\bibinfo {title} {Measurement of particle accelerations in fully
  developed turbulence},\ }\href@noop {} {\bibfield  {journal} {\bibinfo
  {journal} {J. Fluid Mech.}\ }\textbf {\bibinfo {volume} {469}},\ \bibinfo
  {pages} {121} (\bibinfo {year} {2002})}\BibitemShut {NoStop}%
\bibitem [{\citenamefont {Mordant}\ \emph {et~al.}(2001)\citenamefont
  {Mordant}, \citenamefont {Metz}, \citenamefont {Michel},\ and\ \citenamefont
  {Pinton}}]{mordant2001measurement}%
  \BibitemOpen
  \bibfield  {author} {\bibinfo {author} {\bibfnamefont {N.}~\bibnamefont
  {Mordant}}, \bibinfo {author} {\bibfnamefont {P.}~\bibnamefont {Metz}},
  \bibinfo {author} {\bibfnamefont {O.}~\bibnamefont {Michel}},\ and\ \bibinfo
  {author} {\bibfnamefont {J.-F.}\ \bibnamefont {Pinton}},\ }\bibfield  {title}
  {\bibinfo {title} {Measurement of {{Lagrangian}} velocity in fully developed
  turbulence},\ }\href@noop {} {\bibfield  {journal} {\bibinfo  {journal}
  {Phys. Rev. Lett.}\ }\textbf {\bibinfo {volume} {87}},\ \bibinfo {pages}
  {214501} (\bibinfo {year} {2001})}\BibitemShut {NoStop}%
\bibitem [{\citenamefont {Ouellette}\ \emph
  {et~al.}(2006{\natexlab{a}})\citenamefont {Ouellette}, \citenamefont {Xu},
  \citenamefont {Bourgoin},\ and\ \citenamefont
  {Bodenschatz}}]{ouellette2006small}%
  \BibitemOpen
  \bibfield  {author} {\bibinfo {author} {\bibfnamefont {N.~T.}\ \bibnamefont
  {Ouellette}}, \bibinfo {author} {\bibfnamefont {H.}~\bibnamefont {Xu}},
  \bibinfo {author} {\bibfnamefont {M.}~\bibnamefont {Bourgoin}},\ and\
  \bibinfo {author} {\bibfnamefont {E.}~\bibnamefont {Bodenschatz}},\
  }\bibfield  {title} {\bibinfo {title} {Small-scale anisotropy in
  {{{Lagrangian}}} turbulence},\ }\href@noop {} {\bibfield  {journal} {\bibinfo
   {journal} {New J. Phys}\ }\textbf {\bibinfo {volume} {8}},\ \bibinfo {pages}
  {102} (\bibinfo {year} {2006}{\natexlab{a}})}\BibitemShut {NoStop}%
\bibitem [{\citenamefont {Toschi}\ and\ \citenamefont
  {Bodenschatz}(2009)}]{toschi2009lagrangian}%
  \BibitemOpen
  \bibfield  {author} {\bibinfo {author} {\bibfnamefont {F.}~\bibnamefont
  {Toschi}}\ and\ \bibinfo {author} {\bibfnamefont {E.}~\bibnamefont
  {Bodenschatz}},\ }\bibfield  {title} {\bibinfo {title} {{{Lagrangian}}
  properties of particles in turbulence},\ }\href@noop {} {\bibfield  {journal}
  {\bibinfo  {journal} {Annu. Rev. Fluid Mech.}\ }\textbf {\bibinfo {volume}
  {41}},\ \bibinfo {pages} {375} (\bibinfo {year} {2009})}\BibitemShut
  {NoStop}%
\bibitem [{\citenamefont {Ayyalasomayajula}\ \emph {et~al.}(2006)\citenamefont
  {Ayyalasomayajula}, \citenamefont {Gylfason}, \citenamefont {Collins},
  \citenamefont {Bodenschatz},\ and\ \citenamefont
  {Warhaft}}]{ayyalasomayajula2006lagrangian}%
  \BibitemOpen
  \bibfield  {author} {\bibinfo {author} {\bibfnamefont {S.}~\bibnamefont
  {Ayyalasomayajula}}, \bibinfo {author} {\bibfnamefont {A.}~\bibnamefont
  {Gylfason}}, \bibinfo {author} {\bibfnamefont {L.~R.}\ \bibnamefont
  {Collins}}, \bibinfo {author} {\bibfnamefont {E.}~\bibnamefont
  {Bodenschatz}},\ and\ \bibinfo {author} {\bibfnamefont {Z.}~\bibnamefont
  {Warhaft}},\ }\bibfield  {title} {\bibinfo {title} {{{Lagrangian}}
  measurements of inertial particle accelerations in grid generated wind tunnel
  turbulence},\ }\href@noop {} {\bibfield  {journal} {\bibinfo  {journal}
  {Phys. Rev. Lett.}\ }\textbf {\bibinfo {volume} {97}},\ \bibinfo {pages}
  {144507} (\bibinfo {year} {2006})}\BibitemShut {NoStop}%
\bibitem [{\citenamefont {Huck}\ \emph {et~al.}(2019)\citenamefont {Huck},
  \citenamefont {Machicoane},\ and\ \citenamefont {Volk}}]{huck2019lagrangian}%
  \BibitemOpen
  \bibfield  {author} {\bibinfo {author} {\bibfnamefont {P.~D.}\ \bibnamefont
  {Huck}}, \bibinfo {author} {\bibfnamefont {N.}~\bibnamefont {Machicoane}},\
  and\ \bibinfo {author} {\bibfnamefont {R.}~\bibnamefont {Volk}},\ }\bibfield
  {title} {\bibinfo {title} {{{Lagrangian}} acceleration timescales in
  anisotropic turbulence},\ }\href@noop {} {\bibfield  {journal} {\bibinfo
  {journal} {Phys. Rev. Fluids}\ }\textbf {\bibinfo {volume} {4}},\ \bibinfo
  {pages} {064606} (\bibinfo {year} {2019})}\BibitemShut {NoStop}%
\bibitem [{\citenamefont {Huck}\ \emph {et~al.}(2017)\citenamefont {Huck},
  \citenamefont {Machicoane},\ and\ \citenamefont {Volk}}]{huck2017production}%
  \BibitemOpen
  \bibfield  {author} {\bibinfo {author} {\bibfnamefont {P.~D.}\ \bibnamefont
  {Huck}}, \bibinfo {author} {\bibfnamefont {N.}~\bibnamefont {Machicoane}},\
  and\ \bibinfo {author} {\bibfnamefont {R.}~\bibnamefont {Volk}},\ }\bibfield
  {title} {\bibinfo {title} {Production and dissipation of turbulent
  fluctuations close to a stagnation point},\ }\href@noop {} {\bibfield
  {journal} {\bibinfo  {journal} {Phys. Rev. Fluids}\ }\textbf {\bibinfo
  {volume} {2}},\ \bibinfo {pages} {084601} (\bibinfo {year}
  {2017})}\BibitemShut {NoStop}%
\bibitem [{\citenamefont {Angriman}\ \emph {et~al.}(2021)\citenamefont
  {Angriman}, \citenamefont {Cobelli}, \citenamefont {Bourgoin}, \citenamefont
  {Huisman}, \citenamefont {Volk},\ and\ \citenamefont
  {Mininni}}]{angriman2021broken}%
  \BibitemOpen
  \bibfield  {author} {\bibinfo {author} {\bibfnamefont {S.}~\bibnamefont
  {Angriman}}, \bibinfo {author} {\bibfnamefont {P.~J.}\ \bibnamefont
  {Cobelli}}, \bibinfo {author} {\bibfnamefont {M.}~\bibnamefont {Bourgoin}},
  \bibinfo {author} {\bibfnamefont {S.~G.}\ \bibnamefont {Huisman}}, \bibinfo
  {author} {\bibfnamefont {R.}~\bibnamefont {Volk}},\ and\ \bibinfo {author}
  {\bibfnamefont {P.~D.}\ \bibnamefont {Mininni}},\ }\bibfield  {title}
  {\bibinfo {title} {Broken mirror symmetry of tracer’s trajectories in
  turbulence},\ }\href@noop {} {\bibfield  {journal} {\bibinfo  {journal}
  {Phys. Rev. Lett.}\ }\textbf {\bibinfo {volume} {127}},\ \bibinfo {pages}
  {254502} (\bibinfo {year} {2021})}\BibitemShut {NoStop}%
\bibitem [{\citenamefont {{De La Torre}}\ and\ \citenamefont
  {Burguete}(2007)}]{bib:deLaTorre2007_PRL}%
  \BibitemOpen
  \bibfield  {author} {\bibinfo {author} {\bibfnamefont {A.}~\bibnamefont {{De
  La Torre}}}\ and\ \bibinfo {author} {\bibfnamefont {J.}~\bibnamefont
  {Burguete}},\ }\bibfield  {title} {\bibinfo {title} {{Slow dynamics in a
  turbulent von K{\'{a}}rm{\'{a}}n swirling flow}},\ }\href
  {https://doi.org/10.1103/PhysRevLett.99.054101} {\bibfield  {journal}
  {\bibinfo  {journal} {Phys. Rev. Lett.}\ }\textbf {\bibinfo {volume} {99}},\
  \bibinfo {pages} {3} (\bibinfo {year} {2007})},\ \Eprint
  {https://arxiv.org/abs/0702151} {arXiv:0702151 [physics]} \BibitemShut
  {NoStop}%
\bibitem [{\citenamefont {Machicoane}\ \emph {et~al.}(2016)\citenamefont
  {Machicoane}, \citenamefont {L{\'{o}}pez-Caballero}, \citenamefont {Fiabane},
  \citenamefont {Pinton}, \citenamefont {Bourgoin}, \citenamefont {Burguete},\
  and\ \citenamefont {Volk}}]{bib:machicoane2016_PRE}%
  \BibitemOpen
  \bibfield  {author} {\bibinfo {author} {\bibfnamefont {N.}~\bibnamefont
  {Machicoane}}, \bibinfo {author} {\bibfnamefont {M.}~\bibnamefont
  {L{\'{o}}pez-Caballero}}, \bibinfo {author} {\bibfnamefont {L.}~\bibnamefont
  {Fiabane}}, \bibinfo {author} {\bibfnamefont {J.~F.}\ \bibnamefont {Pinton}},
  \bibinfo {author} {\bibfnamefont {M.}~\bibnamefont {Bourgoin}}, \bibinfo
  {author} {\bibfnamefont {J.}~\bibnamefont {Burguete}},\ and\ \bibinfo
  {author} {\bibfnamefont {R.}~\bibnamefont {Volk}},\ }\bibfield  {title}
  {\bibinfo {title} {{Stochastic dynamics of particles trapped in turbulent
  flows}},\ }\href {https://doi.org/10.1103/PhysRevE.93.023118} {\bibfield
  {journal} {\bibinfo  {journal} {Phys. Rev. E}\ }\textbf {\bibinfo {volume}
  {93}},\ \bibinfo {pages} {1} (\bibinfo {year} {2016})},\ \Eprint
  {https://arxiv.org/abs/1512.07760} {arXiv:1512.07760} \BibitemShut {NoStop}%
\bibitem [{\citenamefont {Variano}\ and\ \citenamefont
  {Cowen}(2008)}]{variano2008random}%
  \BibitemOpen
  \bibfield  {author} {\bibinfo {author} {\bibfnamefont {E.~A.}\ \bibnamefont
  {Variano}}\ and\ \bibinfo {author} {\bibfnamefont {E.~A.}\ \bibnamefont
  {Cowen}},\ }\bibfield  {title} {\bibinfo {title} {A random-jet-stirred
  turbulence tank},\ }\href@noop {} {\bibfield  {journal} {\bibinfo  {journal}
  {J. Fluid Mech.}\ }\textbf {\bibinfo {volume} {604}},\ \bibinfo {pages}
  {1–32} (\bibinfo {year} {2008})}\BibitemShut {NoStop}%
\bibitem [{\citenamefont {Laplace}(2022)}]{laplace2022phd}%
  \BibitemOpen
  \bibfield  {author} {\bibinfo {author} {\bibfnamefont {B.}~\bibnamefont
  {Laplace}},\ }\emph {\bibinfo {title} {Étude expérimentale de la
  sédimentation de particules inertielles en turbulence}},\ \href@noop {}
  {Ph.D. thesis},\ \bibinfo  {school} {\'Ecole Normale Supérieure de Lyon}
  (\bibinfo {year} {2022})\BibitemShut {NoStop}%
\bibitem [{\citenamefont {Carter}\ \emph {et~al.}(2016)\citenamefont {Carter},
  \citenamefont {Petersen}, \citenamefont {Amili},\ and\ \citenamefont
  {Coletti}}]{carter2016generating}%
  \BibitemOpen
  \bibfield  {author} {\bibinfo {author} {\bibfnamefont {D.}~\bibnamefont
  {Carter}}, \bibinfo {author} {\bibfnamefont {A.}~\bibnamefont {Petersen}},
  \bibinfo {author} {\bibfnamefont {O.}~\bibnamefont {Amili}},\ and\ \bibinfo
  {author} {\bibfnamefont {F.}~\bibnamefont {Coletti}},\ }\bibfield  {title}
  {\bibinfo {title} {Generating and controlling homogeneous air turbulence
  using random jet arrays},\ }\href@noop {} {\bibfield  {journal} {\bibinfo
  {journal} {Exp. Fluids}\ }\textbf {\bibinfo {volume} {57}},\ \bibinfo {pages}
  {189} (\bibinfo {year} {2016})}\BibitemShut {NoStop}%
\bibitem [{\citenamefont {Zimmermann}\ \emph {et~al.}(2010)\citenamefont
  {Zimmermann}, \citenamefont {Xu}, \citenamefont {Gasteuil}, \citenamefont
  {Bourgoin}, \citenamefont {Volk}, \citenamefont {Pinton}, \citenamefont
  {Bodenschatz},\ and\ \citenamefont {{International Collaboration for
  Turbulence Research}}}]{zimmermann2010lagrangian}%
  \BibitemOpen
  \bibfield  {author} {\bibinfo {author} {\bibfnamefont {R.}~\bibnamefont
  {Zimmermann}}, \bibinfo {author} {\bibfnamefont {H.}~\bibnamefont {Xu}},
  \bibinfo {author} {\bibfnamefont {Y.}~\bibnamefont {Gasteuil}}, \bibinfo
  {author} {\bibfnamefont {M.}~\bibnamefont {Bourgoin}}, \bibinfo {author}
  {\bibfnamefont {R.}~\bibnamefont {Volk}}, \bibinfo {author} {\bibfnamefont
  {J.-F.}\ \bibnamefont {Pinton}}, \bibinfo {author} {\bibfnamefont
  {E.}~\bibnamefont {Bodenschatz}},\ and\ \bibinfo {author} {\bibnamefont
  {{International Collaboration for Turbulence Research}}},\ }\bibfield
  {title} {\bibinfo {title} {The {{{Lagrangian}}} exploration module: An
  apparatus for the study of statistically homogeneous and isotropic
  turbulence},\ }\href@noop {} {\bibfield  {journal} {\bibinfo  {journal} {Rev.
  Sci. Instrum.}\ }\textbf {\bibinfo {volume} {81}},\ \bibinfo {pages} {055112}
  (\bibinfo {year} {2010})}\BibitemShut {NoStop}%
\bibitem [{\citenamefont {Fiabane}\ \emph {et~al.}(2012)\citenamefont
  {Fiabane}, \citenamefont {Zimmermann}, \citenamefont {Volk}, \citenamefont
  {Pinton},\ and\ \citenamefont {Bourgoin}}]{fiabane2012clustering}%
  \BibitemOpen
  \bibfield  {author} {\bibinfo {author} {\bibfnamefont {L.}~\bibnamefont
  {Fiabane}}, \bibinfo {author} {\bibfnamefont {R.}~\bibnamefont {Zimmermann}},
  \bibinfo {author} {\bibfnamefont {R.}~\bibnamefont {Volk}}, \bibinfo {author}
  {\bibfnamefont {J.-F.}\ \bibnamefont {Pinton}},\ and\ \bibinfo {author}
  {\bibfnamefont {M.}~\bibnamefont {Bourgoin}},\ }\bibfield  {title} {\bibinfo
  {title} {Clustering of finite-size particles in turbulence},\ }\href@noop {}
  {\bibfield  {journal} {\bibinfo  {journal} {Phys. Rev. E}\ }\textbf {\bibinfo
  {volume} {86}},\ \bibinfo {pages} {035301} (\bibinfo {year}
  {2012})}\BibitemShut {NoStop}%
\bibitem [{\citenamefont {Qureshi}\ \emph {et~al.}(2007)\citenamefont
  {Qureshi}, \citenamefont {Bourgoin}, \citenamefont {Baudet}, \citenamefont
  {Cartellier},\ and\ \citenamefont {Gagne}}]{qureshi2007turbulent}%
  \BibitemOpen
  \bibfield  {author} {\bibinfo {author} {\bibfnamefont {N.~M.}\ \bibnamefont
  {Qureshi}}, \bibinfo {author} {\bibfnamefont {M.}~\bibnamefont {Bourgoin}},
  \bibinfo {author} {\bibfnamefont {C.}~\bibnamefont {Baudet}}, \bibinfo
  {author} {\bibfnamefont {A.}~\bibnamefont {Cartellier}},\ and\ \bibinfo
  {author} {\bibfnamefont {Y.}~\bibnamefont {Gagne}},\ }\bibfield  {title}
  {\bibinfo {title} {Turbulent transport of material particles: An experimental
  study of finite size effects},\ }\href@noop {} {\bibfield  {journal}
  {\bibinfo  {journal} {Phys. Rev. Lett.}\ }\textbf {\bibinfo {volume} {99}}
  (\bibinfo {year} {2007})}\BibitemShut {NoStop}%
\bibitem [{\citenamefont {Brown}\ \emph {et~al.}(2009)\citenamefont {Brown},
  \citenamefont {Warhaft},\ and\ \citenamefont {Voth}}]{brown2009acceleration}%
  \BibitemOpen
  \bibfield  {author} {\bibinfo {author} {\bibfnamefont {R.~D.}\ \bibnamefont
  {Brown}}, \bibinfo {author} {\bibfnamefont {Z.}~\bibnamefont {Warhaft}},\
  and\ \bibinfo {author} {\bibfnamefont {G.~A.}\ \bibnamefont {Voth}},\
  }\bibfield  {title} {\bibinfo {title} {Acceleration statistics of neutrally
  buoyant spherical particles in intense turbulence},\ }\href@noop {}
  {\bibfield  {journal} {\bibinfo  {journal} {Phys. Rev. Lett.}\ }\textbf
  {\bibinfo {volume} {103}},\ \bibinfo {pages} {194501} (\bibinfo {year}
  {2009})}\BibitemShut {NoStop}%
\bibitem [{\citenamefont {Calzavarini}\ \emph {et~al.}(2009)\citenamefont
  {Calzavarini}, \citenamefont {Volk}, \citenamefont {Bourgoin}, \citenamefont
  {L\'ev\^eque}, \citenamefont {Pinton},\ and\ \citenamefont
  {Toschi}}]{calzavarini2009acceleration}%
  \BibitemOpen
  \bibfield  {author} {\bibinfo {author} {\bibfnamefont {E.}~\bibnamefont
  {Calzavarini}}, \bibinfo {author} {\bibfnamefont {R.}~\bibnamefont {Volk}},
  \bibinfo {author} {\bibfnamefont {M.}~\bibnamefont {Bourgoin}}, \bibinfo
  {author} {\bibfnamefont {E.}~\bibnamefont {L\'ev\^eque}}, \bibinfo {author}
  {\bibfnamefont {J.-F.}\ \bibnamefont {Pinton}},\ and\ \bibinfo {author}
  {\bibfnamefont {F.}~\bibnamefont {Toschi}},\ }\bibfield  {title} {\bibinfo
  {title} {Acceleration statistics of finite-sized particles in turbulent flow:
  the role of {Fax\'en} forces},\ }\href@noop {} {\bibfield  {journal}
  {\bibinfo  {journal} {J. Fluid Mech.}\ }\textbf {\bibinfo {volume} {630}},\
  \bibinfo {pages} {179} (\bibinfo {year} {2009})}\BibitemShut {NoStop}%
\bibitem [{\citenamefont {Volk}\ \emph {et~al.}(2011)\citenamefont {Volk},
  \citenamefont {Calzavarini}, \citenamefont {L\'ev\^eque},\ and\ \citenamefont
  {Pinton}}]{volk2011dynamics}%
  \BibitemOpen
  \bibfield  {author} {\bibinfo {author} {\bibfnamefont {R.}~\bibnamefont
  {Volk}}, \bibinfo {author} {\bibfnamefont {E.}~\bibnamefont {Calzavarini}},
  \bibinfo {author} {\bibfnamefont {E.}~\bibnamefont {L\'ev\^eque}},\ and\
  \bibinfo {author} {\bibfnamefont {J.-F.}\ \bibnamefont {Pinton}},\ }\bibfield
   {title} {\bibinfo {title} {Dynamics of inertial particles in a turbulent von
  {K\'arm\'an} flow},\ }\href@noop {} {\bibfield  {journal} {\bibinfo
  {journal} {J. Fluid Mech.}\ }\textbf {\bibinfo {volume} {668}},\ \bibinfo
  {pages} {223} (\bibinfo {year} {2011})}\BibitemShut {NoStop}%
\bibitem [{git(2021)}]{githubptvlyon}%
  \BibitemOpen
  \href@noop {} {\bibinfo {title} {{4D-PTV} package for turbulence, {Lyon}}},\
  \bibinfo {howpublished} {\url{https://github.com/turbulencelyon/4d-ptv}}
  (\bibinfo {year} {2021})\BibitemShut {NoStop}%
\bibitem [{\citenamefont {Machicoane}\ \emph {et~al.}(2019)\citenamefont
  {Machicoane}, \citenamefont {Aliseda}, \citenamefont {Volk},\ and\
  \citenamefont {Bourgoin}}]{machicoane2019simplified}%
  \BibitemOpen
  \bibfield  {author} {\bibinfo {author} {\bibfnamefont {N.}~\bibnamefont
  {Machicoane}}, \bibinfo {author} {\bibfnamefont {A.}~\bibnamefont {Aliseda}},
  \bibinfo {author} {\bibfnamefont {R.}~\bibnamefont {Volk}},\ and\ \bibinfo
  {author} {\bibfnamefont {M.}~\bibnamefont {Bourgoin}},\ }\bibfield  {title}
  {\bibinfo {title} {A simplified and versatile calibration method for
  multi-camera optical systems in {3D} particle imaging},\ }\href@noop {}
  {\bibfield  {journal} {\bibinfo  {journal} {Rev. Sci. Instrum.}\ }\textbf
  {\bibinfo {volume} {90}},\ \bibinfo {pages} {035112} (\bibinfo {year}
  {2019})}\BibitemShut {NoStop}%
\bibitem [{\citenamefont {Bourgoin}\ and\ \citenamefont
  {Huisman}(2020)}]{bourgoin2020using}%
  \BibitemOpen
  \bibfield  {author} {\bibinfo {author} {\bibfnamefont {M.}~\bibnamefont
  {Bourgoin}}\ and\ \bibinfo {author} {\bibfnamefont {S.~G.}\ \bibnamefont
  {Huisman}},\ }\bibfield  {title} {\bibinfo {title} {Using ray-traversal for
  {3D} particle matching in the context of particle tracking velocimetry in
  fluid mechanics},\ }\href@noop {} {\bibfield  {journal} {\bibinfo  {journal}
  {Rev. Sci. Instrum.}\ }\textbf {\bibinfo {volume} {91}},\ \bibinfo {pages}
  {085105} (\bibinfo {year} {2020})}\BibitemShut {NoStop}%
\bibitem [{\citenamefont {Ohmi}\ and\ \citenamefont
  {Li}(2000)}]{ohmi2000particle}%
  \BibitemOpen
  \bibfield  {author} {\bibinfo {author} {\bibfnamefont {K.}~\bibnamefont
  {Ohmi}}\ and\ \bibinfo {author} {\bibfnamefont {H.-Y.}\ \bibnamefont {Li}},\
  }\bibfield  {title} {\bibinfo {title} {Particle-tracking velocimetry with new
  algorithms},\ }\href@noop {} {\bibfield  {journal} {\bibinfo  {journal}
  {Meas. Sci. Technol.}\ }\textbf {\bibinfo {volume} {11}},\ \bibinfo {pages}
  {603} (\bibinfo {year} {2000})}\BibitemShut {NoStop}%
\bibitem [{\citenamefont {Veenman}\ \emph {et~al.}(2001)\citenamefont
  {Veenman}, \citenamefont {Reinders},\ and\ \citenamefont
  {Backer}}]{veenman2001resolving}%
  \BibitemOpen
  \bibfield  {author} {\bibinfo {author} {\bibfnamefont {C.~J.}\ \bibnamefont
  {Veenman}}, \bibinfo {author} {\bibfnamefont {M.~J.~T.}\ \bibnamefont
  {Reinders}},\ and\ \bibinfo {author} {\bibfnamefont {E.}~\bibnamefont
  {Backer}},\ }\bibfield  {title} {\bibinfo {title} {Resolving motion
  correspondence for densely moving points},\ }\href@noop {} {\bibfield
  {journal} {\bibinfo  {journal} {IEEE T. Pattern Anal.}\ }\textbf {\bibinfo
  {volume} {23}},\ \bibinfo {pages} {54} (\bibinfo {year} {2001})}\BibitemShut
  {NoStop}%
\bibitem [{\citenamefont {Veenman}\ \emph {et~al.}(2003)\citenamefont
  {Veenman}, \citenamefont {Reinders},\ and\ \citenamefont
  {Backer}}]{veenman2003establishing}%
  \BibitemOpen
  \bibfield  {author} {\bibinfo {author} {\bibfnamefont {C.~J.}\ \bibnamefont
  {Veenman}}, \bibinfo {author} {\bibfnamefont {M.~J.~T.}\ \bibnamefont
  {Reinders}},\ and\ \bibinfo {author} {\bibfnamefont {E.}~\bibnamefont
  {Backer}},\ }\bibfield  {title} {\bibinfo {title} {Establishing motion
  correspondence using extended temporal scope},\ }\href@noop {} {\bibfield
  {journal} {\bibinfo  {journal} {Artif. Intell.}\ }\textbf {\bibinfo {volume}
  {145}},\ \bibinfo {pages} {227} (\bibinfo {year} {2003})}\BibitemShut
  {NoStop}%
\bibitem [{\citenamefont {Ouellette}\ \emph
  {et~al.}(2006{\natexlab{b}})\citenamefont {Ouellette}, \citenamefont {Xu},\
  and\ \citenamefont {Bodenschatz}}]{ouellette2006quantitative}%
  \BibitemOpen
  \bibfield  {author} {\bibinfo {author} {\bibfnamefont {N.~T.}\ \bibnamefont
  {Ouellette}}, \bibinfo {author} {\bibfnamefont {H.}~\bibnamefont {Xu}},\ and\
  \bibinfo {author} {\bibfnamefont {E.}~\bibnamefont {Bodenschatz}},\
  }\bibfield  {title} {\bibinfo {title} {A quantitative study of
  three-dimensional {{{Lagrangian}}} particle tracking algorithms},\
  }\href@noop {} {\bibfield  {journal} {\bibinfo  {journal} {Exp. Fluids}\
  }\textbf {\bibinfo {volume} {40}},\ \bibinfo {pages} {301} (\bibinfo {year}
  {2006}{\natexlab{b}})}\BibitemShut {NoStop}%
\bibitem [{\citenamefont {Viggiano}\ \emph {et~al.}(2021)\citenamefont
  {Viggiano}, \citenamefont {Basset}, \citenamefont {Solovitz}, \citenamefont
  {Barois}, \citenamefont {Gibert}, \citenamefont {Mordant}, \citenamefont
  {Chevillard}, \citenamefont {Volk}, \citenamefont {Bourgoin},\ and\
  \citenamefont {Cal}}]{viggiano2021lagrangian}%
  \BibitemOpen
  \bibfield  {author} {\bibinfo {author} {\bibfnamefont {B.}~\bibnamefont
  {Viggiano}}, \bibinfo {author} {\bibfnamefont {T.}~\bibnamefont {Basset}},
  \bibinfo {author} {\bibfnamefont {S.}~\bibnamefont {Solovitz}}, \bibinfo
  {author} {\bibfnamefont {T.}~\bibnamefont {Barois}}, \bibinfo {author}
  {\bibfnamefont {M.}~\bibnamefont {Gibert}}, \bibinfo {author} {\bibfnamefont
  {N.}~\bibnamefont {Mordant}}, \bibinfo {author} {\bibfnamefont
  {L.}~\bibnamefont {Chevillard}}, \bibinfo {author} {\bibfnamefont
  {R.}~\bibnamefont {Volk}}, \bibinfo {author} {\bibfnamefont {M.}~\bibnamefont
  {Bourgoin}},\ and\ \bibinfo {author} {\bibfnamefont {R.~B.}\ \bibnamefont
  {Cal}},\ }\bibfield  {title} {\bibinfo {title} {{{Lagrangian}} diffusion
  properties of a free shear turbulent jet},\ }\href@noop {} {\bibfield
  {journal} {\bibinfo  {journal} {J. Fluid Mech.}\ }\textbf {\bibinfo {volume}
  {918}},\ \bibinfo {pages} {A25} (\bibinfo {year} {2021})}\BibitemShut
  {NoStop}%
\bibitem [{\citenamefont {Basset}\ \emph {et~al.}(2022)\citenamefont {Basset},
  \citenamefont {Viggiano}, \citenamefont {Barois}, \citenamefont {Gibert},
  \citenamefont {Mordant}, \citenamefont {Cal}, \citenamefont {Volk},\ and\
  \citenamefont {Bourgoin}}]{basset2022entrainment}%
  \BibitemOpen
  \bibfield  {author} {\bibinfo {author} {\bibfnamefont {T.}~\bibnamefont
  {Basset}}, \bibinfo {author} {\bibfnamefont {B.}~\bibnamefont {Viggiano}},
  \bibinfo {author} {\bibfnamefont {T.}~\bibnamefont {Barois}}, \bibinfo
  {author} {\bibfnamefont {M.}~\bibnamefont {Gibert}}, \bibinfo {author}
  {\bibfnamefont {N.}~\bibnamefont {Mordant}}, \bibinfo {author} {\bibfnamefont
  {R.~B.}\ \bibnamefont {Cal}}, \bibinfo {author} {\bibfnamefont
  {R.}~\bibnamefont {Volk}},\ and\ \bibinfo {author} {\bibfnamefont
  {M.}~\bibnamefont {Bourgoin}},\ }\bibfield  {title} {\bibinfo {title}
  {Entrainment, diffusion and effective compressibility in a self-similar
  turbulent jet},\ }\href@noop {} {\bibfield  {journal} {\bibinfo  {journal}
  {J. Fluid Mech.}\ }\textbf {\bibinfo {volume} {947}},\ \bibinfo {pages} {A29}
  (\bibinfo {year} {2022})}\BibitemShut {NoStop}%
\bibitem [{\citenamefont {Mordant}\ \emph
  {et~al.}(2004{\natexlab{a}})\citenamefont {Mordant}, \citenamefont
  {Crawford},\ and\ \citenamefont {Bodenschatz}}]{mordant2004experimental2}%
  \BibitemOpen
  \bibfield  {author} {\bibinfo {author} {\bibfnamefont {N.}~\bibnamefont
  {Mordant}}, \bibinfo {author} {\bibfnamefont {A.~M.}\ \bibnamefont
  {Crawford}},\ and\ \bibinfo {author} {\bibfnamefont {E.}~\bibnamefont
  {Bodenschatz}},\ }\bibfield  {title} {\bibinfo {title} {Experimental
  {{{Lagrangian}}} acceleration probability density function measurement},\
  }\href@noop {} {\bibfield  {journal} {\bibinfo  {journal} {Physica D}\
  }\textbf {\bibinfo {volume} {193}},\ \bibinfo {pages} {245} (\bibinfo {year}
  {2004}{\natexlab{a}})}\BibitemShut {NoStop}%
\bibitem [{\citenamefont {Ouellette}(2006)}]{ouellette2006phd}%
  \BibitemOpen
  \bibfield  {author} {\bibinfo {author} {\bibfnamefont {N.~T.}\ \bibnamefont
  {Ouellette}},\ }\emph {\bibinfo {title} {Probing the statistical structure of
  turbulence with measurements of tracer particle tracks}},\ \href@noop {}
  {Ph.D. thesis},\ \bibinfo  {school} {Cornell University} (\bibinfo {year}
  {2006})\BibitemShut {NoStop}%
\bibitem [{\citenamefont {Machicoane}\ \emph
  {et~al.}(2017{\natexlab{a}})\citenamefont {Machicoane}, \citenamefont
  {Huck},\ and\ \citenamefont {Volk}}]{machicoane2017estimating}%
  \BibitemOpen
  \bibfield  {author} {\bibinfo {author} {\bibfnamefont {N.}~\bibnamefont
  {Machicoane}}, \bibinfo {author} {\bibfnamefont {P.~D.}\ \bibnamefont
  {Huck}},\ and\ \bibinfo {author} {\bibfnamefont {R.}~\bibnamefont {Volk}},\
  }\bibfield  {title} {\bibinfo {title} {Estimating two-point statistics from
  derivatives of a signal containing noise: Application to auto-correlation
  functions of turbulent {{{Lagrangian}}} tracks},\ }\href@noop {} {\bibfield
  {journal} {\bibinfo  {journal} {Rev. Sci. Instrum.}\ }\textbf {\bibinfo
  {volume} {88}},\ \bibinfo {pages} {065113} (\bibinfo {year}
  {2017}{\natexlab{a}})}\BibitemShut {NoStop}%
\bibitem [{\citenamefont {Machicoane}\ \emph
  {et~al.}(2017{\natexlab{b}})\citenamefont {Machicoane}, \citenamefont
  {L\'opez-Caballero}, \citenamefont {Bourgoin}, \citenamefont {Aliseda},\ and\
  \citenamefont {Volk}}]{machicoane2017multi}%
  \BibitemOpen
  \bibfield  {author} {\bibinfo {author} {\bibfnamefont {N.}~\bibnamefont
  {Machicoane}}, \bibinfo {author} {\bibfnamefont {M.}~\bibnamefont
  {L\'opez-Caballero}}, \bibinfo {author} {\bibfnamefont {M.}~\bibnamefont
  {Bourgoin}}, \bibinfo {author} {\bibfnamefont {A.}~\bibnamefont {Aliseda}},\
  and\ \bibinfo {author} {\bibfnamefont {R.}~\bibnamefont {Volk}},\ }\bibfield
  {title} {\bibinfo {title} {A multi-time-step noise reduction method for
  measuring velocity statistics from particle tracking velocimetry},\
  }\href@noop {} {\bibfield  {journal} {\bibinfo  {journal} {Meas. Sci.
  Technol.}\ }\textbf {\bibinfo {volume} {28}},\ \bibinfo {pages} {107002}
  (\bibinfo {year} {2017}{\natexlab{b}})}\BibitemShut {NoStop}%
\bibitem [{\citenamefont {Wilczek}\ \emph {et~al.}(2011)\citenamefont
  {Wilczek}, \citenamefont {Daitche},\ and\ \citenamefont
  {Friedrich}}]{bib:wilczek2011}%
  \BibitemOpen
  \bibfield  {author} {\bibinfo {author} {\bibfnamefont {M.}~\bibnamefont
  {Wilczek}}, \bibinfo {author} {\bibfnamefont {A.}~\bibnamefont {Daitche}},\
  and\ \bibinfo {author} {\bibfnamefont {R.}~\bibnamefont {Friedrich}},\
  }\bibfield  {title} {\bibinfo {title} {On the velocity distribution in
  homogeneous isotropic turbulence: Correlations and deviations from
  {{Gaussianity}}},\ }\href {https://doi.org/10.1017/jfm.2011.39} {\bibfield
  {journal} {\bibinfo  {journal} {Journal of Fluid Mechanics}\ }\textbf
  {\bibinfo {volume} {676}},\ \bibinfo {pages} {191} (\bibinfo {year}
  {2011})},\ \Eprint {https://arxiv.org/abs/1107.0139} {arXiv:1107.0139
  [physics]} \BibitemShut {NoStop}%
\bibitem [{\citenamefont {Huisman}\ \emph {et~al.}(2013)\citenamefont
  {Huisman}, \citenamefont {Lohse},\ and\ \citenamefont
  {Sun}}]{bib:huisman2013}%
  \BibitemOpen
  \bibfield  {author} {\bibinfo {author} {\bibfnamefont {S.~G.}\ \bibnamefont
  {Huisman}}, \bibinfo {author} {\bibfnamefont {D.}~\bibnamefont {Lohse}},\
  and\ \bibinfo {author} {\bibfnamefont {C.}~\bibnamefont {Sun}},\ }\bibfield
  {title} {\bibinfo {title} {Statistics of turbulent fluctuations in
  counter-rotating taylor-couette flows},\ }\href
  {https://doi.org/10.1103/PhysRevE.88.063001} {\bibfield  {journal} {\bibinfo
  {journal} {Phys. Rev. E}\ }\textbf {\bibinfo {volume} {88}},\ \bibinfo
  {pages} {063001} (\bibinfo {year} {2013})}\BibitemShut {NoStop}%
\bibitem [{\citenamefont {Sy}\ \emph {et~al.}(2021)\citenamefont {Sy},
  \citenamefont {Diribarne}, \citenamefont {Rousset}, \citenamefont {Gibert},\
  and\ \citenamefont {Bourgoin}}]{bib:sy2021}%
  \BibitemOpen
  \bibfield  {author} {\bibinfo {author} {\bibfnamefont {F.}~\bibnamefont
  {Sy}}, \bibinfo {author} {\bibfnamefont {P.}~\bibnamefont {Diribarne}},
  \bibinfo {author} {\bibfnamefont {B.}~\bibnamefont {Rousset}}, \bibinfo
  {author} {\bibfnamefont {M.}~\bibnamefont {Gibert}},\ and\ \bibinfo {author}
  {\bibfnamefont {M.}~\bibnamefont {Bourgoin}},\ }\bibfield  {title} {\bibinfo
  {title} {Multiscale energy budget of inertially driven turbulence in normal
  and superfluid helium},\ }\href
  {https://doi.org/10.1103/PhysRevFluids.6.064604} {\bibfield  {journal}
  {\bibinfo  {journal} {Phys. Rev. Fluids}\ }\textbf {\bibinfo {volume} {6}},\
  \bibinfo {pages} {064604} (\bibinfo {year} {2021})}\BibitemShut {NoStop}%
\bibitem [{\citenamefont {Lumley}\ and\ \citenamefont
  {Newman}(1977)}]{lumley1977return}%
  \BibitemOpen
  \bibfield  {author} {\bibinfo {author} {\bibfnamefont {J.~L.}\ \bibnamefont
  {Lumley}}\ and\ \bibinfo {author} {\bibfnamefont {G.~R.}\ \bibnamefont
  {Newman}},\ }\bibfield  {title} {\bibinfo {title} {The return to isotropy of
  homogeneous turbulence},\ }\href@noop {} {\bibfield  {journal} {\bibinfo
  {journal} {J. Fluid Mech.}\ }\textbf {\bibinfo {volume} {82}},\ \bibinfo
  {pages} {161–178} (\bibinfo {year} {1977})}\BibitemShut {NoStop}%
\bibitem [{\citenamefont {Lumley}(1979)}]{lumley1979computational}%
  \BibitemOpen
  \bibfield  {author} {\bibinfo {author} {\bibfnamefont {J.~L.}\ \bibnamefont
  {Lumley}},\ }\bibfield  {title} {\bibinfo {title} {Computational modeling of
  turbulent flows}\ }(\bibinfo  {publisher} {Elsevier},\ \bibinfo {year}
  {1979})\ pp.\ \bibinfo {pages} {123--176}\BibitemShut {NoStop}%
\bibitem [{\citenamefont {Pope}(2000)}]{pope2000turbulent}%
  \BibitemOpen
  \bibfield  {author} {\bibinfo {author} {\bibfnamefont {S.~B.}\ \bibnamefont
  {Pope}},\ }\href@noop {} {\emph {\bibinfo {title} {Turbulent Flows}}}\
  (\bibinfo  {publisher} {Cambridge University Press},\ \bibinfo {year}
  {2000})\BibitemShut {NoStop}%
\bibitem [{\citenamefont {Sreenivasan}(1995)}]{bib:sreenivasan1995_PoF}%
  \BibitemOpen
  \bibfield  {author} {\bibinfo {author} {\bibfnamefont {K.~R.}\ \bibnamefont
  {Sreenivasan}},\ }\bibfield  {title} {\bibinfo {title} {{On the universality
  of the Kolmogorov constant}},\ }\href@noop {} {\bibfield  {journal} {\bibinfo
   {journal} {Phys. Fluids}\ }\textbf {\bibinfo {volume} {7}},\ \bibinfo
  {pages} {2778} (\bibinfo {year} {1995})}\BibitemShut {NoStop}%
\bibitem [{\citenamefont
  {Kolmogorov}(1941{\natexlab{b}})}]{kolmogorov1941dissiptation}%
  \BibitemOpen
  \bibfield  {author} {\bibinfo {author} {\bibfnamefont {A.~N.}\ \bibnamefont
  {Kolmogorov}},\ }\bibfield  {title} {\bibinfo {title} {Dissipation of energy
  in isotropic turbulence},\ }\href@noop {} {\bibfield  {journal} {\bibinfo
  {journal} {Dokl. Akad. Nauk SSSR}\ }\textbf {\bibinfo {volume} {32}},\
  \bibinfo {pages} {19} (\bibinfo {year} {1941}{\natexlab{b}})}\BibitemShut
  {NoStop}%
\bibitem [{\citenamefont {Frisch}(1995)}]{frisch1995turbulence}%
  \BibitemOpen
  \bibfield  {author} {\bibinfo {author} {\bibfnamefont {U.}~\bibnamefont
  {Frisch}},\ }\href@noop {} {\emph {\bibinfo {title} {Turbulence: The Legacy
  of A.N. Kolmogorov}}}\ (\bibinfo  {publisher} {Cambridge University Press},\
  \bibinfo {year} {1995})\BibitemShut {NoStop}%
\bibitem [{\citenamefont {Vassilicos}(2015)}]{vassilicos2015dissipation}%
  \BibitemOpen
  \bibfield  {author} {\bibinfo {author} {\bibfnamefont {J.~C.}\ \bibnamefont
  {Vassilicos}},\ }\bibfield  {title} {\bibinfo {title} {Dissipation in
  turbulent flows},\ }\href@noop {} {\bibfield  {journal} {\bibinfo  {journal}
  {Annu. Rev. Fluid Mech.}\ }\textbf {\bibinfo {volume} {47}},\ \bibinfo
  {pages} {95} (\bibinfo {year} {2015})}\BibitemShut {NoStop}%
\bibitem [{\citenamefont {Batchelor}(1951)}]{batchelor1951pressure}%
  \BibitemOpen
  \bibfield  {author} {\bibinfo {author} {\bibfnamefont {G.~K.}\ \bibnamefont
  {Batchelor}},\ }\bibfield  {title} {\bibinfo {title} {Pressure fluctuations
  in isotropic turbulence},\ }\href@noop {} {\bibfield  {journal} {\bibinfo
  {journal} {Math. Proc. Cambridge}\ }\textbf {\bibinfo {volume} {47}},\
  \bibinfo {pages} {359–374} (\bibinfo {year} {1951})}\BibitemShut {NoStop}%
\bibitem [{\citenamefont {Grossmann}\ \emph {et~al.}(1997)\citenamefont
  {Grossmann}, \citenamefont {Lohse},\ and\ \citenamefont
  {Reeh}}]{grossmann1997application}%
  \BibitemOpen
  \bibfield  {author} {\bibinfo {author} {\bibfnamefont {S.}~\bibnamefont
  {Grossmann}}, \bibinfo {author} {\bibfnamefont {D.}~\bibnamefont {Lohse}},\
  and\ \bibinfo {author} {\bibfnamefont {A.}~\bibnamefont {Reeh}},\ }\bibfield
  {title} {\bibinfo {title} {Application of extended self-similarity in
  turbulence},\ }\href@noop {} {\bibfield  {journal} {\bibinfo  {journal}
  {Phys. Rev. E}\ }\textbf {\bibinfo {volume} {56}},\ \bibinfo {pages} {5473}
  (\bibinfo {year} {1997})}\BibitemShut {NoStop}%
\bibitem [{\citenamefont {Yeung}\ and\ \citenamefont
  {Pope}(1989)}]{yeung1989lagrangian}%
  \BibitemOpen
  \bibfield  {author} {\bibinfo {author} {\bibfnamefont {P.~K.}\ \bibnamefont
  {Yeung}}\ and\ \bibinfo {author} {\bibfnamefont {S.~B.}\ \bibnamefont
  {Pope}},\ }\bibfield  {title} {\bibinfo {title} {{{Lagrangian}} statistics
  from direct numerical simulations of isotropic turbulence},\ }\href@noop {}
  {\bibfield  {journal} {\bibinfo  {journal} {J. Fluid Mech.}\ }\textbf
  {\bibinfo {volume} {207}},\ \bibinfo {pages} {531} (\bibinfo {year}
  {1989})}\BibitemShut {NoStop}%
\bibitem [{\citenamefont {Sawford}(1991)}]{sawford1991reynolds}%
  \BibitemOpen
  \bibfield  {author} {\bibinfo {author} {\bibfnamefont {B.~L.}\ \bibnamefont
  {Sawford}},\ }\bibfield  {title} {\bibinfo {title} {Reynolds number effects
  in {{{Lagrangian}}} stochastic models of turbulent dispersion},\ }\href@noop
  {} {\bibfield  {journal} {\bibinfo  {journal} {Phys. Fluids A}\ }\textbf
  {\bibinfo {volume} {3}},\ \bibinfo {pages} {1577} (\bibinfo {year}
  {1991})}\BibitemShut {NoStop}%
\bibitem [{\citenamefont {Viggiano}\ \emph {et~al.}(2020)\citenamefont
  {Viggiano}, \citenamefont {Friedrich}, \citenamefont {Volk}, \citenamefont
  {Bourgoin}, \citenamefont {Cal},\ and\ \citenamefont
  {Chevillard}}]{viggiano2020modelling}%
  \BibitemOpen
  \bibfield  {author} {\bibinfo {author} {\bibfnamefont {B.}~\bibnamefont
  {Viggiano}}, \bibinfo {author} {\bibfnamefont {J.}~\bibnamefont {Friedrich}},
  \bibinfo {author} {\bibfnamefont {R.}~\bibnamefont {Volk}}, \bibinfo {author}
  {\bibfnamefont {M.}~\bibnamefont {Bourgoin}}, \bibinfo {author}
  {\bibfnamefont {R.~B.}\ \bibnamefont {Cal}},\ and\ \bibinfo {author}
  {\bibfnamefont {L.}~\bibnamefont {Chevillard}},\ }\bibfield  {title}
  {\bibinfo {title} {Modelling {{{Lagrangian}}} velocity and acceleration in
  turbulent flows as infinitely differentiable stochastic processes},\
  }\href@noop {} {\bibfield  {journal} {\bibinfo  {journal} {J. Fluid Mech.}\
  }\textbf {\bibinfo {volume} {900}},\ \bibinfo {pages} {A27} (\bibinfo {year}
  {2020})}\BibitemShut {NoStop}%
\bibitem [{Note1()}]{Note1}%
  \BibitemOpen
  \bibinfo {note} {The present definition uses an arithmetic mean instead of a
  geometric mean, which proved to be efficient for computing $S_2^L(\tau )$
  \cite {dumont2021phd}. However the two estimates of the velocity variance
  give here close results and the second order structure function does not
  change much when changing the way the average is computed}\BibitemShut
  {NoStop}%
\bibitem [{\citenamefont {Dumont}(2021)}]{dumont2021phd}%
  \BibitemOpen
  \bibfield  {author} {\bibinfo {author} {\bibfnamefont {D.}~\bibnamefont
  {Dumont}},\ }\emph {\bibinfo {title} {Étude des échanges énergétiques en
  convection thermique turbulente}},\ \href@noop {} {Ph.D. thesis},\ \bibinfo
  {school} {\'Ecole Normale Supérieure de Lyon} (\bibinfo {year}
  {2021})\BibitemShut {NoStop}%
\bibitem [{\citenamefont {Yeung}\ \emph {et~al.}(2006)\citenamefont {Yeung},
  \citenamefont {Pope},\ and\ \citenamefont {Sawford}}]{yeung2006reynolds}%
  \BibitemOpen
  \bibfield  {author} {\bibinfo {author} {\bibfnamefont {P.~K.}\ \bibnamefont
  {Yeung}}, \bibinfo {author} {\bibfnamefont {S.~B.}\ \bibnamefont {Pope}},\
  and\ \bibinfo {author} {\bibfnamefont {B.~L.}\ \bibnamefont {Sawford}},\
  }\bibfield  {title} {\bibinfo {title} {Reynolds number dependence of
  {{{Lagrangian}}} statistics in large numerical simulations of isotropic
  turbulence},\ }\href@noop {} {\bibfield  {journal} {\bibinfo  {journal} {J.
  Turbul.}\ }\textbf {\bibinfo {volume} {7}},\ \bibinfo {pages} {N58} (\bibinfo
  {year} {2006})}\BibitemShut {NoStop}%
\bibitem [{\citenamefont {Biferale}\ \emph {et~al.}(2008)\citenamefont
  {Biferale}, \citenamefont {Bodenschatz}, \citenamefont {Cencini},
  \citenamefont {Lanotte}, \citenamefont {Ouellette}, \citenamefont {Toschi},\
  and\ \citenamefont {Xu}}]{biferale2008lagrangian}%
  \BibitemOpen
  \bibfield  {author} {\bibinfo {author} {\bibfnamefont {L.}~\bibnamefont
  {Biferale}}, \bibinfo {author} {\bibfnamefont {E.}~\bibnamefont
  {Bodenschatz}}, \bibinfo {author} {\bibfnamefont {M.}~\bibnamefont
  {Cencini}}, \bibinfo {author} {\bibfnamefont {A.~S.}\ \bibnamefont
  {Lanotte}}, \bibinfo {author} {\bibfnamefont {N.~T.}\ \bibnamefont
  {Ouellette}}, \bibinfo {author} {\bibfnamefont {F.}~\bibnamefont {Toschi}},\
  and\ \bibinfo {author} {\bibfnamefont {H.}~\bibnamefont {Xu}},\ }\bibfield
  {title} {\bibinfo {title} {{{Lagrangian}} structure functions in turbulence:
  {A} quantitative comparison between experiment and direct numerical
  simulation},\ }\href@noop {} {\bibfield  {journal} {\bibinfo  {journal}
  {Phys. Fluids}\ }\textbf {\bibinfo {volume} {20}},\ \bibinfo {pages} {065103}
  (\bibinfo {year} {2008})}\BibitemShut {NoStop}%
\bibitem [{\citenamefont {Sawford}\ and\ \citenamefont
  {Yeung}(2011)}]{sawford2011kolmogorov}%
  \BibitemOpen
  \bibfield  {author} {\bibinfo {author} {\bibfnamefont {B.~L.}\ \bibnamefont
  {Sawford}}\ and\ \bibinfo {author} {\bibfnamefont {P.~K.}\ \bibnamefont
  {Yeung}},\ }\bibfield  {title} {\bibinfo {title} {Kolmogorov similarity
  scaling for one-particle {{{Lagrangian}}} statistics},\ }\href@noop {}
  {\bibfield  {journal} {\bibinfo  {journal} {Phys. Fluids}\ }\textbf {\bibinfo
  {volume} {23}},\ \bibinfo {pages} {091704} (\bibinfo {year}
  {2011})}\BibitemShut {NoStop}%
\bibitem [{\citenamefont {Calzavarini}\ \emph {et~al.}(2012)\citenamefont
  {Calzavarini}, \citenamefont {Volk}, \citenamefont {Lévèque}, \citenamefont
  {Pinton},\ and\ \citenamefont {Toschi}}]{CALZAVARINI2012237}%
  \BibitemOpen
  \bibfield  {author} {\bibinfo {author} {\bibfnamefont {E.}~\bibnamefont
  {Calzavarini}}, \bibinfo {author} {\bibfnamefont {R.}~\bibnamefont {Volk}},
  \bibinfo {author} {\bibfnamefont {E.}~\bibnamefont {Lévèque}}, \bibinfo
  {author} {\bibfnamefont {J.-F.}\ \bibnamefont {Pinton}},\ and\ \bibinfo
  {author} {\bibfnamefont {F.}~\bibnamefont {Toschi}},\ }\bibfield  {title}
  {\bibinfo {title} {Impact of trailing wake drag on the statistical properties
  and dynamics of finite-sized particle in turbulence},\ }\href@noop {}
  {\bibfield  {journal} {\bibinfo  {journal} {Physica D: Nonlinear Phenomena}\
  }\textbf {\bibinfo {volume} {241}},\ \bibinfo {pages} {237} (\bibinfo {year}
  {2012})}\BibitemShut {NoStop}%
\bibitem [{\citenamefont {Sawford}\ and\ \citenamefont
  {Pinton}(2013)}]{sawford2013lagrangian}%
  \BibitemOpen
  \bibfield  {author} {\bibinfo {author} {\bibfnamefont {B.~L.}\ \bibnamefont
  {Sawford}}\ and\ \bibinfo {author} {\bibfnamefont {J.-F.}\ \bibnamefont
  {Pinton}},\ }\bibinfo {title} {{Ten Chapters in Turbulence}}\ (\bibinfo
  {publisher} {Cambridge University Press},\ \bibinfo {year} {2013})\ Chap.\
  \bibinfo {chapter} {A {{Lagrangian}} View of Turbulent Dispersion and
  Mixing}\BibitemShut {NoStop}%
\bibitem [{\citenamefont {Bec}\ and\ \citenamefont
  {Vallée}(2024)}]{Bec_Vallee_2024}%
  \BibitemOpen
  \bibfield  {author} {\bibinfo {author} {\bibfnamefont {J.}~\bibnamefont
  {Bec}}\ and\ \bibinfo {author} {\bibfnamefont {R.}~\bibnamefont {Vallée}},\
  }\bibfield  {title} {\bibinfo {title} {Homogeneous turbophoresis of heavy
  inertial particles in turbulent flow},\ }\href
  {https://doi.org/10.1017/jfm.2024.980} {\bibfield  {journal} {\bibinfo
  {journal} {J. Fluid Mech.}\ }\textbf {\bibinfo {volume} {999}},\ \bibinfo
  {pages} {A83} (\bibinfo {year} {2024})}\BibitemShut {NoStop}%
\bibitem [{\citenamefont {Lanotte}\ \emph {et~al.}(2013)\citenamefont
  {Lanotte}, \citenamefont {Biferale}, \citenamefont {Boffetta},\ and\
  \citenamefont {and}}]{lanotte2013new}%
  \BibitemOpen
  \bibfield  {author} {\bibinfo {author} {\bibfnamefont {A.}~\bibnamefont
  {Lanotte}}, \bibinfo {author} {\bibfnamefont {L.}~\bibnamefont {Biferale}},
  \bibinfo {author} {\bibfnamefont {G.}~\bibnamefont {Boffetta}},\ and\
  \bibinfo {author} {\bibfnamefont {F.~T.}\ \bibnamefont {and}},\ }\bibfield
  {title} {\bibinfo {title} {A new assessment of the second-order moment of
  {{Lagrangian}} velocity increments in turbulence},\ }\href
  {https://doi.org/10.1080/14685248.2013.839882} {\bibfield  {journal}
  {\bibinfo  {journal} {J. Turbul.}\ }\textbf {\bibinfo {volume} {14}},\
  \bibinfo {pages} {34} (\bibinfo {year} {2013})}\BibitemShut {NoStop}%
\bibitem [{\citenamefont {Monin}\ and\ \citenamefont
  {Yaglom}(1975)}]{monin1975statistical}%
  \BibitemOpen
  \bibfield  {author} {\bibinfo {author} {\bibfnamefont {A.~S.}\ \bibnamefont
  {Monin}}\ and\ \bibinfo {author} {\bibfnamefont {A.~M.}\ \bibnamefont
  {Yaglom}},\ }\href@noop {} {\emph {\bibinfo {title} {Statistical Fluid
  Mechanics: Mechanics of Turbulence, Volume 2}}}\ (\bibinfo  {publisher} {MIT
  Press},\ \bibinfo {year} {1975})\BibitemShut {NoStop}%
\bibitem [{\citenamefont {{La Porta}}\ \emph {et~al.}(2001)\citenamefont {{La
  Porta}}, \citenamefont {Voth}, \citenamefont {Crawford}, \citenamefont
  {Alexander},\ and\ \citenamefont {Bodenschatz}}]{laporta2001fluid}%
  \BibitemOpen
  \bibfield  {author} {\bibinfo {author} {\bibfnamefont {A.}~\bibnamefont {{La
  Porta}}}, \bibinfo {author} {\bibfnamefont {G.~A.}\ \bibnamefont {Voth}},
  \bibinfo {author} {\bibfnamefont {A.~M.}\ \bibnamefont {Crawford}}, \bibinfo
  {author} {\bibfnamefont {J.}~\bibnamefont {Alexander}},\ and\ \bibinfo
  {author} {\bibfnamefont {E.}~\bibnamefont {Bodenschatz}},\ }\bibfield
  {title} {\bibinfo {title} {Fluid particle accelerations in fully developed
  turbulence},\ }\href@noop {} {\bibfield  {journal} {\bibinfo  {journal}
  {Nature}\ }\textbf {\bibinfo {volume} {409}},\ \bibinfo {pages} {1017}
  (\bibinfo {year} {2001})}\BibitemShut {NoStop}%
\bibitem [{\citenamefont {Vedula}\ and\ \citenamefont
  {Yeung}(1999)}]{vedula1999similarity}%
  \BibitemOpen
  \bibfield  {author} {\bibinfo {author} {\bibfnamefont {P.}~\bibnamefont
  {Vedula}}\ and\ \bibinfo {author} {\bibfnamefont {P.~K.}\ \bibnamefont
  {Yeung}},\ }\bibfield  {title} {\bibinfo {title} {Similarity scaling of
  acceleration and pressure statistics in numerical simulations of isotropic
  turbulence},\ }\href@noop {} {\bibfield  {journal} {\bibinfo  {journal}
  {Phys. Fluids}\ }\textbf {\bibinfo {volume} {11}},\ \bibinfo {pages} {1208}
  (\bibinfo {year} {1999})}\BibitemShut {NoStop}%
\bibitem [{\citenamefont {Lee}\ \emph {et~al.}(2015)\citenamefont {Lee},
  \citenamefont {Gylfason}, \citenamefont {Perlekar},\ and\ \citenamefont
  {Toschi}}]{lee_2015}%
  \BibitemOpen
  \bibfield  {author} {\bibinfo {author} {\bibfnamefont {C.-M.}\ \bibnamefont
  {Lee}}, \bibinfo {author} {\bibfnamefont {A.}~\bibnamefont {Gylfason}},
  \bibinfo {author} {\bibfnamefont {P.}~\bibnamefont {Perlekar}},\ and\
  \bibinfo {author} {\bibfnamefont {F.}~\bibnamefont {Toschi}},\ }\bibfield
  {title} {\bibinfo {title} {Inertial particle acceleration in strained
  turbulence},\ }\href@noop {} {\bibfield  {journal} {\bibinfo  {journal} {J.
  Fluid Mech.}\ }\textbf {\bibinfo {volume} {785}},\ \bibinfo {pages} {31}
  (\bibinfo {year} {2015})}\BibitemShut {NoStop}%
\bibitem [{\citenamefont {Mordant}\ \emph
  {et~al.}(2004{\natexlab{b}})\citenamefont {Mordant}, \citenamefont
  {Crawford},\ and\ \citenamefont {Bodenschatz}}]{mordant2004threedimensional}%
  \BibitemOpen
  \bibfield  {author} {\bibinfo {author} {\bibfnamefont {N.}~\bibnamefont
  {Mordant}}, \bibinfo {author} {\bibfnamefont {A.~M.}\ \bibnamefont
  {Crawford}},\ and\ \bibinfo {author} {\bibfnamefont {E.}~\bibnamefont
  {Bodenschatz}},\ }\bibfield  {title} {\bibinfo {title} {Three-dimensional
  structure of the {{{Lagrangian}}} acceleration in turbulent flows},\
  }\href@noop {} {\bibfield  {journal} {\bibinfo  {journal} {Phys. Rev. Lett.}\
  }\textbf {\bibinfo {volume} {93}},\ \bibinfo {pages} {214501} (\bibinfo
  {year} {2004}{\natexlab{b}})}\BibitemShut {NoStop}%
\bibitem [{\citenamefont {Volk}\ \emph {et~al.}(2008)\citenamefont {Volk},
  \citenamefont {Calzavarini}, \citenamefont {Verhille}, \citenamefont {Lohse},
  \citenamefont {Mordant}, \citenamefont {Pinton},\ and\ \citenamefont
  {Toschi}}]{volk2008acceleration}%
  \BibitemOpen
  \bibfield  {author} {\bibinfo {author} {\bibfnamefont {R.}~\bibnamefont
  {Volk}}, \bibinfo {author} {\bibfnamefont {E.}~\bibnamefont {Calzavarini}},
  \bibinfo {author} {\bibfnamefont {G.}~\bibnamefont {Verhille}}, \bibinfo
  {author} {\bibfnamefont {D.}~\bibnamefont {Lohse}}, \bibinfo {author}
  {\bibfnamefont {N.}~\bibnamefont {Mordant}}, \bibinfo {author} {\bibfnamefont
  {J.-F.}\ \bibnamefont {Pinton}},\ and\ \bibinfo {author} {\bibfnamefont
  {F.}~\bibnamefont {Toschi}},\ }\bibfield  {title} {\bibinfo {title}
  {Acceleration of heavy and light particles in turbulence: Comparison between
  experiments and direct numerical simulations},\ }\href@noop {} {\bibfield
  {journal} {\bibinfo  {journal} {Physica D}\ }\textbf {\bibinfo {volume}
  {237}},\ \bibinfo {pages} {2084} (\bibinfo {year} {2008})}\BibitemShut
  {NoStop}%
\end{thebibliography}

%apsrev4-2.bst 2019-01-14 (MD) hand-edited version of apsrev4-1.bst
%Control: key (0)
%Control: author (8) initials jnrlst
%Control: editor formatted (1) identically to author
%Control: production of article title (0) allowed
%Control: page (0) single
%Control: year (1) truncated
%Control: production of eprint (0) enabled
%

\end{document}